\documentclass[prd,nofootinbib,twocolumn,superscriptaddress,preprintnumbers]{revtex4-1}
\usepackage{epsfig}
\usepackage{graphicx}
\usepackage{palatino}
\usepackage[english]{babel}
\usepackage{hyphenat}
\usepackage{amsmath}

\usepackage{amssymb}
\usepackage{mathtools}
\usepackage{tcolorbox}
\usepackage{mathrsfs}
\usepackage{slashed}
\usepackage{epstopdf}
\usepackage{xcolor}
\usepackage{booktabs}
\definecolor{lcolor}{rgb}{0.,0.0,0.}
\definecolor{citcolor}{rgb}{0,0.,0.5}
\usepackage[breaklinks,colorlinks,urlcolor=blue,citecolor=blue,linkcolor=blue]{hyperref}
\usepackage{multirow}
\usepackage{ltablex}
\usepackage{enumerate}
\newcommand{\beq}{\begin{eqnarray}}
\newcommand{\eeq}{\end{eqnarray}}

\def\dd{{\rm d}}

\newcommand{\bem}{\begin{multline}}
\newcommand{\eem}{\end{multline}}
\newcommand{\beg}{\begin{gather}}
\newcommand{\eeg}{\end{gather}}

\newcommand{\nn}{\nonumber\\}

\newcommand{\ben}{\begin{eqnarray*}}
\newcommand{\een}{\end{eqnarray*}}

\newcommand{\secn}[1]{Section~1}
\newcommand{\appn}[1]{Appendix~1}

\long\def\comment#1{ }

\def\jet{\text{jet}}

\def\and{\quad\text{and}\quad}

\begin{document}

\title{Bottom-up approach to describe groomed jet data in heavy-ion collisions}

\author{Liliana Apolinário}
 \email{liliana@lip.pt}
\affiliation{%
 LIP, Av. Prof. Gama Pinto, 2, P-1649-003 Lisboa, Portugal
}
\affiliation{
Instituto Superior Técnico (IST), Universidade de Lisboa,
Av. Rovisco Pais 1, P-1049-001 Lisboa, Portugal
}%

\author{Diogo Costa}
 \email{diogo.costa@ugr.es}

\author{Alba Soto-Ontoso}
\email{aontoso@ugr.es}
\affiliation{Departamento de Física Teórica y del Cosmos, Universidad de Granada, Campus de Fuentenueva, E-18071 Granada, Spain}

\setlength{\belowcaptionskip}{-10pt}

\renewcommand{\arraystretch}{1.4}

\begin{abstract}
The theoretical interpretation of jet observables in heavy-ion collisions is a complex task due to the intricate interplay of perturbative and non-perturbative effects. One way to reduce this complexity is to groom away soft, wide-angle radiation so that perturbative dynamics dominates. Even in this simplified scenario, there are competing explanations for the physical origin of the measured medium-induced modifications. In this paper, we present a minimal approach to compute groomed substructure observables. The core idea is to treat medium effects as an effective energy shift of the hard, vacuum-like substructure. This energy shift includes a gradual onset of colour decoherence effects and thus depends on the jet substructure itself. We first study a NLO-exact dijet configuration in vacuum and apply radiative energy-loss to the two subjets. We find that this minimal setup already captures the narrowing trend of groomed observables but it's not able to quantitatively describe the existing data. Next, we match the NLO matrix-element to a leading-logarithm accurate parton shower and perform a clustering algorithm to recover a two-prong system to which we again apply the energy-loss distribution. Despite its simplicity, the model results in a very good theory-to-data agreement (within $10\%$) for a broad range of observables including both ALICE and ATLAS kinematics. We also examine the discriminating power of groomed jet data in terms of colour decoherence effects and find that substructure-dependent energy loss yields an overall better agreement. 
\end{abstract}

\maketitle

\section{Introduction}\label{sec:Intro}
Ultra-relativistic heavy-ion collisions at the LHC create extreme conditions of temperature and energy density under which quarks and gluons become deconfined, forming the Quark–Gluon Plasma (QGP). Over the past two decades, multiple experimental signatures, including collective flow and suppression of quarkonia states and high transverse momentum hadrons, have confirmed the existence of this strongly interacting medium~\cite{Busza:2018rrf}. Among these probes, jets stand out as calibrated perturbative objects that traverse the QGP and encode, in their final-state structure, the interplay between partonic dynamics and medium properties~\cite{Wang:2025lct}. Early studies focused on inclusive jet suppression ($R_{AA}$)~\cite{ALICE:2013dpt,ATLAS:2014ipv,CMS:2016uxf,STAR:2020xiv} and dijet momentum imbalance~\cite{ATLAS:2010isq,CMS:2011iwn,STAR:2016dfv}, but these observables integrate over the complex radiation pattern of jets and are therefore not directly sensitive to the dynamics of parton–medium interactions, although they serve the purpose of quantifying out-of-cone radiation in terms of energy loss. Jet substructure observables, and in particular those employing grooming algorithms, provide a differential view of the jet evolution: they expose splittings with particular characteristics of interest inside the jet, e.g. `earliest', hardest $k_t$ or heavy-flavoured tagged, and suppress contamination from the underlying event and soft particles, offering a cleaner window into how the medium modifies the parton shower. The study of jet substructure in heavy-ion collisions has thus emerged as a powerful probe of QCD at high temperature~\cite{Cunqueiro:2021wls,Apolinario:2022vzg,Apolinario:2024equ}. 

Grooming algorithms such as SoftDrop~\cite{Dasgupta:2013ihk,Larkoski:2014wba} or Dynamical grooming~\cite{Mehtar-Tani:2019rrk} identify two subjets in the clustering sequence that satisfy a given `hardness' condition. All emissions at angles larger than this declustering step are groomed away. First measurements of groomed jets in heavy-ion collisions focused on a broad range of properties of this pair of subjets such as their momentum balance, opening angle or  mass~\cite{CMS:2017qlm,ALICE:2019ykw,STAR:2021kjt,ALargeIonColliderExperiment:2021mqf,ALICE:2024fip,ALICE:2024jtb}. All these measurements revealed a strongly quenched jet core with respect to the vacuum baseline. In other words, there is an enhancement of splittings at small opening angles relative to proton-proton collisions. A second generation of measurements is trying to find out the physical origin of this narrowing by correlating the kinematics of the two-prong system with other properties of the event such as the hard-scattering process itself~\cite{CMS:2024zjn} or other jet observables~\cite{ATLAS:2022vii,ATLAS:2023hso,ATLAS:2025lfb,CMS:2025gdw}. For instance, recent CMS data~\cite{CMS:2025gdw} has shown that the aforementioned narrowing gradually disappears when selecting sufficiently hard (large transverse momentum) prongs. Thus, for very energetic splittings, jet suppression is largely independent of the internal substructure~\cite{Cunqueiro:2023vxl}. 

From a theoretical perspective, several distinct physical mechanisms can influence groomed jet substructure in heavy-ion collisions, and current experimental precision does not yet allow a clean separation of their relative contributions. For instance, quark- and gluon-initiated jets are expected to lose energy in the QGP differently due to their distinct fragmentation patterns and overall colour charge. This can modify the effective quark/gluon composition of the reconstructed jet sample and thereby generate an apparent narrowing of groomed jets~\cite{Spousta:2015fca,Ringer:2019rfk}. Medium-induced modifications of the splitting dynamics themselves can also affect groomed observables~\cite{Chien:2016led}. In addition, medium response and recoil transport energy to large angles or low momenta, indirectly reshaping the reconstructed jet substructure~\cite{Wang:2022yrp,Milhano:2017nzm}. Finally, colour-coherence effects and the finite resolution of the medium can bias the relative suppression of narrow and wide jets through an angle-dependent pattern of energy loss~\cite{Mehtar-Tani:2010ebp,Mehtar-Tani:2011hma,Casalderrey-Solana:2011ule,Mehtar-Tani:2011vlz,Casalderrey-Solana:2012evi,Apolinario:2014csa,Abreu:2024wka}. In practice, these mechanisms may coexist, and disentangling their interplay remains an open challenge.

A possible interpretation, often invoked in analytic estimates~\cite{Mehtar-Tani:2016aco,Caucal:2021cfb,Pablos:2022mrx} and jet quenching Monte Carlo models (such as \texttt{JetMed}~\cite{Caucal:2018dla,Caucal:2019uvr}, the Hybrid Strong/Weak Coupling Model~\cite{Casalderrey-Solana:2014bpa,Hulcher:2017cpt,Casalderrey-Solana:2019ubu,Kudinoor:2025gao}, and \texttt{JetScape}~\cite{JETSCAPE:2023hqn}) to describe groomed jet measurements in heavy-ion collisions, is that of an early, vacuum-like regime of the jet evolution. Subsequent interactions with the QGP reduce the jet transverse momentum through energy loss and medium-induced broadening. In this scenario, selection biases and colour-coherence effects can already generate an apparent narrowing of the groomed core without requiring explicit modifications of the splitting kernel~\cite{Mehtar-Tani:2016aco,Caucal:2021cfb,Pablos:2022mrx}. However, the interpretation of groomed-jet measurements in terms of these mechanisms is only as reliable as the underlying vacuum baseline. In particular, omitting higher-order corrections can bias the vacuum baseline and thus distort the apparent magnitude of medium-induced effects. Before attributing residual differences between data and theory to quenching mechanisms, it is therefore essential to at least quantify the impact of next-to-leading order (NLO) corrections on the vacuum baseline, and to establish the level of perturbative accuracy required for a robust comparison to groomed-jet observables.

In this paper, we address this question by following a bottom-up strategy. Namely, we compute groomed substructure observables by dressing hard, vacuum-like jet substructure with radiative energy loss. We build our vacuum skeleton by matching exact NLO hard matrix elements to a leading-logarithmic (LL) parton shower, using  \texttt{POWHEG-BOX}\,\cite{Nason:2004rx,Alioli:2010xd,Alioli:2010xa} interfaced with \texttt{PYTHIA8}\,\cite{Bierlich:2022pfr}. Then, after clustering and grooming, we apply radiative energy loss to the two-prong system according to the semi-analytic calculation of the in-medium energy loss probability distribution presented in Ref.~\cite{Mehtar-Tani:2017ypq}.  Thus, in this framework, medium effects enter only through an energy-loss probability that depends on the opening angle of the splitting, without explicit modifications of the splitting kernel. This setup is not intended as a full-fledged event generator, but rather as a controlled environment in which to (i) quantify the size of NLO corrections on groomed-jet observables and on the extracted medium-induced effects (e.g, via a transport coefficient $\hat{q}$), and (ii) test the role of colour coherence by contrasting the energy-loss calculation of Ref.~\cite{Mehtar-Tani:2017ypq} with an energy-loss scenario in which the amount of energy lost uniquely depends on the medium properties and the colour charge of the jet initiator, i.e., it does not depend on the jet substructure. For that, we present theory-to-data comparisons for a suite of groomed observables for both ALICE and ATLAS kinematics.

This paper is organized as follows. We begin by defining the observables that we compute together with the foundations of our bottom-up approach in Section~\ref{sec:model}. The proton-proton baseline is studied in Sec.~\ref{sec:vacuum} including a discussion on the size of NLO corrections. Quantitative results for heavy-ion collisions can be found in Sec.~\ref{sec:medium}. We end up with a brief summary of our results and some comments on how to systematically improve the calculation in Sec.~\ref{sec:conclusions}. Technical details can be found in Appendices~\ref{app:Psing_derivation} and \ref{app:simulation}.

\section{Groomed jet susbtructure in heavy-ions}
\label{sec:model}
Our goal is to calculate groomed substructure observables in heavy-ion collisions encoding medium effects in an energy loss probability distribution. Before entering into the precise definition of the
model, let us briefly comment on the definition of the observables that we compute and their kinematic regime. For the medium description we adopt the brick approximation in which the medium is homogenous, isotropic and static with a fixed length $L=4$ fm, which roughly corresponds to a central PbPb collision~\cite{Mehtar-Tani:2021fud}. 
\subsection{Observables definition}
Events are clustered with the anti-$k_t$ algorithm~\cite{Cacciari:2008gp,Cacciari:2011ma} using a jet radius of $R$ within some fiducial cuts on transverse momentum $p_{t,{\rm jet}}$ and rapidity $y_{\rm jet}$. This subset of jets is then reclustered with the Cambridge/Aachen algorithm~\cite{Dokshitzer:1997in,Wobisch:1998wt} so as to reorder the branching sequence in decreasing angles. The SoftDrop (SD) algorithm is subsequently applied to the reclustered jet, until it finds the first pair of subjets, $j_1$ and $j_2$, with transverse momenta $p_{t,j_1}$ and $p_{t,j_2}$ that satisfy the grooming condition
\begin{equation}\label{eq:zg}
z_g \equiv \frac{\min(p_{t,j_1},p_{t,j_2})}{p_{t,j_1}+p_{t,j_2}} > z_{\rm cut}\, .
\end{equation}
Once the SD condition is satisfied, the two subjets at that position in the angular-ordered tree, $j_1$ and $j_2$, are used to compute a series of observables. We consider the splitting fraction itself, see Eq.~\eqref{eq:zg},
the groomed opening angle:
\begin{equation}\label{eq:thetag}
    \theta_g = \frac{\sqrt{(y_i-y_j)^2 + (\phi_i-\phi_j)^2}}{R} \equiv \frac{r_g}{R} \, ,
\end{equation} 
with $y_i$ and $\phi_i$ denoting rapidity and azimuthal angle of each subjet, and the groomed jet mass:
\begin{equation}\label{eq:mg}
    m_g = \sqrt{E^2_{\rm{g,jet}} - p^2_{\rm{g,jet}}} \, ,
\end{equation}
where $E_{\rm{g,jet}}$ and $p_{\rm{g,jet}}$ are the energy and total momentum of the groomed jet, respectively. Finally, the groomed angularities are defined as
\begin{equation}\label{eq:lambdag}
    \lambda^{\kappa}_{\alpha, g}=\!\sum_{i \in j_1,j_2} \left( \dfrac{p_{t,i}}{p_{t,{\rm jet}}} \right)^\kappa \left(\dfrac{\sqrt{(y_i-y_{\rm jet})^2+(\phi_i-\phi_{\rm jet})^2}}{R} \right)^\alpha \, ,
\end{equation}
where the sum runs over the constituents of subjets $j_1$ and $j_2$. We fix $\kappa=1$ in the rest of the manuscript but explore a few values of $\alpha$. The fiducial cuts for the two experimental configurations that we study are presented in table \ref{table:fiducial_cuts}.
\begin{table}[t]
	\centering\small
	\begin{tabular}{|c|c|c|c|c|}
		\hline
		Dataset & Particles & $p_{t,\rm jet}$ [GeV] & (Pseudo) Rapidity& R  \\
		\hline \text{ALICE~\cite{ALargeIonColliderExperiment:2021mqf,ALICE:2024jtb}} & \text{charged} & $[60, 80]$ & $|\eta_{\rm jet}| < 0.7$  & 0.2 \\ 
        ATLAS~\cite{ATLAS:2022vii} & \text{inclusive} & $[158, 315]$ &  $|y_{\rm jet}|< 2.1$ & $0.4$ \\
        \hline
	\end{tabular}
	\caption{\label{table:fiducial_cuts} Kinematic configurations of the experimental data studied in this paper.}
\end{table}
\subsection{Basic model description}
\label{sec:fixed-order}
Let us denote $\chi$ any of the observables described in the previous section. The SD condition given by Eq.~\eqref{eq:zg} tends to select splittings with short formation times (hard emissions), $t_f\sim 2/(zp_t\theta)$, since for typical values of $z_{\rm cut}=0.2$, $p_{t,\rm{jet}}=200$ GeV and $R=0.4$ we find $t_f<1$ fm. The characteristic timescale of medium effects is given by the so-called decoherence time $t_d \sim [4/(\hat q\theta^2)]^{1/3}$, where $\hat q$ is the so-called quenching parameter. In this work we focus on the limit in which $t_f\ll t_d$, i.e., when the time-scale of medium modifications is parametrically larger than the production of the hard pair. Based on this separation of scales we then assume that the SD splitting happens instantaneously and as in vacuum. This approximation allows us to isolate the role of medium-induced energy loss without entangling it with the detailed structure of the parton shower, which is known to affect the magnitude of jet suppression~\cite{Andres:2024egc}\footnote{A complete treatment of in-medium shower evolution lies beyond the scope of this minimal model.}. Thus, at lowest order in perturbative QCD, $\chi$ can be computed from the $\mathcal{O}(\alpha_s)$ dijet cross-section, corresponding to a $pp\to jj$ configuration containing at most one emission. We first cluster the resulting three-parton final state, which produces jets with no more than two constituents within our cuts. To study groomed-jet substructure, we retain only those jets with exactly two partons so that the SD procedure acts on a two-prong configuration. Each jet is then tagged according to its net flavour: jets containing one quark (or antiquark) are labelled as quark-initiated, while those with zero net flavour are labelled as gluon-initiated. This way, we associate each hard-scattering configuration to either a $gg$ or $qg$ dipole configuration. All along we disregard $q\bar q$ and $qq'$ dipoles since they are subleading in the SD phase space relevant for this study. 

Let us ignore additional vacuum radiation off this QCD dipole and discuss the medium modification of the two-parton system. The only medium effect that we consider is radiative energy loss mediated by soft medium-induced gluons in the large number of colours (large-$N_c$) approximation. The kinematics of the hard splitting is then unaltered, but the total yield of events is modified by energy loss. The corresponding calculation for the energy loss distribution of the two-parton system was presented in Ref.~\cite{Mehtar-Tani:2017ypq}. This calculation has never been applied to jet subtructure observables and we do so for the first time in this paper. Following Ref.~\cite{Mehtar-Tani:2017ypq}, the energy loss probability distribution for a dipole with opening angle $\theta_{12}$ triggered by a parent parton in colour representation $\text{R}$ is given by
\begin{align}\label{eq:P2R}
    \nonumber
    P_2^\text{R}(\varepsilon,\theta_{12}, L) &= \int_0^\infty \dd \varepsilon_1 \int_0^\infty \dd\varepsilon_2 P_1^\text{R}(\varepsilon_1,L) P_\text{sing}(\varepsilon_2, \theta_{12},L)\nn 
    & \times \delta(\varepsilon - \varepsilon_1 - \varepsilon_2),
\end{align}
where $P^\text{R}_1$ is the energy loss distribution of a single parton given by
\begin{equation}\label{eq:P1}
P_1^{\text R}(\varepsilon, L) = \sqrt{2 \frac{\omega_s^{\text R}}{\varepsilon^3}}\exp\left[ -2 \pi \frac{\omega_s^{\text R}}{\varepsilon}\right], \quad \omega_s^{\text R} = \frac{\alpha^{\rm med}_s C_\text{R}}{2\pi}\hat q L^2\, ,
\end{equation}
where $C_{\text R}=C_{\text A}=3$ for gluon-initiated splittings and $C_{\text R}=C_{\text F}=C_{\text A}/2$ for quark-initiated (using the large-$N_c$ limit). We fix the in-medium strong coupling to $\alpha^{\rm med}_s=0.24$~\cite{Caucal:2019uvr}. 
The $P_\text{sing}$ term in Eq.~\eqref{eq:P2R} describes the energy loss distribution of a colour-singlet dipole and reads\footnote{Note that we have managed to reduce the number of numerical integrations compared to Ref.~\cite{Mehtar-Tani:2017ypq} as detailed in Appendix~\ref{app:Psing_derivation}.}
\begin{align}\label{eq:Psing-simple}
    P_\text{sing}(\varepsilon,\theta_{12},L) &= P_1^\text{A}(\varepsilon, L) 
    - 
    \int_0^L \dd t \, P_1^\text{A}(\varepsilon, L-t)F(\varepsilon, L-t)\nn 
    & \times \left[ 1- \Delta_\text{med}(t,\theta_{12})\right]\, ,
\end{align}
where $F(\varepsilon, L-t)$ is given in Eq.~\eqref{eq:f-def}.

The physical interpretation of Eqs.~\eqref{eq:P2R} and \eqref{eq:Psing-simple} is quite transparent. The $P_\text{sing}$ term is controlled by the so-called decoherence parameter $\Delta_{\rm med} (t,\theta_{12})$ defined in Eq.~\eqref{eq:deltaMed} and bounded between $0$ and $1$. This term is proportional to the ratio between $\theta_{12}$ and a characteristic angle $\theta_d\sim (\hat q t^3)^{-1/2}$ where $t$ is the propagation time of the dipole inside the medium. The maximum propagation time set by the medium length $L$ yields a minimum value of $\theta_d$, known as critical angle, i.e.,
\begin{equation}\label{eq:thetac-def}
\theta_d\leq \theta_c \sim (\hat q L^3)^{-1/2}\, .
\end{equation} 
If $\theta_{12}\gg \theta_d$ the decoherence factor approaches $1$ and $P_\text{sing}\approx P_1^\text{A}$. Plugging this result into Eq.~\eqref{eq:P2R} we find that, in this scenario of broad splittings, the antenna loses energy as two-independent colour charges as given by the product $P_1^\text{R}P_1^\text{A}$. This limit corresponds to the so-called incoherent regime. In turn, when $\theta_{12}\ll \theta_d$ the decoherence parameter evaluates to $\Delta_{\rm med} (t,\theta_{12})\approx 0$ so that $P_\text{sing}\approx \delta(\varepsilon)$. In this coherent regime the medium only resolves the global colour charge of the antenna and thus Eq.~\eqref{eq:P2R} reduces to $P_2^\text{R}\approx P_1^\text{R}$. The physical interpretation of these limiting regimes in terms of colour randomization by the medium has been extensively discussed in the jet quenching literature~\cite{Mehtar-Tani:2010ebp,Mehtar-Tani:2011hma,Casalderrey-Solana:2011ule,Mehtar-Tani:2011vlz,Casalderrey-Solana:2012evi,Apolinario:2014csa,Abreu:2024wka}.

The impact of energy loss on any groomed observable $\chi$ can be assessed at this fixed-order level by computing the corresponding in-medium differential distribution as 
\begin{align}
    \label{eq:v-med}
    \frac{\dd \sigma_{\rm med}}{\dd \chi\dd p_{t,{\rm jet}}}&=\int_0^\infty\dd\varepsilon \, \frac{\dd\sigma^{\text R}_{\rm vac}}{\dd \chi\dd (p_{t,\rm jet}+\varepsilon)} P_2^{\text R}(\varepsilon,\theta_{12},L)\, .
\end{align}
\begin{figure}
\centering
\includegraphics[width=\columnwidth]{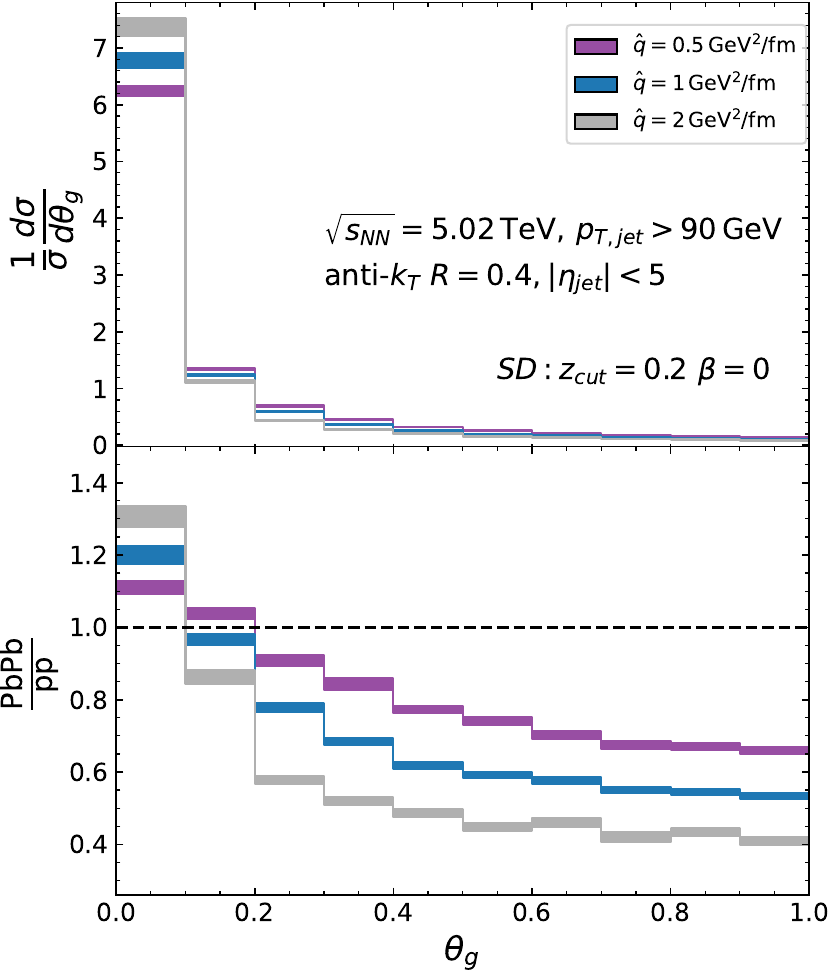}
\caption{Top: Self-normalized in-medium $\theta_g$-distribution for different values of $\hat q$. The vacuum cross-section was computed at NLO using \texttt{MADGRAPH} (fixed-order mode)~\cite{Alwall:2014hca} and then shifted in $p_t$ using the energy-loss distribution $P^{\text R}_2$ as discussed in the main text. Bottom: ratio to the vacuum baseline. The error band in each line represents an uncertainty estimate as provided by \texttt{MADGRAPH}.} 
\label{fig:thg-nlo}
\end{figure}
That is, we shift the $p_t$ of the vacuum differential cross-section by an amount $\varepsilon$ that depends on the opening angle of the QCD dipole as well as on the colour representation without modifying neither the relative angle nor the energy sharing of the splitting. 

We illustrate this mechanism in Fig.~\ref{fig:thg-nlo}, where the self-normalized version of Eq.~\eqref{eq:v-med} is evaluated for $\chi=\theta_g$, and for different values of $\hat q$, depicted by different coloured bands. The vacuum cross-section is computed at NLO using \texttt{MADGRAPH}~\cite{Alwall:2014hca}~\footnote{Note that we first use \texttt{MADGRAPH} to generate fixed-order results and then switch to \texttt{POWHEG-BOX-V2} when interfacing to \texttt{PYTHIA8} in Sec.~\ref{sec:all-orders}. The reason for using \texttt{MADGRAPH} in this section is simply because it gives access to the flavour structure of the event (needed for our energy loss prescription) while \texttt{POWHEG-BOX-V2} in fixed-order mode does not provide this feature.} and setting $p_{t,\rm jet}>90$ GeV and $|\eta_{\rm jet}|<5$, with jets clustered using the anti-$k_t$ algorithm with $R=0.4$. We choose the \texttt{CT14nlo} PDF set~\cite{Dulat:2015mca}. First, we note that the sharply peaked behaviour of the $\theta_g$-distribution around $\theta_g\sim 0$ is driven by QCD's collinear singularity, i.e., $\rm d\sigma/\rm d\theta_g\sim1/\theta_g$. This collinear sensitivity is due to both our choice of $\beta=0$ in the SD condition and the fixed-order nature of the calculation. Regarding medium effects, we observe a $10\%-30\%$ enhancement of collinear configurations in the first bin of the $\theta_g$-distribution when comparing PbPb and $pp$ collisions (bottom panel of Fig.~\ref{fig:thg-nlo}). The overall magnitude of this modification naturally increases with the value of $\hat q$. In particular, the characteristic angular scale that controls the transition between enhancement and suppression is set by the critical angle, which depends on the $\hat q$ value, $\theta_c \sim (\hat q L^3)^{-1/2}$, see Eq.~\eqref{eq:thetac-def}. Since narrow splittings ($\theta_g \ll \theta_c$) lose less energy than wide ones ($\theta_g \gg \theta_c$), the location at which the medium-to-vacuum ratio transitions from suppression to enhancement is expected to track $\theta_c$. While this effect is  not sharply resolved in Fig.~\ref{fig:thg-nlo}, the results remain fully consistent with the expected parametric trend. We also emphasize that, due to the self-normalization of the distributions, the PbPb-to-$pp$ ratio primarily highlights shape modifications rather than differences in the absolute cross section. The inclusive jet yield in PbPb collisions is, of course, suppressed relative to $pp$ as a consequence of energy loss.

Although not shown in the plot, we have checked that a fully coherent version of energy-loss, obtained by replacing $P^{\rm R}_2$ with $P^{\rm R}_1$ in Eq.~\eqref{eq:v-med}, results into a significantly milder narrowing of the $\theta_g$-distribution. In fact, we find that the medium-to-vacuum ratio for $P^{\rm R}_1$ becomes flat as a function of $\theta_g$, while the PbPb cross-section remains suppressed with respect to vacuum. In general, for any observable, $\chi$, computed within a $p_t$ range and a fully coherent energy-loss prescription $P^{\rm R}_1$, we would only expect a non-monotonic medium modification across observable values if two conditions are satisfied: (i) the jet sample is an admixture of quark and gluon initiated jets, and (ii) the observable's shape depends on the flavour of the jet initiator. Since $P^{\rm A}_1 > P^{\rm F}_1$, the contribution of gluon-initiated jets to the quenched observable diminishes with respect to vacuum. For instance, in the case of Fig.~\ref{fig:thg-nlo}, we found a gluon fraction in vacuum of around $70\%$ for these fiducial cuts. After quenching, this fraction goes down to $60\%-45\%$, depending on the $\hat q$ value. However, the shape of the NLO vacuum baseline $\theta_g$-distribution turned out to be rather independent of the jet-initiator flavour. This explains why $P^{\rm R}_1$ yields an in-medium $\theta_g$-distribution which is equally suppressed with respect to vacuum for all $\theta_g$ values. 

So far, our calculation of groomed observables is based on a vacuum antenna that loses energy. This minimal setup is enough to generate an enhancement of collinear splittings whose size, depending on the value of $\hat q$, can become comparable to what is observed experimentally. However, the previous treatment omits additional vacuum radiation off the energetic antenna legs. These emissions will enter the calculation of the observable as a Sudakov form factor that will resum logarithmically-enhanced contributions. In the next section, we discuss an heuristic extension of Eq.~\eqref{eq:v-med} from the fixed-order scenario to full jets. 

\subsection{From fixed-order to all-orders}
\label{sec:all-orders}
Formally, extending the previous fixed-order discussion to also account for vacuum radiation is far from trivial. One would need to resum vacuum-like emissions which are resolved by the medium and thus contribute to the total energy loss that a jet experiences. This resummation leads to a non-linear evolution equation that has been used to describe jet-$p_t$ suppression in a series of papers using the so-called collimator formalism~\cite{Mehtar-Tani:2017web,Mehtar-Tani:2021fud,Takacs:2021bpv,Mehtar-Tani:2024jtd,Pablos:2025cli}. However, the extension of this framework to groomed substructure observables has not yet been carried out. We therefore adopt a simpler, heuristic approach consistent with the working limits ($t_f\ll t_d$) adopted in this work. The idea is to replace the fixed-order vacuum cross-section by its all-orders version to account for vacuum-like evolution and from it reconstruct an antenna to which we apply $P^\text{R}_{2}$. As such, energy loss is applied as an afterburner, i.e, medium effects do not act on the shower evolution itself; they modify only the final, groomed two-prong kinematics through a transverse-momentum shift of the jet.

In practice, we first match the $\mathcal{O}(\alpha_s)$ vacuum cross section in Eq.~\eqref{eq:v-med} to a parton shower algorithm with leading-logarithmic (LL) accuracy using the \texttt{POWHEG+PYTHIA8} interface. This generates a multi-parton (or multi-hadron) final state that we cluster into jets using the anti-$k_t$ algorithm. We obtain the SoftDrop splitting following the standard procedure. This pair of subjets is then identified as the two offsprings of the antenna to which we need to assign a colour representation. To determine the colour representation $\text{R}$, we extend the coincidence procedure introduced in
Ref.~\cite{Spousta:2015fca} for leading order (LO) hard-scattering matrix elements to NLO. At LO, the jet flavour corresponds to that of the hard-scattering parton, $a$, which minimizes the value of $\Delta R_{a,\rm jet} = \sqrt{(y_a-y_{\rm jet})^2+(\phi_a-\phi_{\rm jet})^2}$, i.e., the angular separation between the hard-scattering parton and the jet that also satisfies $\Delta R < R$. At NLO, where the hard matrix element contains three partons, $(a,b,c)$, we compute their distances to the jet axis ($\Delta R_{a,\rm jet}, \Delta R_{b,\rm jet}, \Delta R_{c,\rm jet}$) and distinguish three cases:
\begin{enumerate}[(i)] 
    \item if two of these distances are smaller than the jet radius, e.g. $\Delta R_{a,\rm jet}$ and $\Delta R_{b,\rm jet}$, we assign the flavour of such partons to the pair of subjets and find the colour representation of the antenna $\text R$ as $a=q,b=g\to \text R=\text F$ and $a=b=g\to \text R=\text A$;
    \item if only one of these distances, e.g. $\Delta R_{a,\rm jet}$, is smaller than the jet radius we identify the colour representation of the antenna to be the flavour of $a$; 
    \item if all distances are larger than the jet radius, we skip the jet (this occurs rarely).
\end{enumerate}
We have checked that this procedure yields quark/gluon fractions compatible with those determined at fixed-order  level. Based on the previous procedure we then apply energy loss to the jet according to $P^\text{R}_{2}$. We emphasize again that the kinematics of the groomed splitting remains unaltered through this procedure and that the only medium modification is an overall transverse momentum shift to the jet applied via $P_2^{\text R}$. 

\begin{figure}
\centering
\includegraphics[width=\columnwidth]{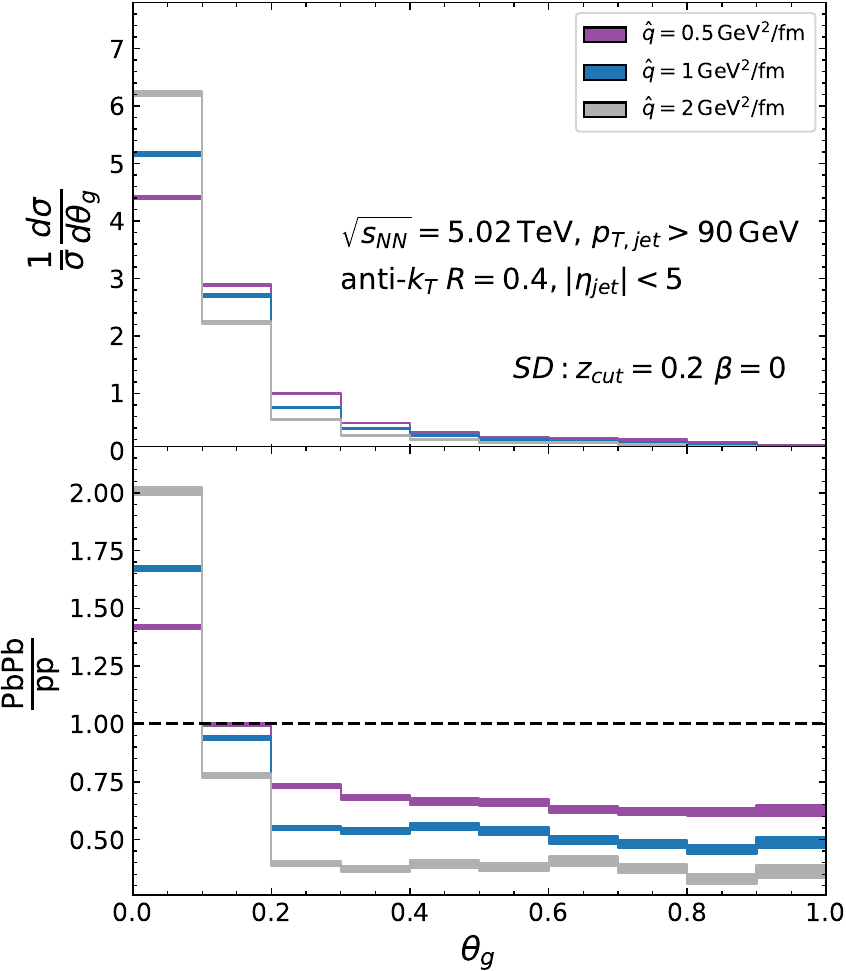}
\caption{Same as Fig.~\ref{fig:thg-nlo} but using \texttt{POWHEG+PYTHIA8} for the vacuum cross section.}
\label{fig:thg-nlops}
\end{figure}

In Fig.~\ref{fig:thg-nlops} we present the $\theta_g$-distribution including the LL-resummed vacuum emissions from the parton shower. Neither hadronization effects, nor underlying event activity is included in the simulation at this stage. The $\theta_g$-distribution at all-orders is broader compared to the fixed-order case shown in Fig.~\ref{fig:thg-nlo} as a result of the Sudakov form factor. 
At the level of the medium-to-vacuum ratio, we observe a similar qualitative trend as in Fig.~\ref{fig:thg-nlo}. Also, the turning point between enhancement and suppression is located at roughly the same values of $\theta_g$ as in the fixed-order case. However, there are quantitative differences when comparing the size of medium modifications. In particular, the collinear enhancement is more pronounced when accounting for parton shower effects.  
\begin{figure*}
\centering
\includegraphics[width=0.33\textwidth]{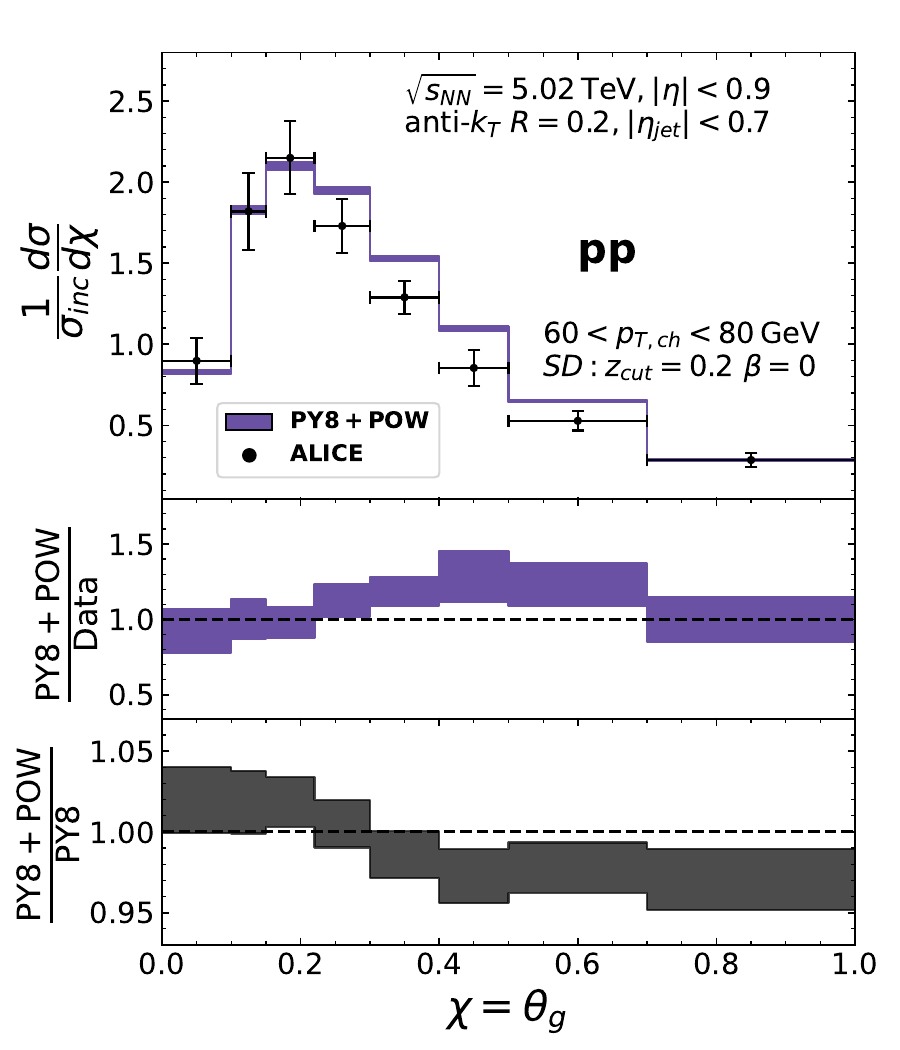}
\includegraphics[width=0.33\textwidth]{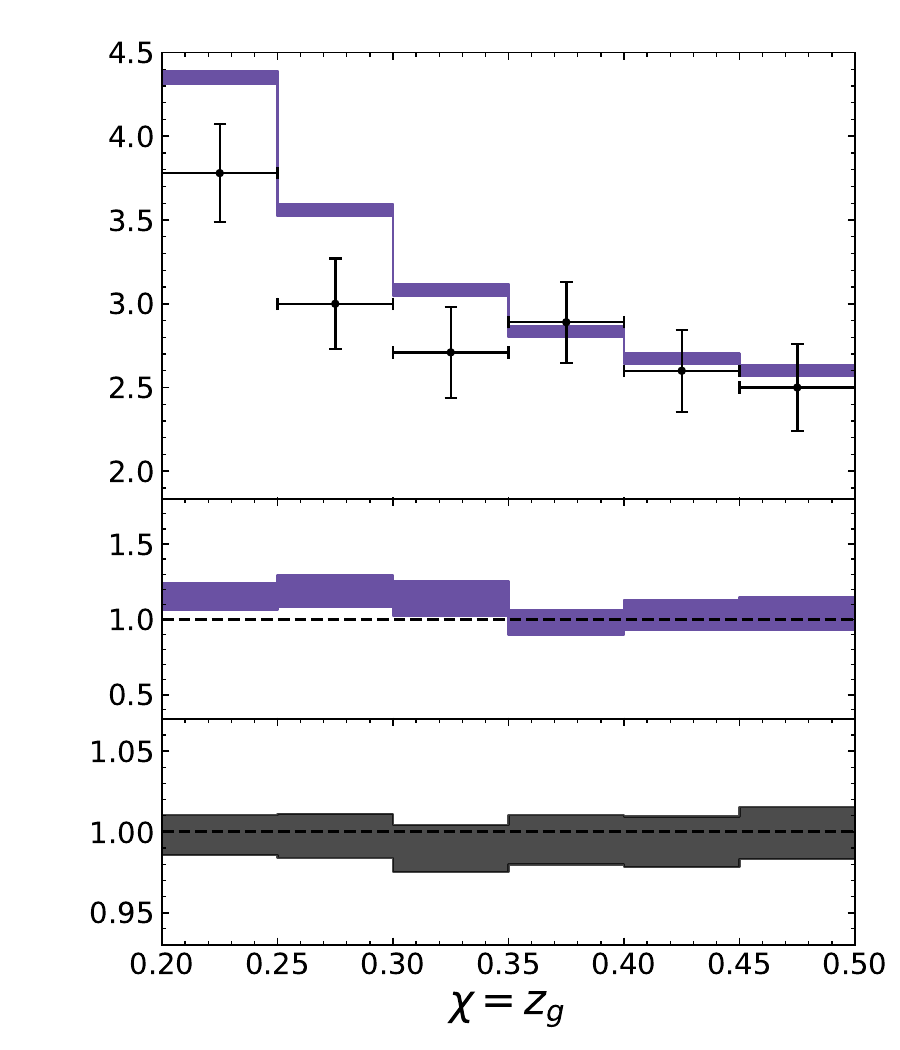}
\includegraphics[width=0.33\textwidth]{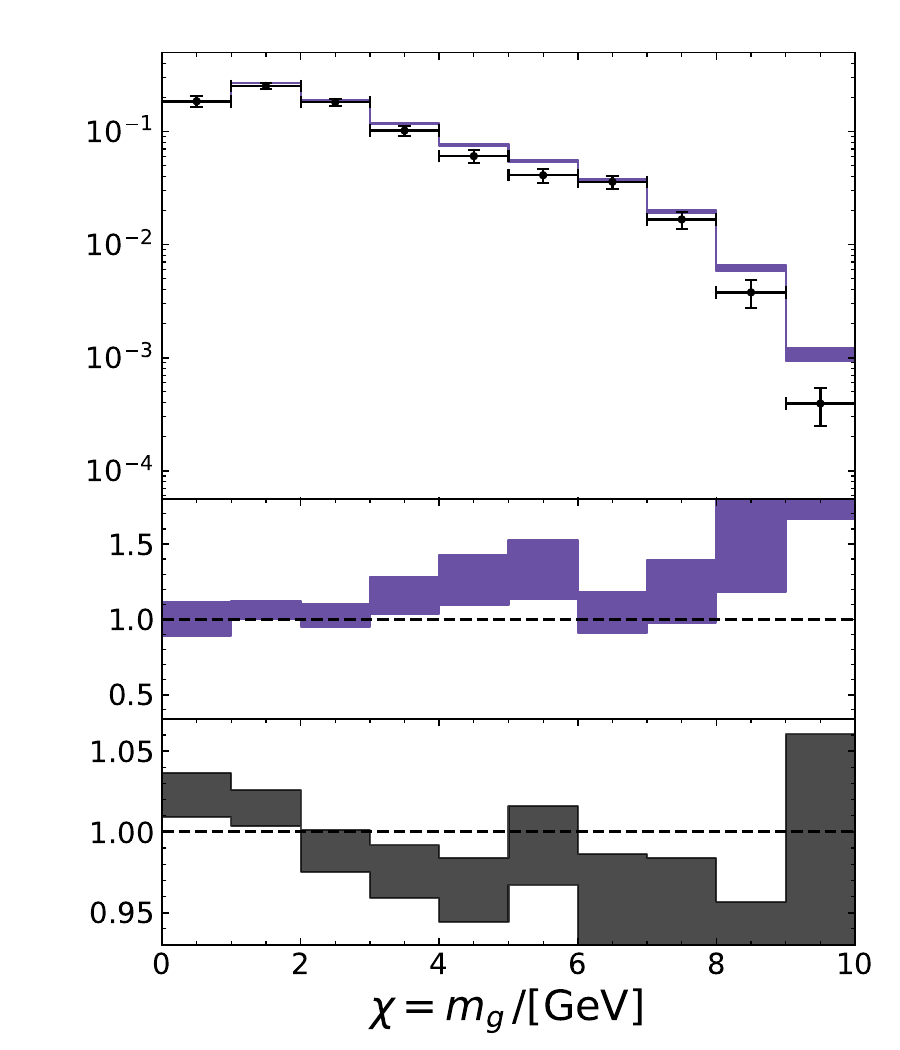} \\
\includegraphics[width=0.33\textwidth]{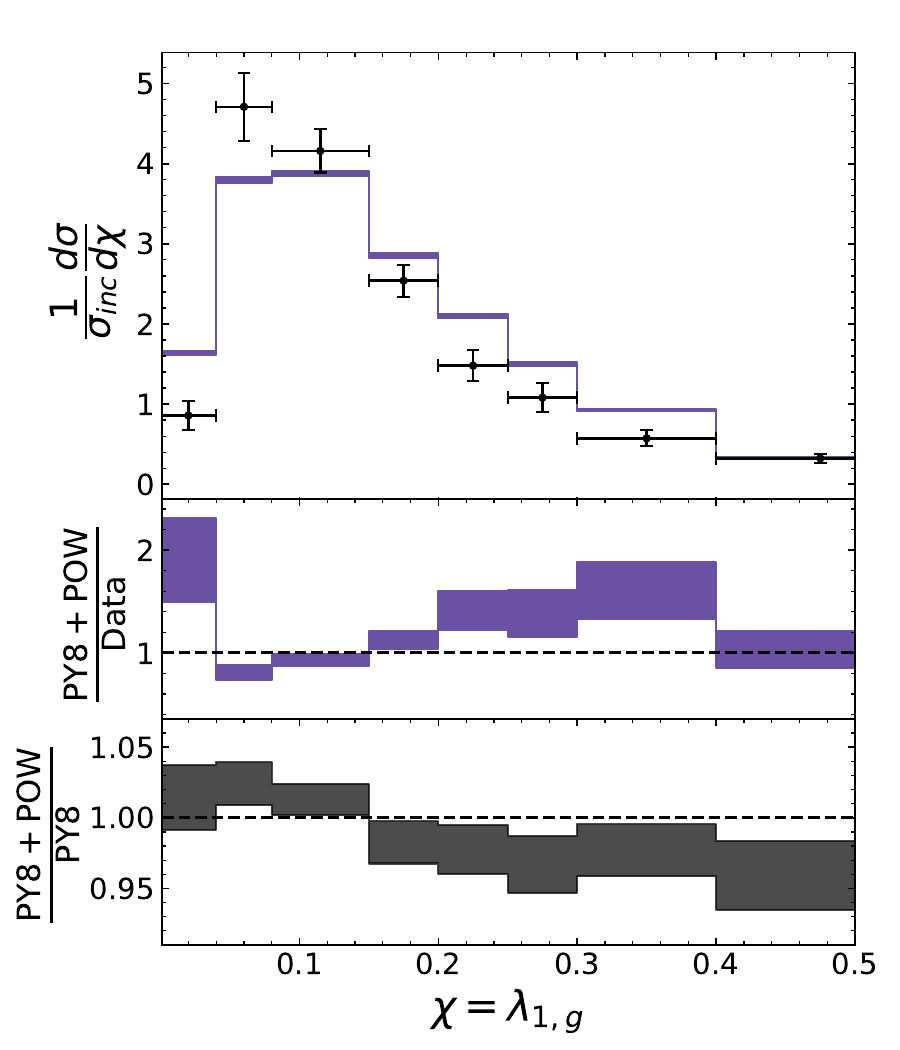}
\includegraphics[width=0.33\textwidth]{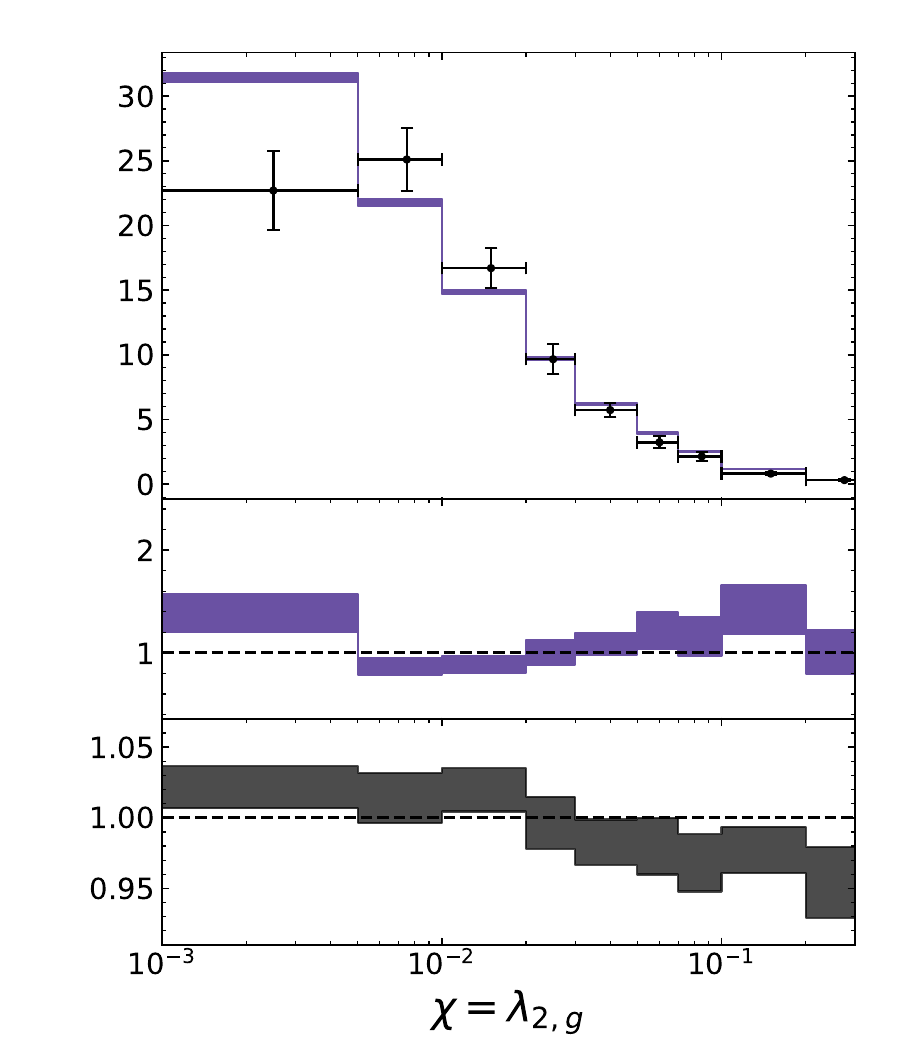}
\includegraphics[width=0.33\textwidth]{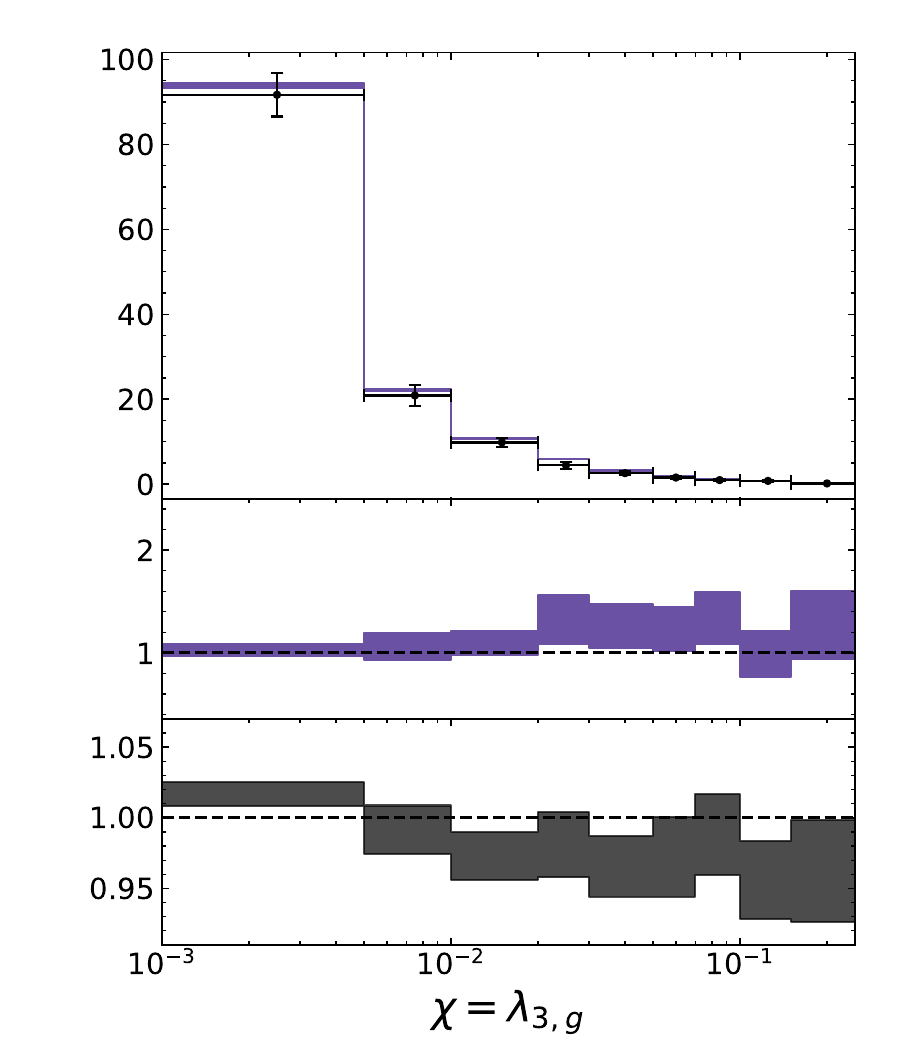}
\caption{\texttt{POWHEG+PYTHIA8} results for $z_g, \theta_g$ and $m_g$ (see Eqs.~\eqref{eq:zg}-\eqref{eq:mg}) (top) and groomed angularities (bottom) for increasing values of $\alpha$ in Eq.~\eqref{eq:lambdag}. In all cases, the main panel shows the distributions compared to experimental data, the middle panel presents the theory-to-data ratio and the lower panel the size of NLO matching corrections. The band in the simulation results corresponds to the standard deviation.}
\label{fig:alice-vac}
\end{figure*}
This is connected to the dependence of the $\theta_g$-spectrum on the flavour of the jet initiator at NLO+LL being more relevant than at NLO. As we discussed in Sec.~\ref{sec:fixed-order}, the shape of the $\theta_g$-distribution at $\mathcal{O}(\alpha_s)$ mildly depends on the colour factor of the jet-initiator. In turn, when accounting for multiple emissions we find that quark-initiated jets have a narrower fragmentation pattern than gluon jets, a well-known result in the literature~\cite{OPAL:1993uun}. As a consequence, the $p_{t,\rm jet}$-selection in the computation of $\theta_g$ (or any other observable) will naturally bias the in-medium $\theta_g$-distribution towards that of high-$p_t$ (mostly) quark-initiated jets. Note that this selection bias effect, that has been extensively discussed in the literature, is present even in a fully coherent energy loss scenario described by $P_1^{\text R}$. Thus, unlike the fixed order case, $P_1^{\text R}$ also yields a narrowing of the $\theta_g$ distribution although quantitatively smaller than $P_2^{\text R}$. Part of the phenomenological discussion in Sec.~\ref{sec:medium} will revolve around whether current experimental data can disentangle colour coherence from selection bias effects. 

In the remaining of this paper we apply this simple model of groomed substructure observables to realistic LHC kinematics and compare to data obtained from both $pp$ and PbPb collisions (see table~\ref{table:fiducial_cuts}). 
\section{Vacuum baseline}
\label{sec:vacuum}
We first study groomed substructure in proton-proton collisions at $\sqrt s=5.02$ TeV. The results were obtained by simulating 10 million events including hadronization and multi-parton interactions. The simulation details can be found in Appendix~\ref{app:simulation}. We split the discussion into moderate-$p_t$ jets as measured by ALICE~\cite{ALargeIonColliderExperiment:2021mqf,ALICE:2024jtb,ALICE-PUBLIC-2024-004} and the high-$p_t$ regime accessed by ATLAS~\cite{ATLAS:2022vii} using the fiducial cuts presented in table~\ref{table:fiducial_cuts}. We note that \texttt{PYTHIA8} simulations were presented in the original experimental papers, so we mainly focus on the role of NLO corrections throughout the discussion. For each observable, we plot the average value together with an uncertainty band computed as the standard deviation.
\subsection{ALICE kinematics}
We compare \texttt{POWHEG+PYTHIA8} simulations and six groomed observables in Fig.~\ref{fig:alice-vac}. For both $\theta_g$ and $z_g$, the NLO+LL baseline reproduces the experimental measurements within 10\% deviations. NLO corrections are negligible across the $z_g$-spectrum. In turn, for $\theta_g$ we observe a $5\%$ effect that makes the distribution narrower and into closer agreement with the data. 

A more noticeable deviation between simulation and data arises for the $m_g$ distribution. The NLO+LL result captures the peak of the distribution and its overall shape, but overpopulates the large-$m_g$ region with respect to the data, by up to a factor of two around $m_g \simeq 10$ GeV. NLO corrections are also sizeable for larger $m_g$, implying that LO would overshoot even more strongly. Since both $\theta_g$ and $z_g$ are well described, the excess at large $m_g$ cannot be attributed to the two-prong splitting kinematics alone. This suggests that, while NLO matching mitigates some limitations of the LL approximation, the groomed mass remains more sensitive to details of the internal momentum distribution within the groomed jet. This deviation was already observed in Ref.~\cite{ALICE-PUBLIC-2024-004} for a wide range of Monte Carlo simulations.

We conclude the discussion with the groomed angularities. For $\alpha=1$, the NLO+LL baseline overshoots the data at small angularities, gradually approaching the data for larger values. As $\alpha$ increases, the overall agreement improves: the overpopulation at low $\lambda_{\alpha,g}$ weakens, and by $\alpha=3$ the NLO+LL result is broadly consistent with the measurements within uncertainties. The observed trend with $\alpha$ reflects the relative weighting of particles inside the jet. At low $\alpha$, the angularity emphasizes the collinear core, and, as $\alpha$ increases, the observable becomes more sensitive to the sparser, wider-angle radiation in the outer jet region, which is better described perturbatively. Consequently, the agreement with data improves systematically with $\alpha$. The size of NLO corrections remains small, below $5\%$ in all cases. This confirms that higher-order fixed-order corrections have a limited impact in this kinematic regime, consistent with the conclusions drawn from the other observables.
\subsection{ATLAS kinematics}
\begin{figure*}
\centering
\includegraphics[width=0.33\textwidth]{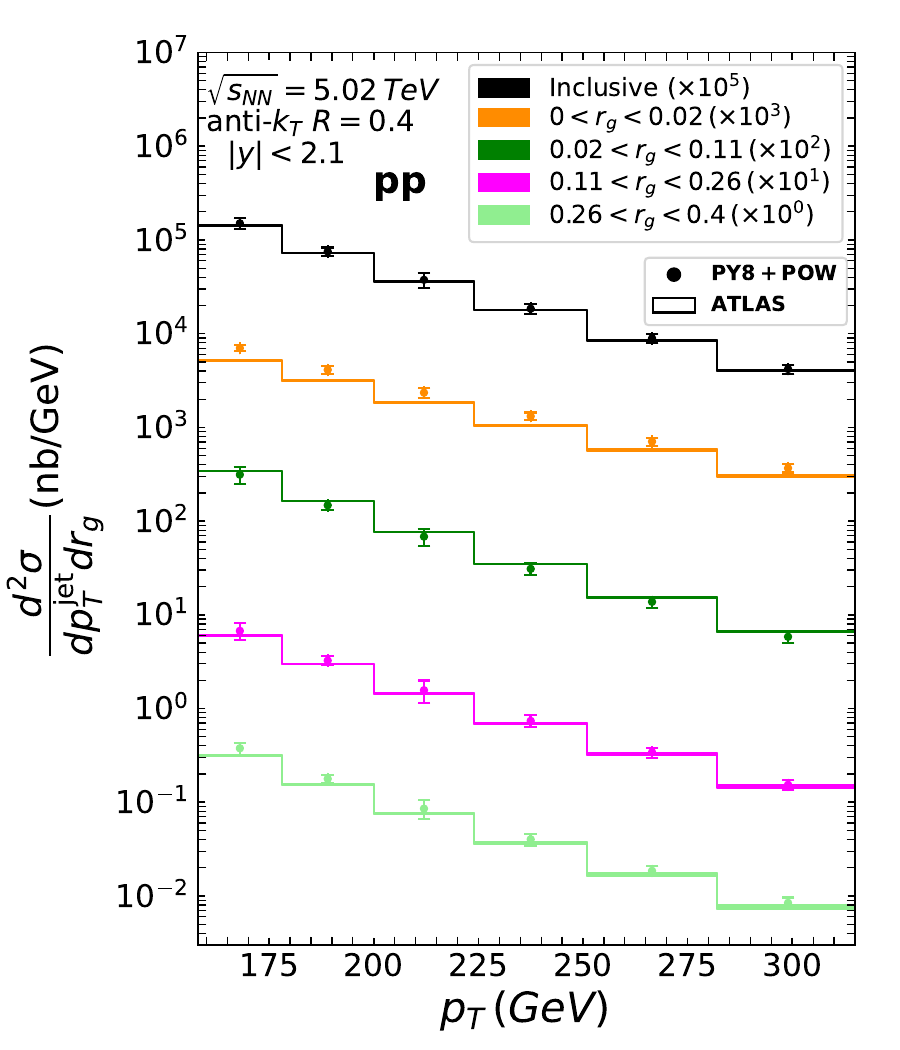}
\includegraphics[width=0.33\textwidth]{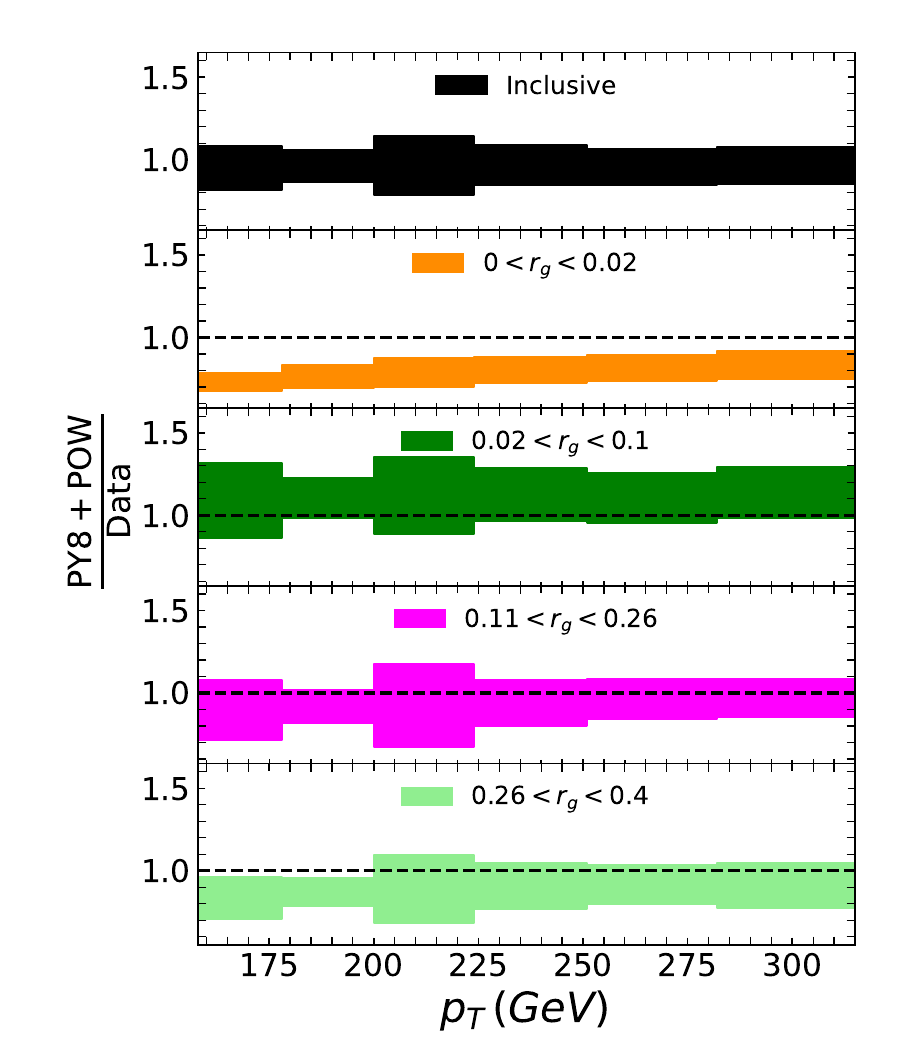}
\includegraphics[width=0.33\textwidth]{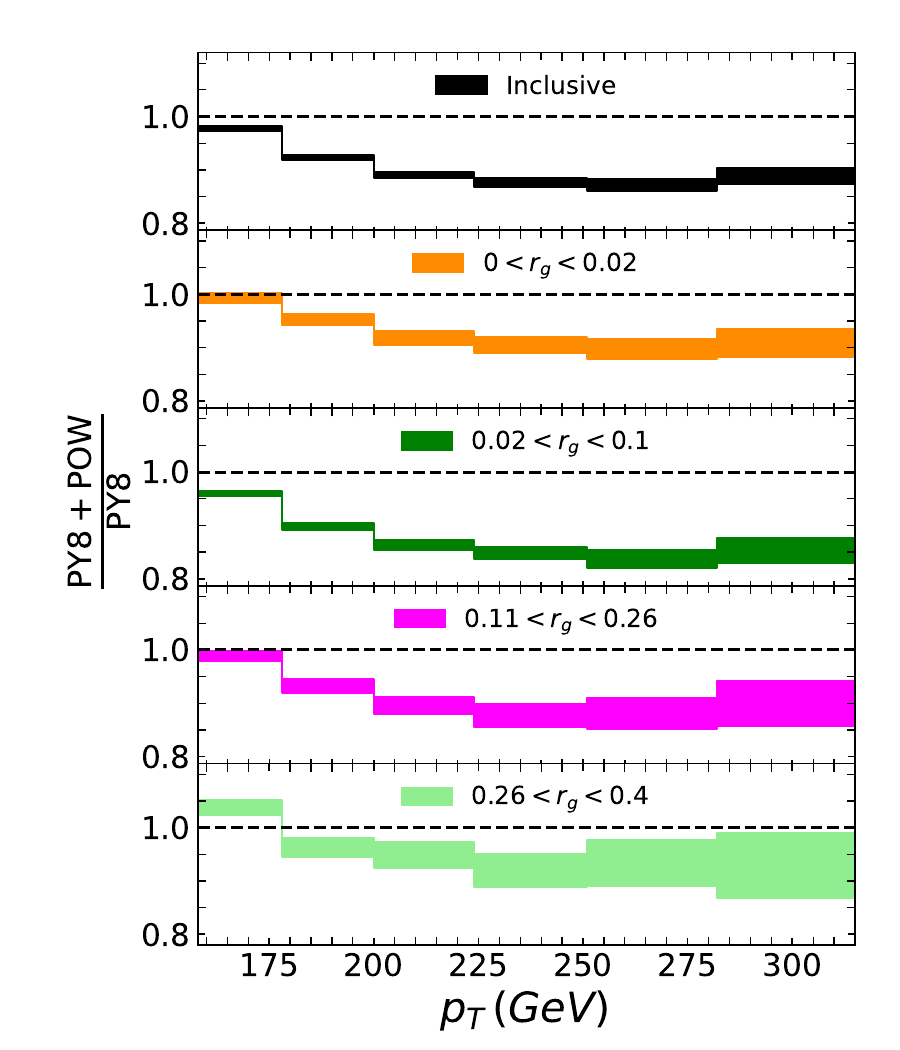} \\
    \includegraphics[width=0.33\textwidth]{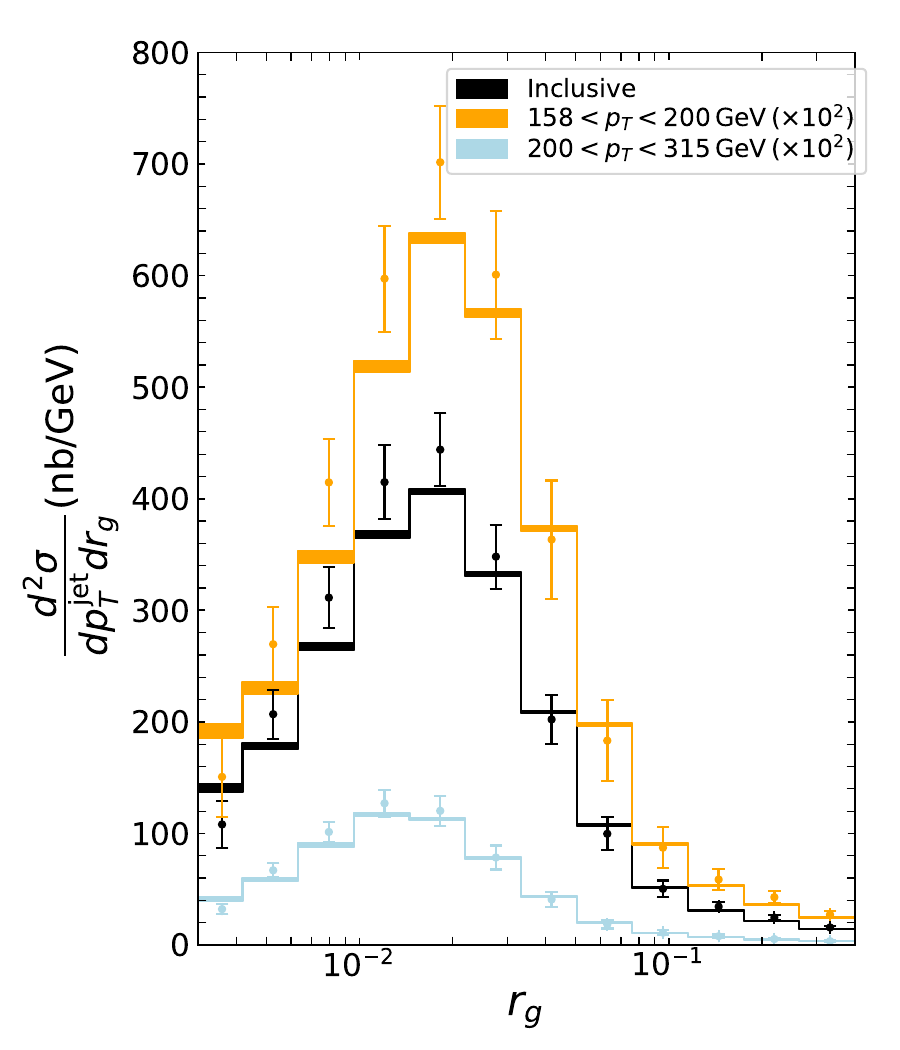}
\includegraphics[width=0.33\textwidth]{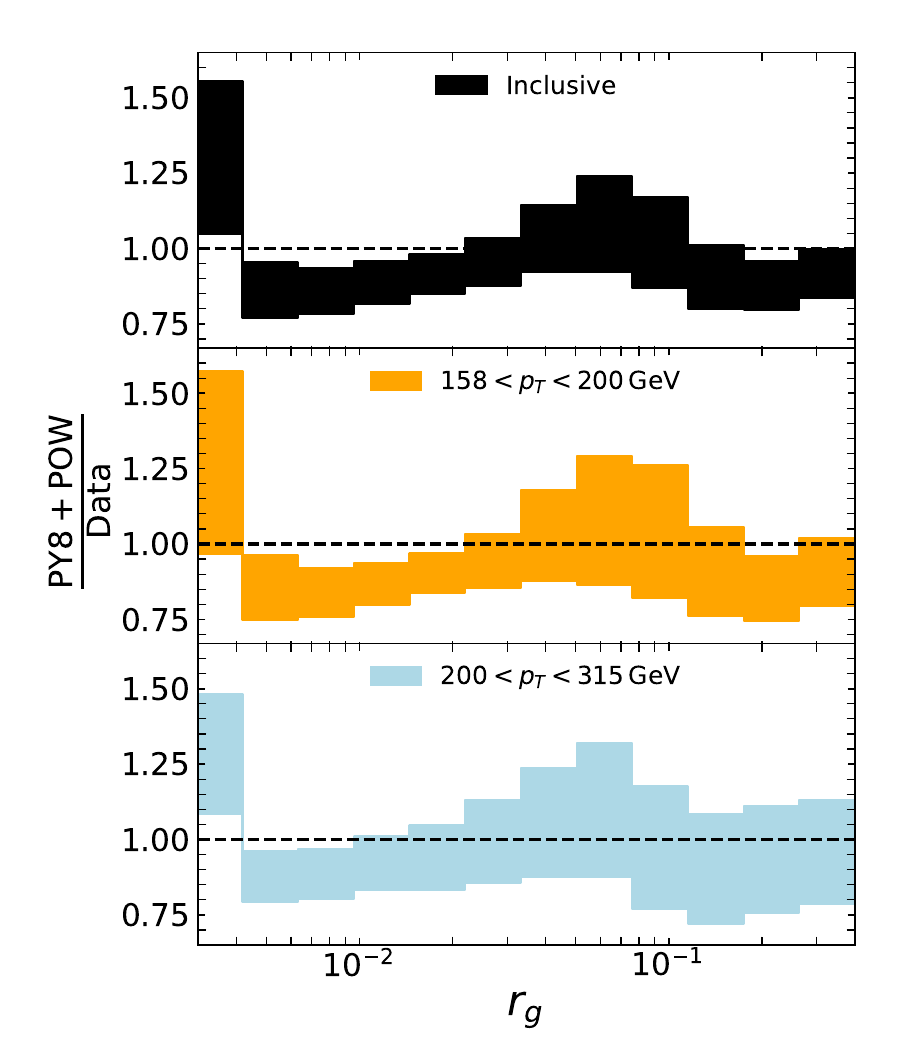}
\includegraphics[width=0.33\textwidth]{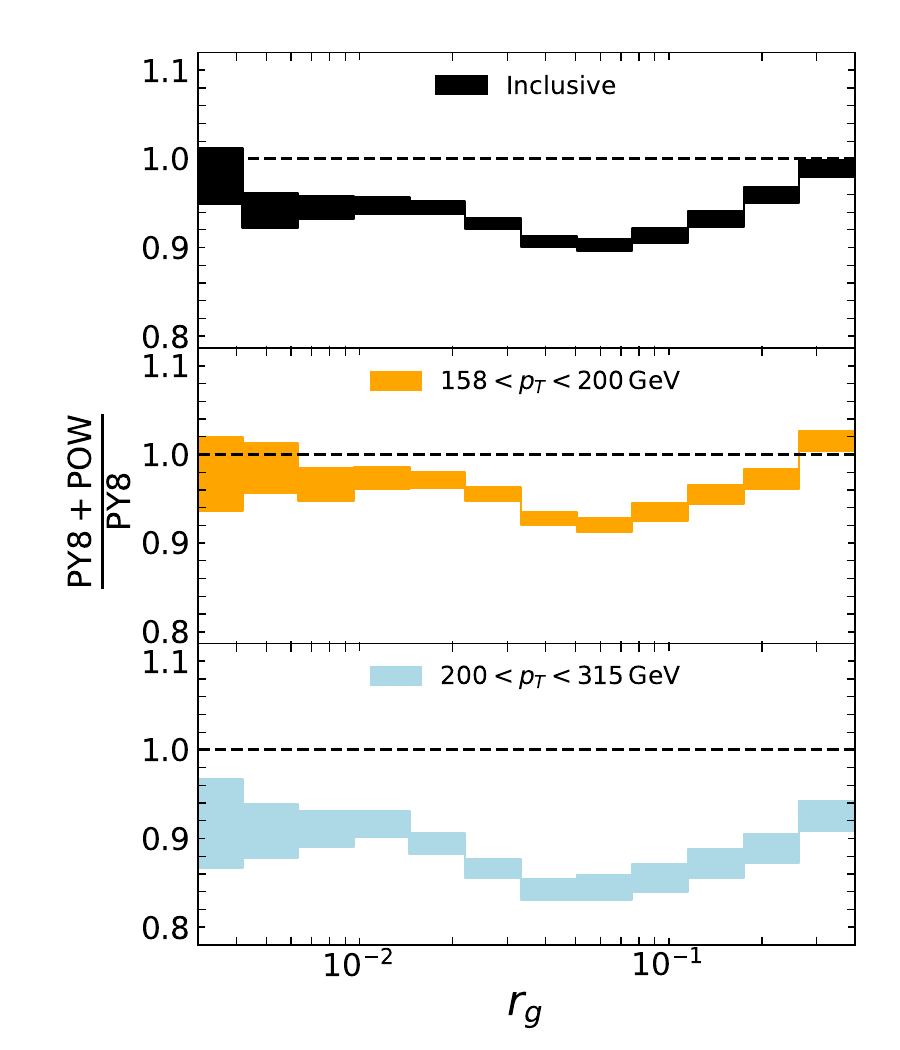}
\caption{\texttt{POWHEG+PYTHIA8} results for the differential cross-section of jets passing the SD condition in: different $r_g$-intervals as a function of jet $p_t$ (top row), and different $p_t$-intervals as a function of $r_g$ (bottom row). In both rows, the leftmost panel shows the distributions compared to experimental data, the middle panel the theory-to-data ratio and the rightmost panel the size of NLO corrections.}
\label{fig:atlas-vac}
\end{figure*}
Figure~\ref{fig:atlas-vac} presents the comparison between NLO+LL simulations and ATLAS data for the double-differential cross-section to produce a jet with a given $p_{t,\rm jet}$ and a given groomed jet radius $r_g$. The analysis presented in Ref.~\cite{ATLAS:2022vii} explored jets up to $p_{t,\rm jet}=1$ TeV in vacuum and up to $p_{t,\rm jet}\!=\!500$ GeV in heavy-ions. For concreteness, we here focus on three $p_t$-intervals: inclusive jets, $158\!<\!p_{t,\rm jet}\!<\!200$ GeV and $200\!<\!p_{t,\rm jet}\!<\!315$ GeV. The top panel shows the jet-$p_t$ spectrum divided into $r_g$ intervals, while the bottom panel displays the $r_g$-distribution itself for different $p_{t,\rm jet}$ selections. Let us first discuss the former. We find that except for very narrow configurations, where $r_g<10^{-2}$, \texttt{POHWEG+PYTHIA8} does an excellent job in describing the experimental data. Consistently, the middle panel of the bottom row of Fig.~\ref{fig:atlas-vac} displays a discrepancy on the first $r_g$-bin for the three $p_t$-intervals while the rest of the distribution falls within the data uncertainties. These two figures also display the clear narrowing of the $r_g$-distribution with increasing jet-$p_t$ as we mentioned in the previous section when discussing the ALICE data. Concerning the size of NLO corrections, we find that they are more pronounced than for ALICE kinematics throughout the $p_t$ and $r_g$-distributions. For instance, we find a 15\% suppression of jets with $r_g>10^{-2}$ for the highest $p_t$ bin under study. In all cases, NLO matching improves the theory-to-data agreement.

We conclude that our NLO+LL setup provides a quantitative description of groomed jet observables within experimental uncertainties for both ALICE and ATLAS kinematics. Equipped with this robust vacuum baseline we now address medium-induced effects.

\section{Comparison to heavy-ion data}
\label{sec:medium}
In this section we present theory-to-data comparisons for groomed jet substructure observables in heavy-ion collisions. Our purpose is twofold: (i) show that our bottom-up approach encoded in Eq.~\eqref{eq:v-med}, built upon a NLO+LL vacuum baseline, achieves a quantitative description of jet data at the LHC, and (ii) quantify the discriminating power of these observables with respect to different energy-loss scenarios. We address the latter point by comparing results where the vacuum baseline is fixed and we change the energy-loss distribution. In particular, we compare the $P^{\rm{R}}_2(\varepsilon,\theta_{12},L)$ energy-loss distribution given by Eq.~\eqref{eq:P2R} and a fully coherent model using $P^{\rm{R}}_1(\varepsilon,L)$ defined in Eq.~\eqref{eq:P1}. Note that the latter, despite not depending on the opening angle of the subjet pair, might still produce an enhancement of collinear splittings due to the jet-$p_t$ selection bias introduced in Secs.~\ref{sec:fixed-order} and \ref{sec:all-orders}.\footnote{A simple exercise of just shifting the vacuum jet-$p_t$ spectrum was also done. By shifting the vacuum baseline by the average energy loss value of the surviving jets (around $10$ GeV), we found a similar behaviour, i.e., enhancement for lower angles and suppression for higher angles.} The vacuum simulation setup is the same as in the previous section, while for the heavy-ion results we use the \texttt{EPPS16} nuclear PDF set~\cite{Eskola:2016oht}. The role of nuclear PDF effects was found to be negligible (less than $5\%$) for these observables.

\subsection{Fixing medium parameters}
\label{sec:medium-param}
Before discussing comparisons to data, let us first explain the methodology that we followed to fix the free parameters of the energy-loss distribution, namely the medium length $L$, the in-medium coupling $\alpha^{\rm med}_s$, and $\hat q$. As mentioned earlier, we decide to fix $L=4$ fm for all scenarios based on the fact that we always compare to data in the $0-10\%$ centrality interval. We also take $\alpha^{\rm med}_s=0.24$ in all simulations. The pending free parameter is the value of $\hat q$ which is obtained by fitting the nuclear modification factor, $R_{AA}$, in the relevant kinematic configurations. Instead of fitting only the central value of $R_{AA}$ we decide to perform two fits corresponding to both the upper and lower bounds of $R_{AA}$ given by the experimental uncertainty. As such, for each observable, we produce theoretical results with the two values of $[\hat q_{\rm min}, \hat q_{\rm max}]$ and treat the difference as a theoretical uncertainty. An important consideration is that, for a given kinematic configuration, the fit depends both on the energy-loss model and the vacuum baseline, since NLO corrections change the jet-$p_t$ spectrum. To quantify all these differences we define the following ratio that we express as a percentage
\begin{equation}
\label{eq:qhatvar}
\delta\hat q_{x} = 100 \, (\frac{\hat q_x}{\hat q}-1)\, ,
\end{equation}
where the denominator corresponds to the $\hat q$-value extracted using our default settings: NLO+LL accuracy for the vacuum baseline and $P_2^{\rm R}$ for the energy loss distribution. The subscript $x$ refers to variations in either (a) the vacuum setting for which $x={\rm vac}$ and we use LO+LL accuracy for the vacuum baseline, or (b) the energy loss model for which $x={\rm E-loss}$ and we use $P_1^{\rm R}$ for the energy loss distribution. 

The extracted $\hat q$ values are compiled in Table~\ref{tab:qhat_fits}. Clearly, the choice of energy-loss distribution is the dominant source of $\hat q$ variations. Yet, the perturbative accuracy of the vacuum baseline yields a non-negligible, $20\%$ effect. To the best of our knowledge, this is the first time that the role of perturbative corrections on the vacuum baseline in $\hat q$-extractions is quantified. Certainly, our extraction is not very sophisticated and we thus look forward to studying the impact of an improved vacuum baseline in full-fledged Bayesian analyses such as the one presented in~\cite{JETSCAPE:2024cqe}. In the rest of this paper we show results using the $\hat q$ values displayed in the second column of Table~\ref{tab:qhat_fits}. 

\begin{table}
\centering
\begin{tabular}{|l|c|c|c|}
\hline
\centering
Dataset & $\hat q$ [GeV$^2$/fm] & $\delta\hat q_{\rm vac}$ & $\delta\hat q_{\rm E-loss}$\\
\hline
ALICE~\cite{ALICE:2023waz} & $[0.13, 0.40]$ & $[20\%, 19\%]$ & $[77\%, 125\%]$ \\
ATLAS~\cite{ATLAS:2022vii} & $[0.18, 0.28]$ & $[14\%, 27\%]$ & $[94\%, 96\%]$\\
\hline
\end{tabular}
\caption{Extracted intervals of $\hat q$ for ALICE and ATLAS kinematic configurations. The last two columns corresponds to the percent change in $\hat q$ when using a different vacuum baseline (third column) or a different energy loss distribution (fourth column) as given by Eq.~\eqref{eq:qhatvar} and explained in the main text.}
\label{tab:qhat_fits}
\end{table}

\subsection{ALICE kinematics}
\label{sec:medium-alice}
\begin{figure*}
\centering
\includegraphics[width=0.33\textwidth]{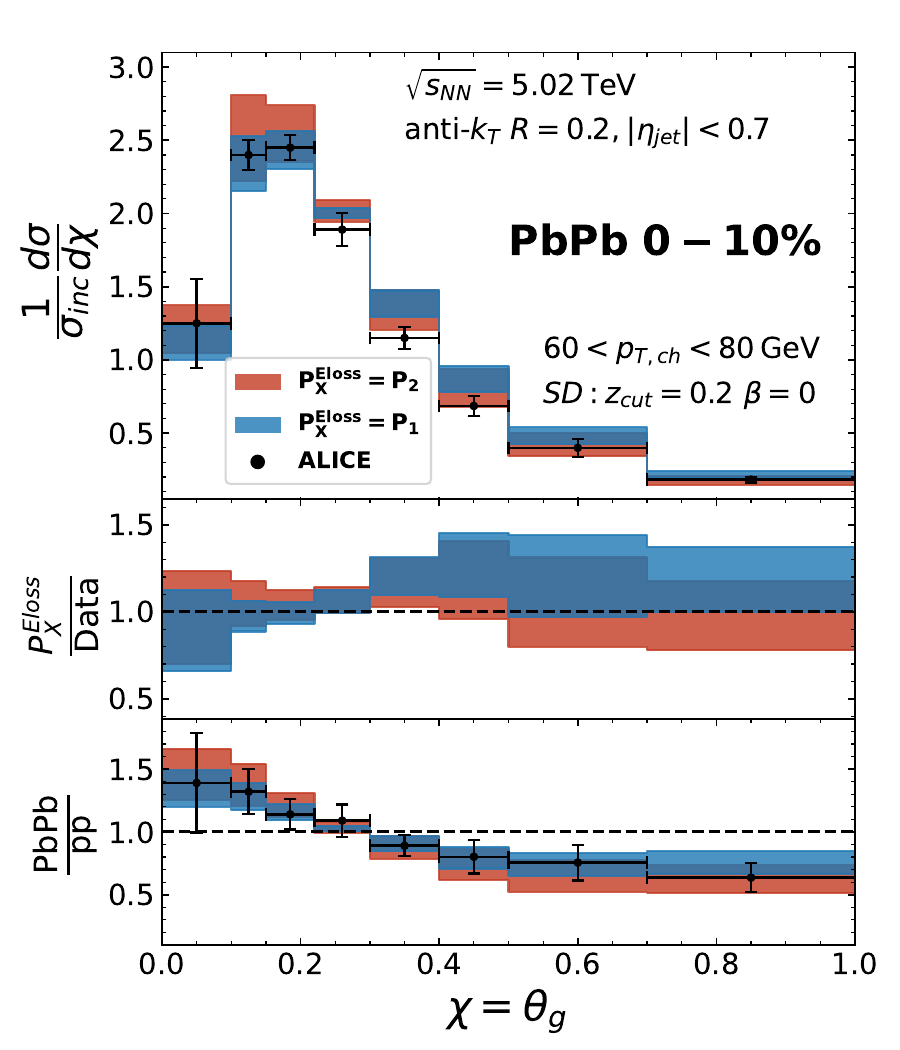}
\includegraphics[width=0.33\textwidth]{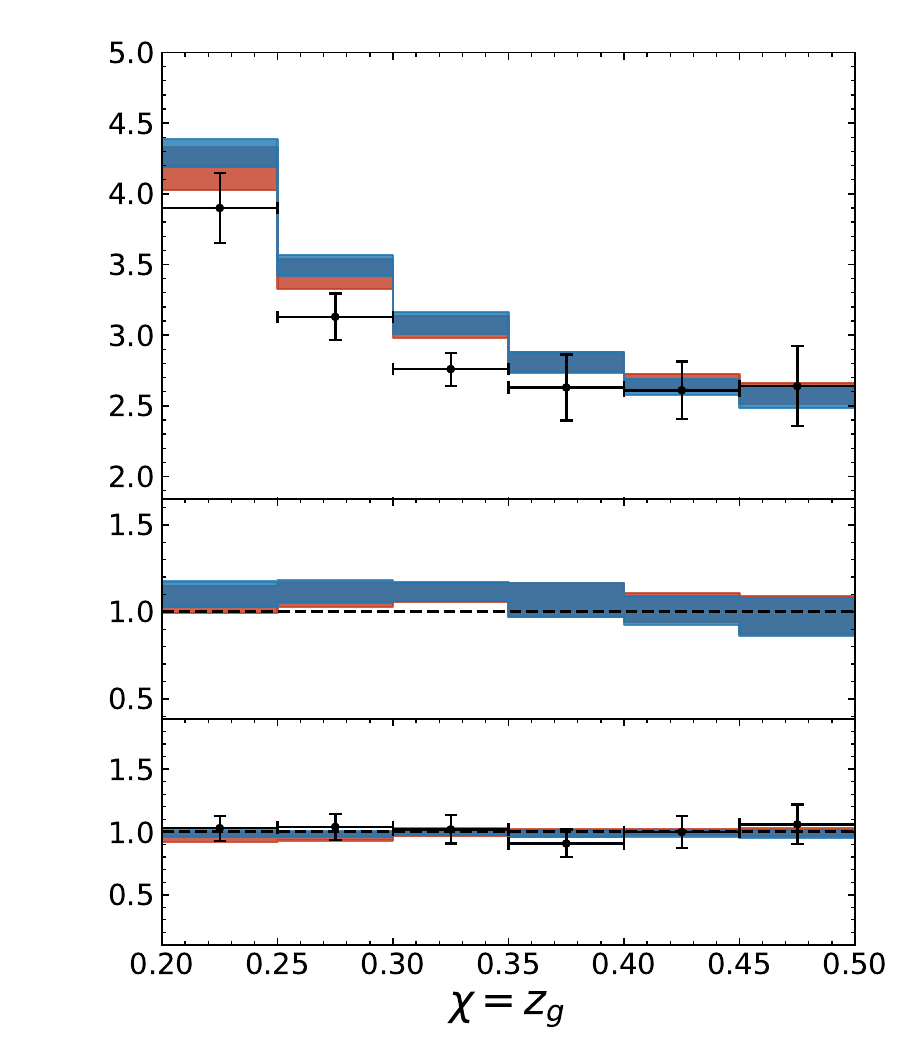}
\includegraphics[width=0.33\textwidth]{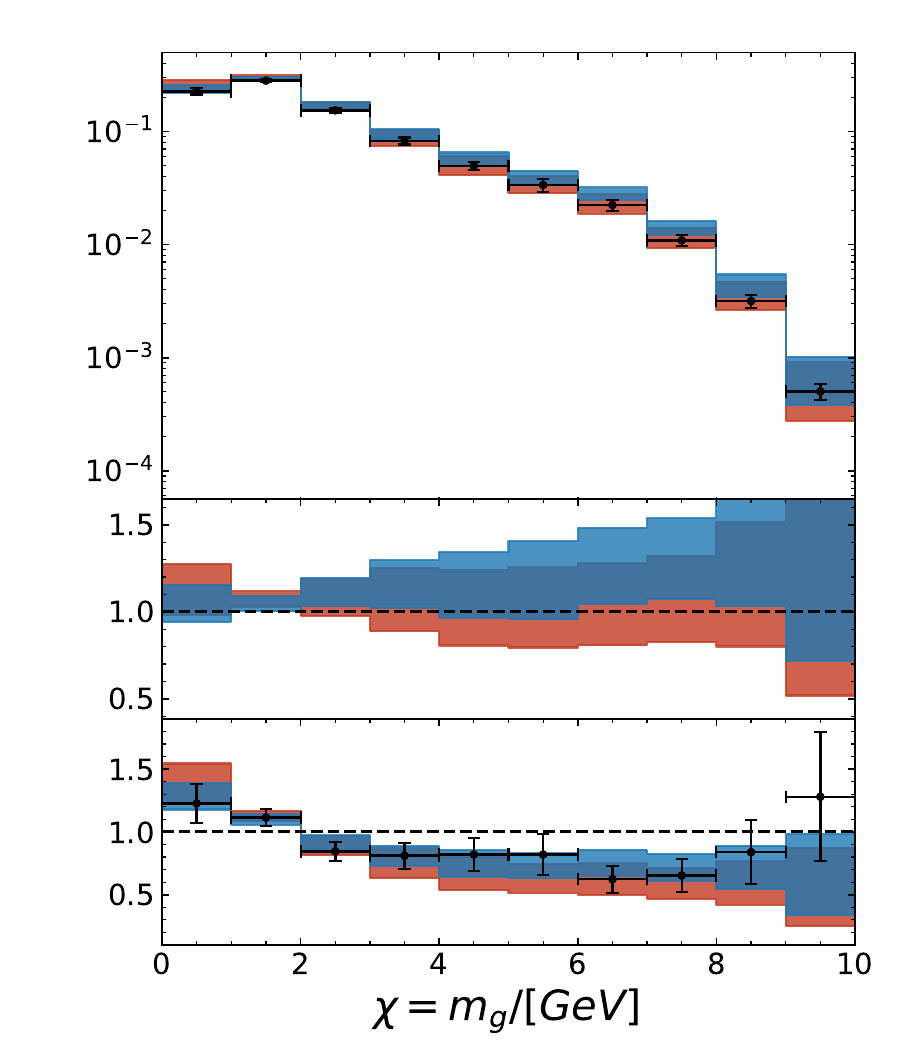} \\
\includegraphics[width=0.33\textwidth]{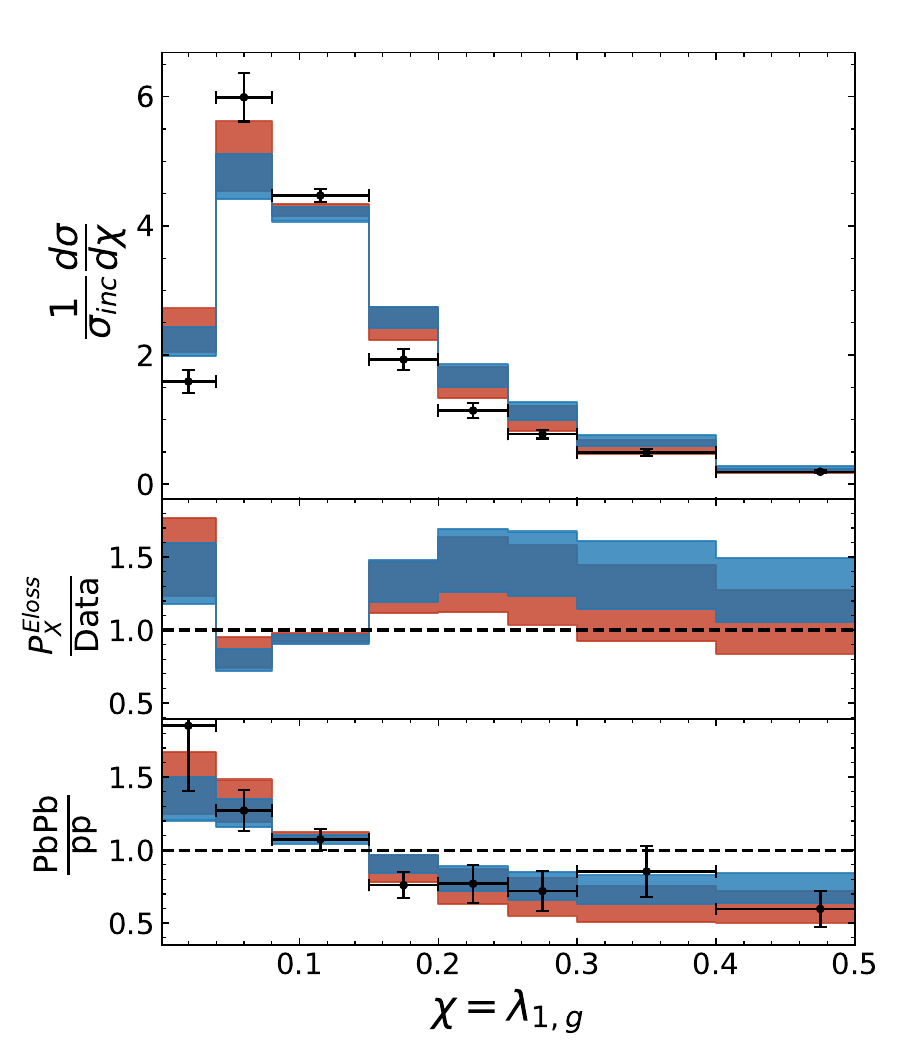}
\includegraphics[width=0.33\textwidth]{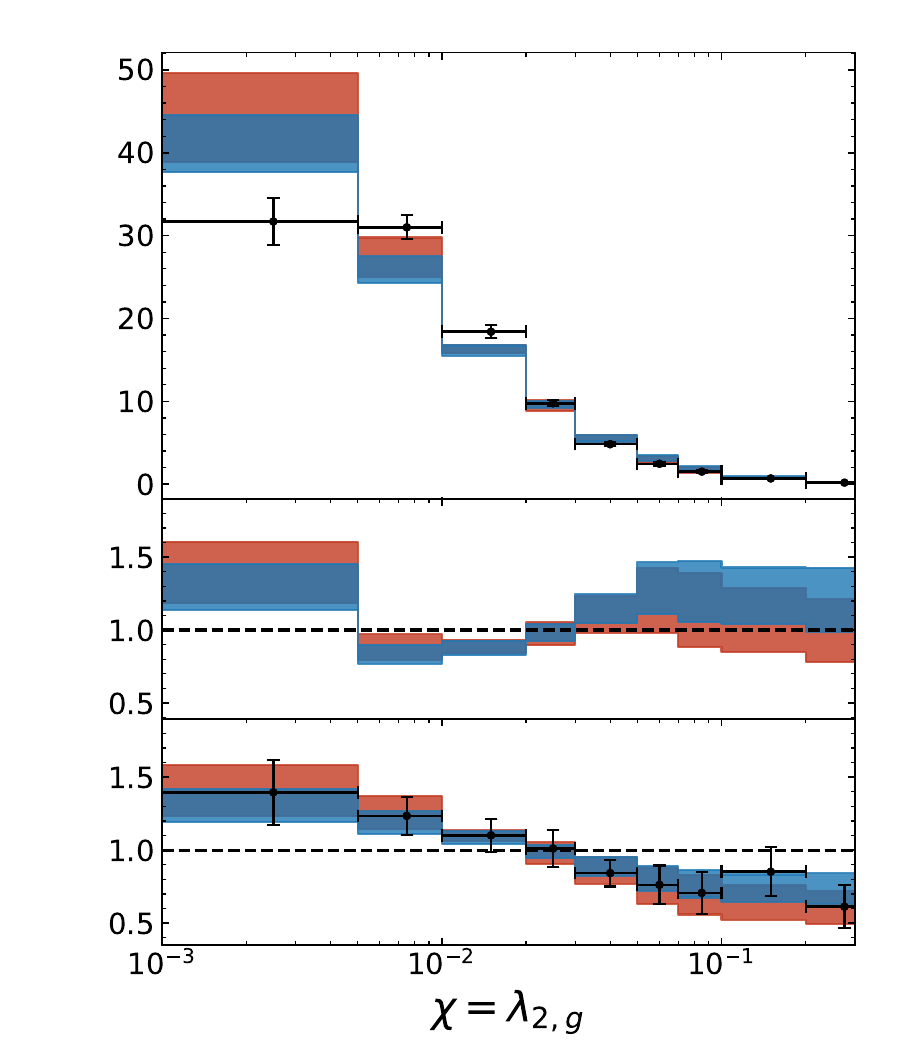}
\includegraphics[width=0.33\textwidth]{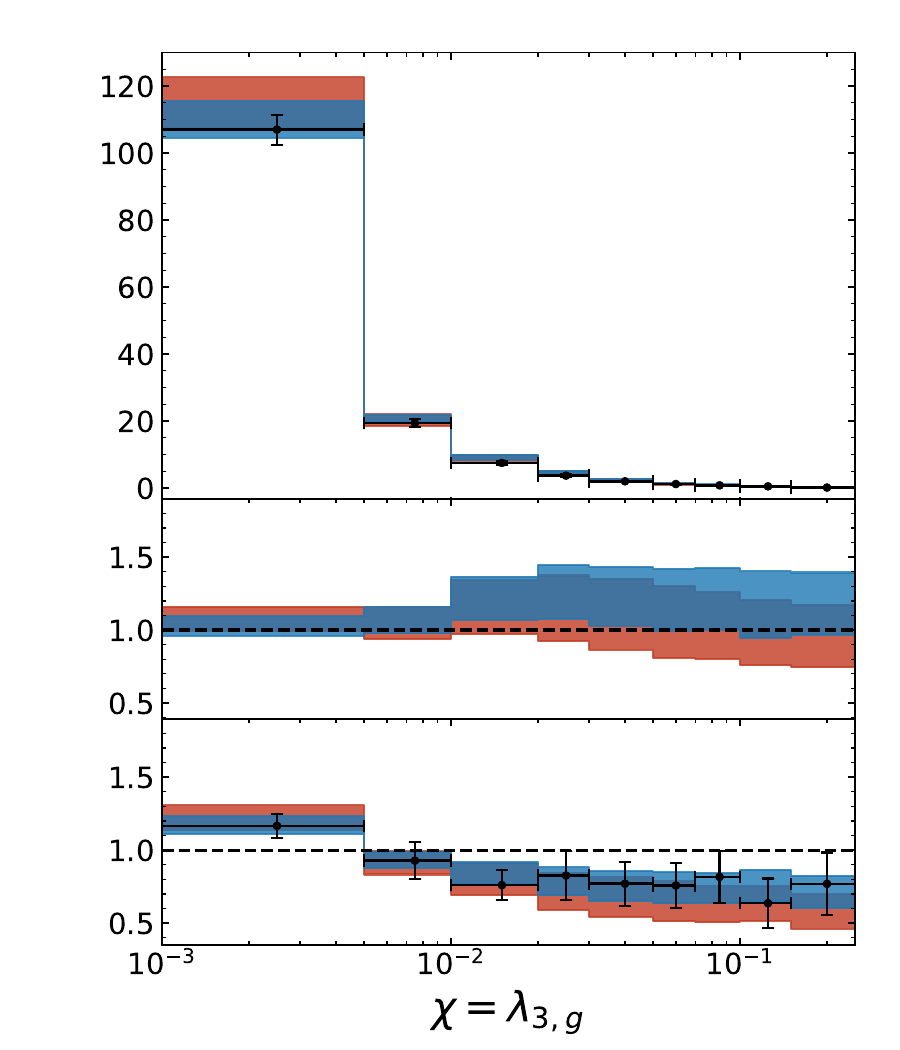}
\caption{In-medium results for the same observables as in Fig.~\ref{fig:alice-vac} using two different energy loss models: $P_2^{\rm R}$ (red), see Eq.~\eqref{eq:P2R}, and $P_1^{\rm R}$ (blue), see Eq.~\eqref{eq:P1}. In all cases, the main panel shows the distributions compared to experimental data in PbPb, the middle panel presents the theory-to-data ratio and the lower panel the medium-to-vacuum ratio. Note that the numerator of the middle and lower panels is the same. The vacuum distribution is computed using the $\texttt{POWHEG+PYTHIA8}$ interface.}
\label{fig:alice-med}
\end{figure*}
The analogue of Fig.~\ref{fig:alice-vac} but for heavy-ion collisions is shown in Fig.~\ref{fig:alice-med}. We find a remarkable agreement between our theoretical results and data across the considered groomed observables at the level of full distributions. This indicates that an accurate NLO+LL vacuum baseline, together with an energy-loss mechanism that accounts for colour decoherence through an explicit dependence on the two-prong opening angle, is sufficient to describe the data. 
When considering the medium-to-vacuum ratio we find even a better agreement. The only noticeable deviation occurs at large groomed mass, where the model slightly underestimates the enhancement seen in data. We attribute this failure to the vacuum baseline since we already saw in Fig.~\ref{fig:alice-vac} that the large-$m_g$ region was not properly described at this level of accuracy. For all observables, except for $z_g$, we observe an enhancement of collinear structures in heavy-ions with respect to proton-proton. This is in qualitative agreement with the academic exercise presented in Figs.~\ref{fig:thg-nlo} and \ref{fig:thg-nlops}, but also in quantitative agreement with data. As discussed above, this enhancement of collinear splittings and corresponding suppression of broad structures is, in this framework, directly related to the behaviour of the decoherence parameter, $\Delta_{\rm med}$ in Eq.~\eqref{eq:Psing-simple}. The groomed angularities offer a useful handle on the interplay between coherent and decoherent regimes thanks to the exponent $\alpha$ in their definition, see Eq.~\eqref{eq:lambdag}. We observe that for increasing $\alpha$ the point where the medium-to-vacuum ratio becomes larger than $1$ shifts towards smaller values of $\lambda_{\alpha,g}$. This shift arises because higher exponents enhance the weight of large-angle splittings with $\theta\gg \theta_c$. As a consequence, the transition from the coherent to the incoherent regime manifests at smaller $\lambda_{\alpha,g}$ as $\alpha$ increases.
Regarding the imprint of colour coherence effects,  we find that the current experimental precision is not yet sufficient to discriminate unambiguously between the two energy-loss scenarios, since both the $P_2^{\text R}$ and $P_1^{\text R}$ results remain compatible with the measurements within uncertainties. However, the main difference between the two energy-loss scenarios is a stronger suppression for large values of the observables when using $P_2^{\text R}$, except for $z_g$ where both models give a medium-to-vacuum ratio compatible with one. This is expected since for $\theta_g\gg \theta_c$, $P_2^{\text R}$ effectively corresponds to the two-prongs of the antenna loosing energy independently, resulting in an overall greater suppression when compared with the $P_1^{\text{R}}$ case, which is independent of the splitting angle.

\subsection{ATLAS kinematics}
\label{sec:medium-atlas}
\begin{figure*}
\centering
\includegraphics[width=0.33\textwidth]{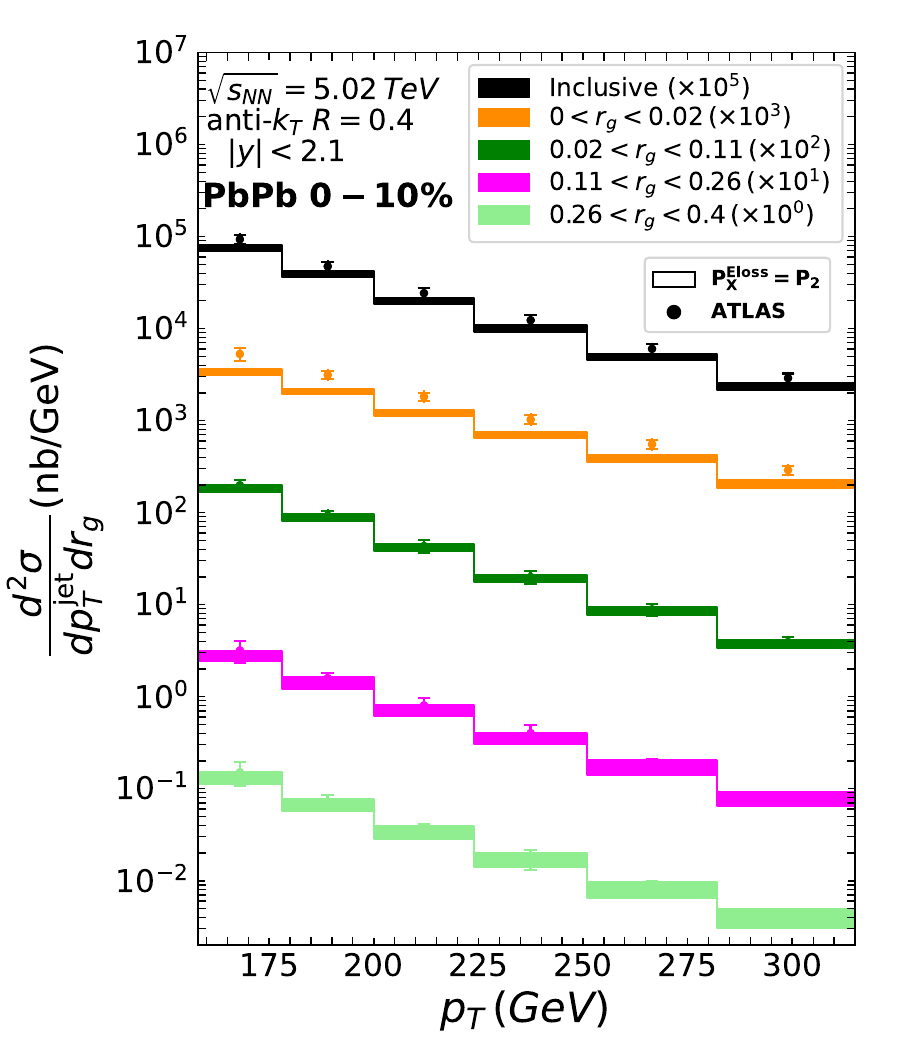}
\includegraphics[width=0.33\textwidth]{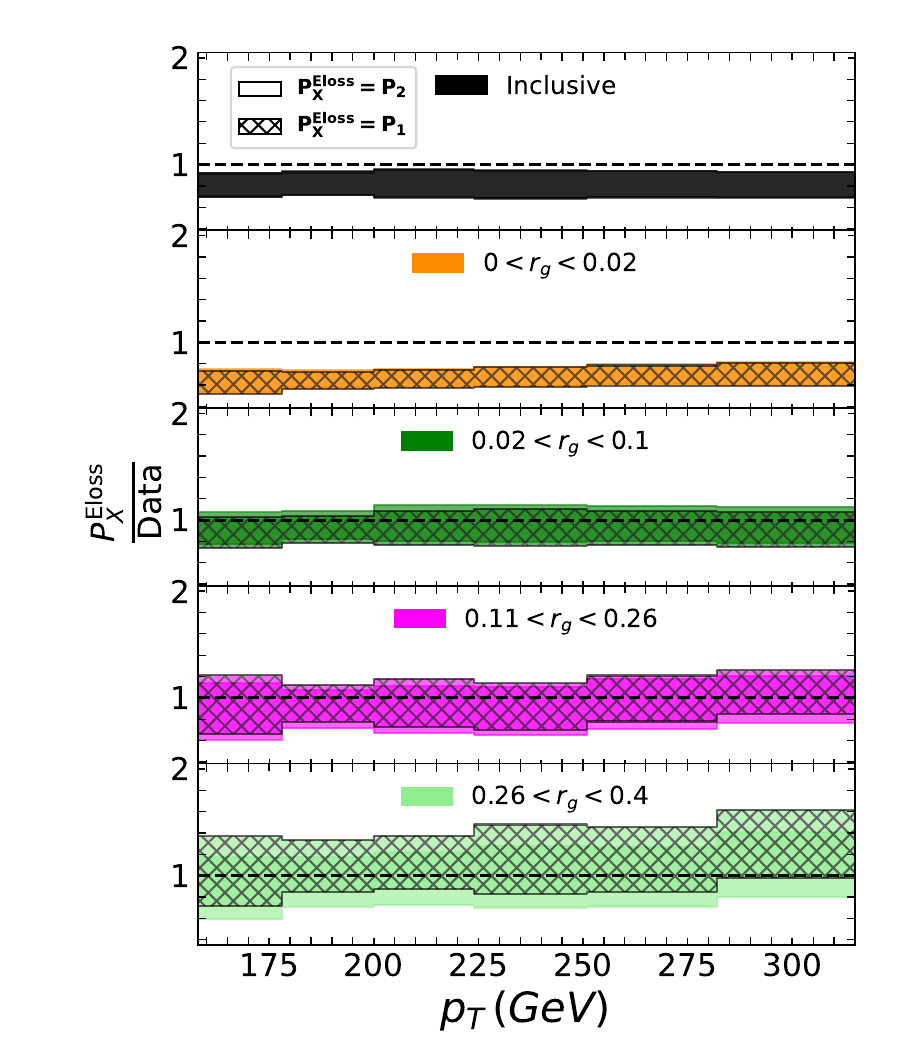}
\includegraphics[width=0.33\textwidth]{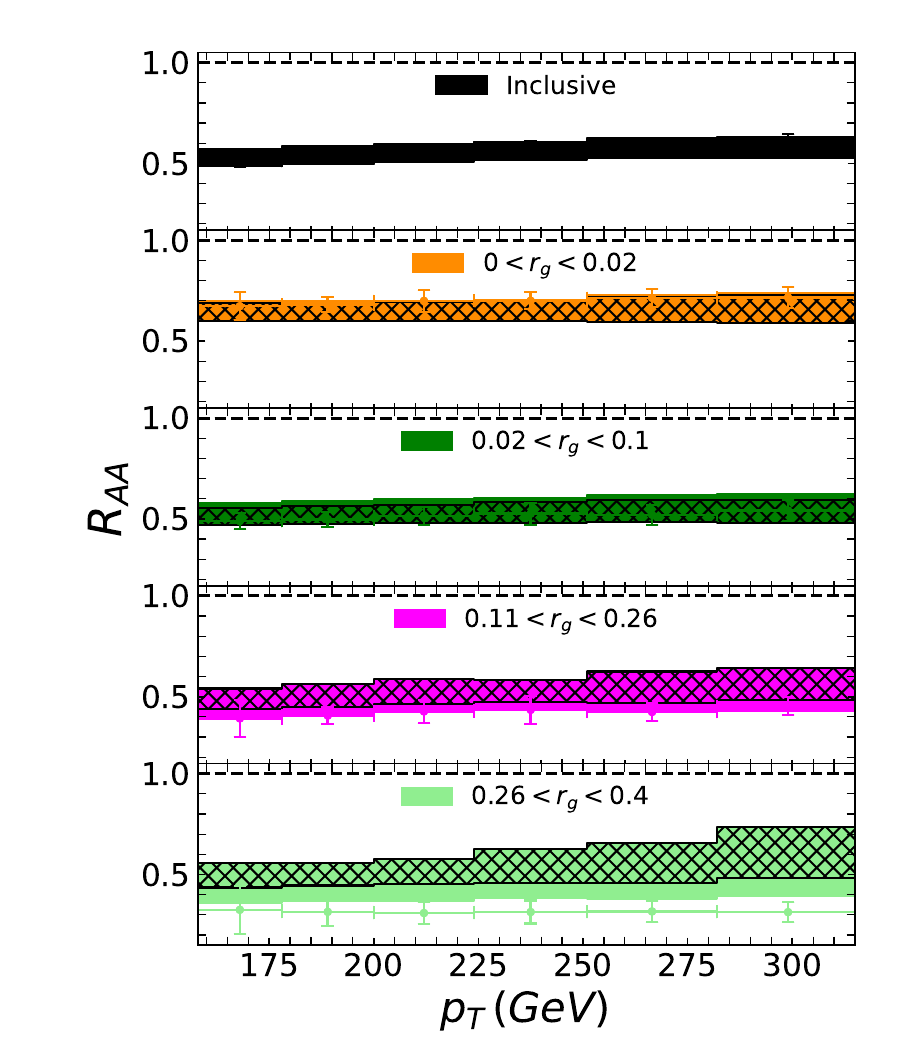} \\
\includegraphics[width=0.33\textwidth]{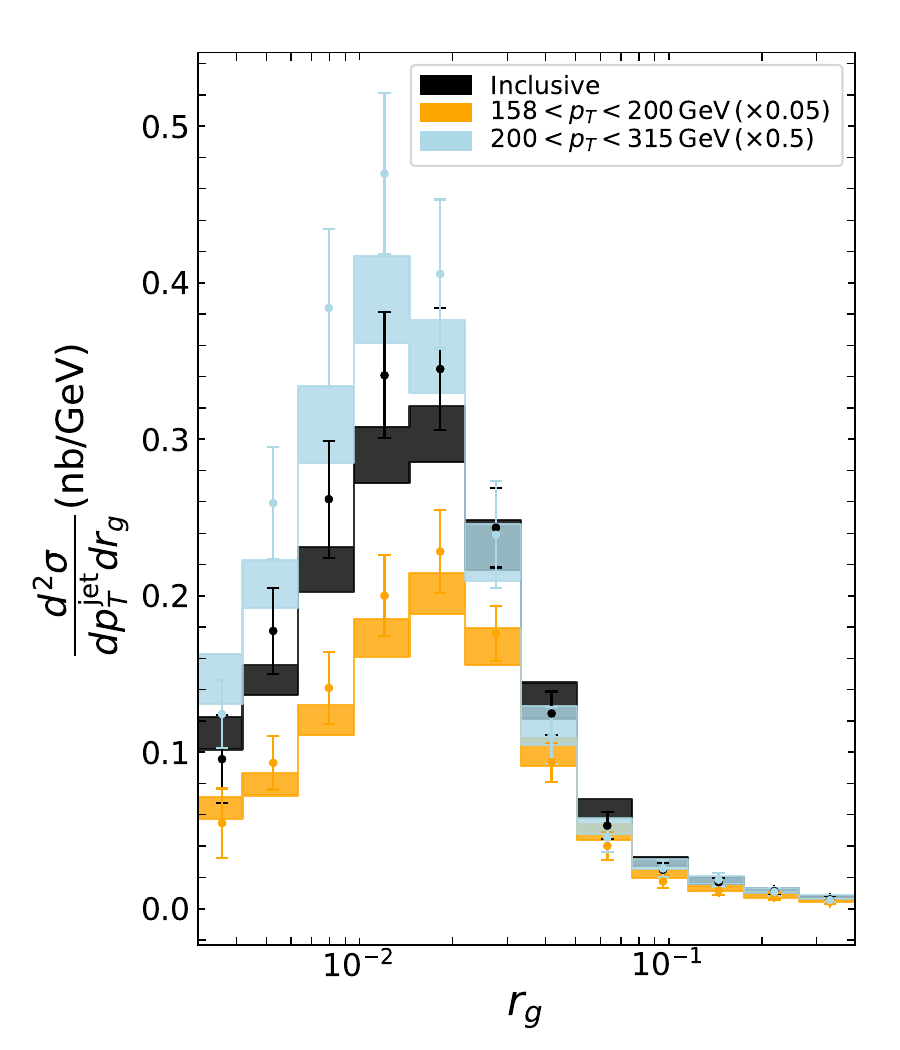}
\includegraphics[width=0.33\textwidth]{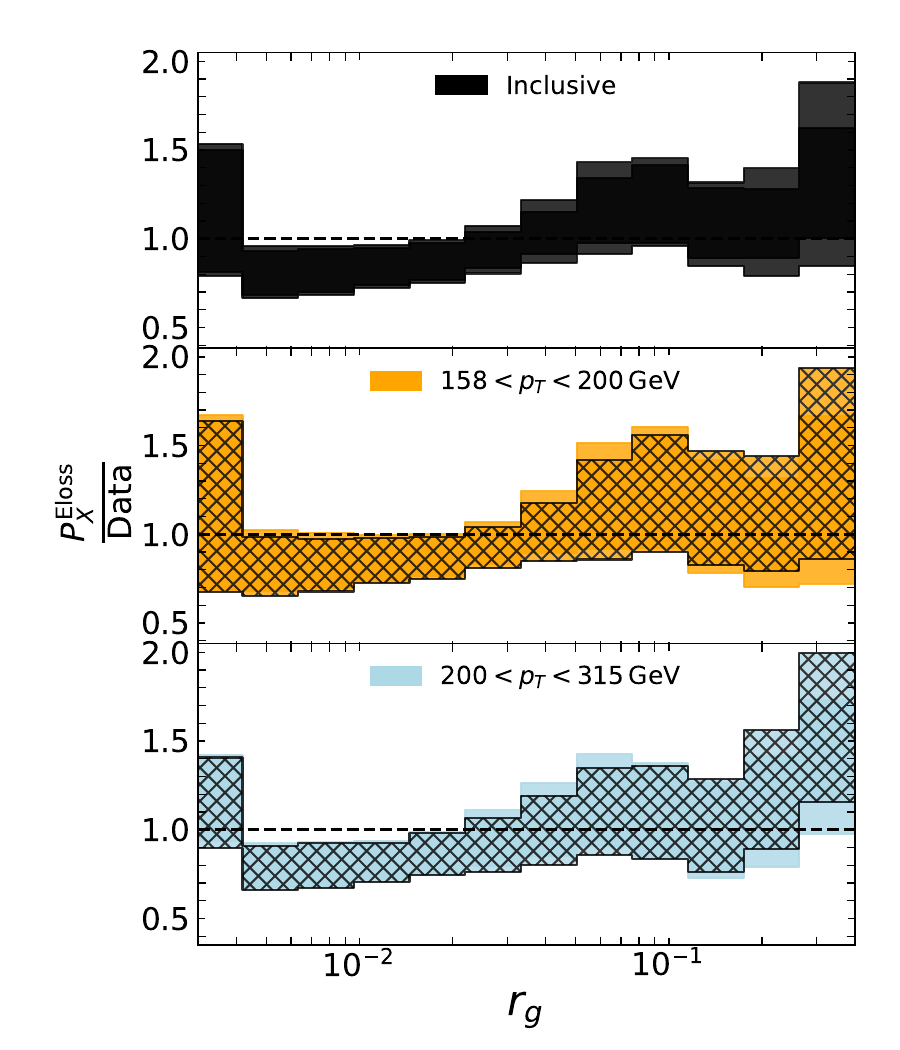}
\includegraphics[width=0.33\textwidth]{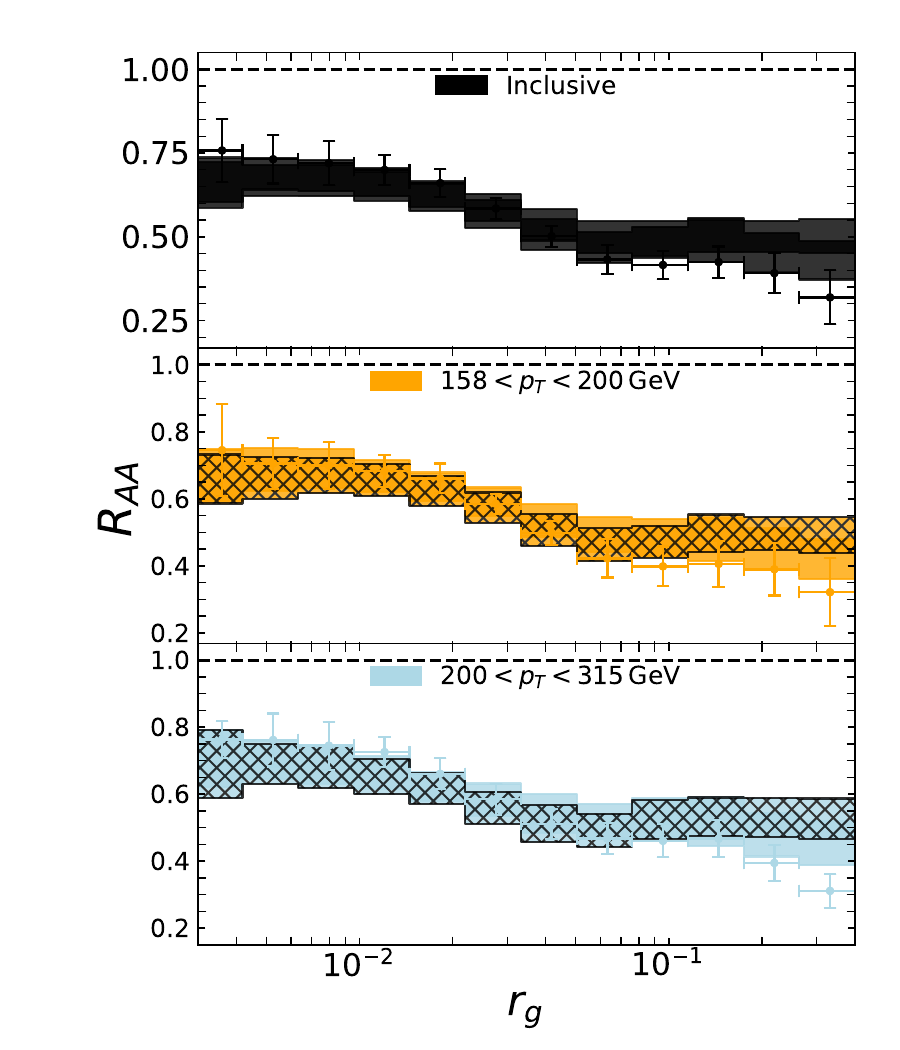}
\caption{In-medium results for the same observables as in Fig.~\ref{fig:atlas-vac}. In both rows, the main panel shows the distributions compared to experimental data in PbPb using the $P^{\text R}_{2}$ energy-loss model, the middle panel presents the theory-to-data ratio for both $P^\text{R}_2$ (solid band) and $P^\text{R}_1$ (chequered band) and the lower panel the medium-to-vacuum ratio again for both energy loss models. The vacuum distribution is computed using the $\texttt{POWHEG+PYTHIA8}$ interface.}
\label{fig:atlas-med}
\end{figure*}
We show results for the double differential distribution to produce jets with certain $(p_{t,\rm jet},r_g)$ values measured by ATLAS in Fig.~\ref{fig:atlas-med}. We again find that our theoretical calculation is able to describe the experimental data with a high-level of precision. This is a highly non-trivial result given that, for instance, the $p_{t,\rm jet}$ differential cross section spans four orders of magnitude and we are able to describe it even when sliced in $r_g$-intervals. The largest deviation occurs for broad jets. Since the calculation is still in agreement with data within experimental uncertainties we postpone the investigation of this potential deviation until its either confirmed or refuted by more precise experimental data. 
The medium-to-vacuum ratio decreases smoothly with $r_g$ thus confirming that collimated structures loose less energy than broader ones. This is compatible with the dynamics encoded in the $P_2^{\text{R}}$ distribution for which jets with $r_g\gg \theta_c$ loose more energy (smaller $R_{AA}$) than jets with $r_g\ll \theta_c$. However, similar to what was observed for the ALICE kinematics, there is also an overall agreement with data when using the $P^{\rm R}_1$ energy-loss distribution. This more differential observable nevertheless reveals some differences between $P_2^{\text{R}}$ and $P_1^{\text{R}}$ theory-to-data in the large-$r_g$ region. For instance, focusing on the $r_g$ distribution we find that the $P^{\rm R}_1$ model misses the $r_g>0.2$ region for all $p_{t,\jet}$-intervals, with the discrepancy being larger for higher-$p_t$ jets. This is also the case when fixing the $r_g$ interval to $0.26<r_g<0.4$ and looking at the $p_{t,\rm jet}$-spectrum. The statistical significance of such deviations is however not sufficient to claim evidence for colour coherence effects. Nevertheless, we think that these types of observables where jet substructure is correlated with global jet properties are a promising avenue for finding univocal signatures of colour coherence effects.

\section{Summary and outlook}
\label{sec:conclusions}
The study of jet substructure observables in heavy-ion collisions started a decade ago with the ambitious goal of unravelling the spatio-temporal evolution of in-medium jets. The experimental jet substructure program has successfully delivered numerous jet substructure measurements based both on the jet clustering tree~\cite{CMS:2017qlm,ALICE:2019ykw,STAR:2021kjt,ALargeIonColliderExperiment:2021mqf,ALICE:2024fip,ALICE:2024jtb} as well as on energy correlations~\cite{CMS:2025ydi}. Recent theory-to-data comparisons have revealed that a clean separation between different medium modifications is rather challenging~\cite{Andres:2022ovj,Cunqueiro:2023vxl,Barata:2023bhh,Barata:2023zqg,Yang:2023dwc,Fu:2024pic,Mehtar-Tani:2024jtd,Xing:2024yrb,Singh:2024vwb,Bossi:2024qho,Barata:2024ukm,Barata:2025uxp,Kudinoor:2025gao,Andres:2025yls,Apolinario:2025vtx}. In this paper we presented a minimal model for a subset of jet substructure observables which are built out of the kinematics of the subjet pair tagged by the SoftDrop algorithm. We showed theory-to-data comparisons for both ALICE and ATLAS kinematics.

Our starting hypothesis was to investigate to which extent can a minimal (and perhaps bold) model, in which the medium effectively shifts the $p_t$ of a vacuum-like SoftDrop pair can quantitatively describe current experimental data. This construction is motivated by a hierarchy of scales based on the kinematic configurations explored at the LHC. In particular, the formation time of the splitting tagged by SoftDrop is shorter than typical medium timescales and we thus ignore medium modifications to the splitting kernel itself. This assumption is adopted here to isolate energy-loss effects, but it does not preclude the known sensitivity of jet quenching and $\hat q$ extractions to the space–time structure of the parton shower~\cite{Andres:2024egc}. Technically, we simulate dijet production in proton-proton collisions matching the NLO matrix element to \texttt{PYTHIA8}'s parton shower using \texttt{POWHEG-BOX}. Note that NLO corrections to the hard-scattering description are often disregarded in heavy-ion studies. As expected, we found that their inclusion yields a better description of the proton-proton distributions. Perhaps more importantly, we found $15-30\%$ variations in the value of $\hat q$ extracted by fitting $R_{AA}$, see Table~\ref{tab:qhat_fits}, when accounting for these higher-order vacuum corrections. We thus conclude that NLO corrections must be taken into account in future extractions of $\hat q$ through global Bayesian analyses. 

The second ingredient of our calculation is the energy-loss distribution that was presented in Ref.~\cite{Mehtar-Tani:2017ypq} which includes a smooth transition between the coherent and incoherent regimes, see Eq.~\eqref{eq:P2R}. This calculation improves on previous jet substructure studies where colour coherence effects were included via a sharp cutoff leading to spiky curves around the critical angle $\theta_c$, see e.g. Ref.~\cite{Caucal:2021cfb}. We applied this energy loss distribution to the SoftDrop pair of subjets and found that this minimal setup is sufficient to obtain a quantitative description of the available groomed substructure measurements from both ALICE and ATLAS. This indicates that, within the present framework and experimental precision, energy-loss–based descriptions acting on a NLO-accurate vacuum baseline provide an adequate account of the observed medium modifications of groomed jet substructure. Note that using a coherent energy loss distribution which does not depend on jet substructure also leads to an overall agreement with experimental data, except for a couple of deviations in the tails of the ATLAS distributions, see Fig.~\ref{fig:atlas-med}. These deviations, however, are not statistically significant and thus, within our framework, both energy loss models give a similarly good description of experimental data within the current precision. We then look forward to the jet substructure program with Run 4 data, where statistical uncertainties are expected to shrink. 

A natural continuation of this work would be to consider an energy loss distribution which is sensitive to the number of resolved emissions in the vacuum cascade. This would require adapting the formalism first introduced in Ref.~\cite{Mehtar-Tani:2017web} to jet substructure observables and would provide a more theoretically-robust alternative to the all-orders computation presented in Sec.~\ref{sec:all-orders}. Another theoretical improvement of the present computation would be to explore IRC-safe definitions of jet-flavour following recent proposals~\cite{Caletti:2022hnc,Czakon:2022wam,Gauld:2022lem,Caola:2023wpj,Behring:2025ilo}. Phenomenologically, it would be interesting to check whether the success of our minimal setup persists when considering jet substructure observables at RHIC energies or for boson-tagged jets.  

Finally, we note that many of the effects discussed in this work are expected to become relevant in jets reconstructed with larger radii. Increasing the jet radius enhances sensitivity to wide-angle radiation, thereby amplifying differences between coherent and decoherent energy-loss regimes. At the same time, larger-$R$ jets can mitigate selection biases, particularly in lighter collision systems where background fluctuations are more manageable. Recent theoretical and phenomenological studies have emphasized that such configurations may offer a cleaner handle on colour coherence effects~\cite{Apolinario:2025fwd,Kudinoor:2025gao,Pablos:2025cli}. Extending groomed substructure measurements to larger jet radii, and systematically exploring their dependence on system size and collision energy, therefore appears as a promising avenue to sharpen the discrimination between competing quenching mechanisms and to further constrain the role of medium resolution in jet evolution.
\acknowledgments
 We thank Dhanush Anil Hangal for clarifying many aspects of ATLAS data. D.C has been supported by MCIN/AEI (10.13039/501100011033) and ERDF (grant PID2022-139466NB-C21), by Consejería de Universidad, Investigación e Innovación, Gobierno de España and Unión Europea – NextGenerationEU under grant AST22 6.5 and by the Ramón y Cajal program under grant RYC2022-037846-I. ASO is supported by the Ramón y Cajal program under grant RYC2022-037846-I and from ERDF (grant PID2024-161668NB-100). This work is also supported by Fundação para a Ciância e a Tecnologia (FCT), under ERC-PT A-Projects ‘Unveiling’, financed by PRR, NextGenerationEU. LA acknowledge support by FCT under contract 2021.03209.CEECIND.
 
\appendix
\section{Simplifying $P_\text{sing}$ with respect to Ref.~\cite{Mehtar-Tani:2017ypq} and sampling details}\label{app:Psing_derivation}
Our starting point is Eq.16 in Ref.~\cite{Mehtar-Tani:2017ypq} which displays the two-prong energy loss probability for a colour singlet antenna and reads:
\begin{align}\label{eq:Psing:yacine}
P_{\text {sing }}(\varepsilon,\theta_{12},L) & =\int_{\varepsilon_1, \varepsilon_2} P_1^\text{F}(\varepsilon_1, L) P_1^\text{F}(\varepsilon_2, L) \delta\left(\varepsilon-\varepsilon_1-\varepsilon_2\right) \nn
 &-2 \int_0^L \mathrm{~d} t \left[1-\Delta_{\text {med}}(t,\theta_{12})\right] \nn
 &\times \int_{\varepsilon_1,\varepsilon_2}P_1^\text{F}\left(\varepsilon_1, L-t\right)P_1^\text{F}\left(\varepsilon_2, L-t\right) \nn
 &\times \int_\omega \Gamma(\omega, t) \delta\left(\varepsilon-\varepsilon_1-\varepsilon_2-\omega\right),
\end{align}
where we have used the shorthand notation $\int_{x} = \int_0^\infty {\rm d}x$ and
\begin{equation}
P_1^\text{F}(\varepsilon, L) = \sqrt{2 \frac{\omega_s^\text{F}}{\varepsilon^3}}\exp\left[ -2 \pi \frac{\omega_s^{\text F}}{\varepsilon}\right], \quad \omega_s^\text{F} = \frac{\alpha_s C_\text{F}}{2\pi}\hat q L^2\, ,
\end{equation}
corresponds to the one-prong energy loss probability distribution for a quark with a characteristic in-medium frequency $\omega_s^\text{F}$. The other two ingredients entering Eq.~\eqref{eq:Psing:yacine} are the decoherence parameter 
\begin{equation}\label{eq:deltaMed}
\Delta_{\text {med}}(t,\theta_{12}) = 1 - \exp\left[ -\frac{1}{12}\hat q \theta^2_{12}t^3\right]\, ,
\end{equation}
which encodes the medium probability to resolve an antenna with opening angle $\theta_{12}$, and the regularized splitting rate
\begin{equation}\label{eq:Gamma}
\Gamma(\omega, t) = \frac{\alpha_s C_\text{F}}{\pi} \sqrt{\frac{\hat q}{\omega^3}} - \delta(\omega)\int_{\omega'} \frac{\alpha_s C_\text{F}}{\pi}\sqrt{\frac{\hat q}{\omega'^3}}\, .
\end{equation}

We can further split Eq.~\eqref{eq:Psing:yacine} into coherent and incoherent contributions, namely:
\begin{equation}
P_{\text {sing }}(\varepsilon,\theta_{12},L) = P_{\text {inc}}(\varepsilon,L) + P_{\text{coh}}(\varepsilon,\theta_{12},L) \, ,
\end{equation}
with the incoherent contribution describing the independent energy loss of the two antenna legs and given by 
\begin{equation}\label{eq:Pinc}
    P_{\text {inc}}(\varepsilon,L)  =\int_{\varepsilon_1,\varepsilon_2} P_1^\text{F}(\varepsilon_1, L) P_1^\text{F}(\varepsilon_2, L) \delta(\varepsilon-\varepsilon_1-\varepsilon_2)
\end{equation}
and the coherent contribution which accounts for interferences reads
\begin{align}\label{eq:Pcoh}
    P_{\text{coh}}(\varepsilon,\theta_{12},L) &=-2 \int_0^L \mathrm{~d} t \left[1-\Delta_{\text {med}}
    (t,\theta_{12})\right]  \nn
    &\times \int_{\varepsilon_1,\varepsilon_2} P_1^\text{F}(\varepsilon_1, L-t)P_1^\text{F}(\varepsilon_2, L-t) \nn
   & \times \int_ \omega \Gamma(\omega, t) \delta(\varepsilon-\varepsilon_1-\varepsilon_2-\omega).
\end{align}

Let us first simplify the incoherent contribution. By performing the $\delta$-integration in Eq.~\eqref{eq:Pinc} one obtains
\begin{equation}
    P_{\text {inc}}(\varepsilon) = \int_{\varepsilon_1} \dfrac{2 \omega_s^\text{F}}{\varepsilon_1^{3/2}(\varepsilon - \varepsilon_1)^{3/2}}\exp\left[-\frac{2\pi\varepsilon \omega_s^\text{F}}{\varepsilon_1(\varepsilon-\varepsilon_1)}  \, \right] \Theta(\varepsilon-\varepsilon_1).
\end{equation}
Changing variables to $y=\varepsilon_1(\varepsilon- \varepsilon_1)$ one obtains $\Theta(\varepsilon-\varepsilon_1) \rightarrow \Theta(\varepsilon \mp \sqrt{(\varepsilon^2 - 4y)})$. Analysing the domain of both branches we have,
\begin{align}
    \varepsilon_1 &= \frac{1}{2}\left[ \varepsilon \pm \sqrt{\varepsilon^2 - 4y}\right] \nn
    &= 
\begin{cases}
\dfrac{1}{2} \left[ \varepsilon - \sqrt{\varepsilon^2 - 4y} \right], &  0 < \varepsilon_1 < \frac{\varepsilon}{2} \\
\\
\dfrac{1}{2} \left[ \varepsilon + \sqrt{\varepsilon^2 - 4y} \right], &  \frac{\varepsilon}{2} < \varepsilon_1 < \varepsilon \, .
\end{cases}
\end{align}
For the first branch, $\varepsilon_1(y=0) = 0$  and  $\varepsilon_1(y=\varepsilon^2/4) = \varepsilon/2$, while for the second branch the integration limits are inverted since $\varepsilon_1(y=0) = \varepsilon$ and $\varepsilon_1(y-\varepsilon^2/4) = \varepsilon/2$. Noting that the Jacobian associated to the change of variables brings an extra factor $d\varepsilon_1/dy = \mp 1 /\sqrt{\varepsilon^2 - 4y}$, we conclude that both branches give the same contribution. We thus have:
\begin{align}
    P_\text{inc}(\varepsilon,L) &= 2 \int_0^{\varepsilon^2/4} \mathrm{~d}y \, \frac{2 \omega_s^F}{y^{3/2} \sqrt{\varepsilon^2 - 4y}} \exp\left(-\frac{2 \pi \varepsilon \omega_s^\text{F}}{y}\right)   \nn
    & = \sqrt{\frac{8\omega_s^\text{F}}{\varepsilon^3}}\exp\left( -\frac{8 \pi \omega_s^\text{F}}{\varepsilon}\right) \nn
    & = P_1^\text{A}(\varepsilon, L),
\end{align}
with $P_1^{\text A}(\varepsilon, L)$ corresponding to the energy loss of a single parton in the adjoint representation, which is consistent with the large-$N_c$ approximation. 

For the coherent contribution we first perform the $\varepsilon_2$-integration using the $\delta$-function in Eq.~\eqref{eq:Pcoh} and obtain:
\begin{align}
\begin{split}
     P_{\text{coh}}(\varepsilon,\theta_{12},L) &=-2 \int_0^L \mathrm{~d} t \left[1-\Delta_{\text {med}}
    (t,\theta_{12})\right]  \nn
    &\times \int_0^\varepsilon\Gamma(\omega, t)\int_0^{\varepsilon-\omega} {\rm d}\varepsilon_1 P_1^{\text F}(\varepsilon_1, L-t) \nn
    &\times P_1^{\text F}(\varepsilon - \varepsilon_1 - \omega, L-t)\, .
\end{split}
\end{align}
The $\varepsilon_1$-integration resembles that of the incoherent contribution and thus we can apply the same logic. Further plugging the definition of $\Gamma(\omega, t)$ as given by Eq.~\eqref{eq:Gamma} we get  
\begin{align}
P_{\text{coh}}(\varepsilon,\theta_{12},L) &=-2 \int_0^L \mathrm{~d} t \left[1-\Delta_{\text {med}}
    (t,\theta_{12})\right]C(t)  \nn
    &\times \left[\underbrace{\int_0^{\varepsilon} \dd\omega \, 
\frac{\exp\left(-\frac{8\pi(L - t)^2 \omega_r^{\text F} }{\varepsilon - \omega}\right)}
{(\varepsilon - \omega)^{3/2} \omega^{3/2}}}_{\beta}\right.\nn
&
\left.- \underbrace{
\frac{\exp\left(-\frac{8\pi(L - t)^2\omega_r^{\text F} }{\varepsilon}\right)}{\varepsilon^{3/2}}
\int_0^{\infty} \frac{{\rm d}\omega'}{{\omega'}^{3/2}}
}_{\gamma}\right]\, ,
\end{align}
where $\omega_r^{\text F} = \alpha_s^2C_{\text F}^2\hat q/(2\pi^2)$, and $C(t) = 4 (L-t) \omega_r^{\text F}$. Note that the $\beta$-term can be rewritten with the following change of variables:
\begin{align}
\label{eq:beta}
    \beta \xrightarrow{x = \varepsilon-\omega} &\int_0^\varepsilon \mathrm{~d}x \frac{\exp(-a/x)}{x^{3/2}(\varepsilon-x)^{3/2}} \nn
&\xrightarrow{u=1/x}\int_{1/\varepsilon}^\infty {\rm d}u \frac{\exp(-au)u^{-1/2}}{(\varepsilon-1/u)}, 
\end{align}
with $a = 8\pi(L-t)^2\omega_r^{\text F}$. The resulting integral in Eq.~\eqref{eq:beta} yields two contributions that we denote $\beta_{F}$ (finite) and $\beta_{D}$ (divergent). The latter is given by
\begin{equation}
    \beta_{D} = -2 \left[ \exp\left(-\frac{8 \pi (L-t)^2\omega_r^{\text F}}{\varepsilon-\omega} \right) \sqrt{\frac{(\varepsilon-\omega)}{\varepsilon^4 \omega}}\right]_{\omega=0}^{\omega=\epsilon}.
\end{equation}
The upper limit of integration gives zero while the lower one gives a divergence that can be combined with $\gamma$ to yield,
\begin{align}
        \lim_{\omega \rightarrow 0} (\beta_{D} + \gamma)& \propto   \lim_{\omega \rightarrow 0} \frac{1}{\sqrt{\omega \varepsilon}}\left[ \sqrt{\varepsilon- \omega} - \sqrt{\varepsilon}\right]\nn
        &=
        -\lim_{\omega \rightarrow 0} \frac{\sqrt{\varepsilon \omega}}{\varepsilon \sqrt{(\varepsilon-\omega)}} = 0 \, .
\end{align}
As for the remaining term of $\beta$ we have:
\begin{equation}
    \beta_{F} \propto \left[G(t, \varepsilon) \text{Erf}\left(\frac{\omega}{\varepsilon-\omega}
    \right) \right]_0^\varepsilon,
\end{equation}
with $\text{Erf}(z) = 2/\sqrt{\pi}\int_0^z {\rm d}t e^{-t^2}$. Combining all terms and constants we arrive at
\begin{align}\label{eq:pcoh-simple}
    P_\text{coh}(\varepsilon,\theta_{12},L)
    &= -
    \int_0^L \dd t P_1^{\text A}(\varepsilon, L-t) F(\varepsilon, L-t)\nn
    & \times \left[ 1- \Delta_\text{med}(t)\right]\, ,
\end{align}
with 
\begin{equation}\label{eq:f-def}
 F(\varepsilon, L-t) = \frac{1}{L-t} - 4\pi \omega_r^{\text A}\frac{L-t}{\varepsilon} \,.
\end{equation}
We have reduced our original expression to a single integration over $t$ which we do numerically. To sample $P^\text{R}_{2}(\varepsilon,\theta_{12},L)$ on an event-by-event level we first discretize and tabulate $P^\text{R}_{2}(\varepsilon,\theta_{12},L)$ as a function of $\theta_{12}$ using steps of $0.01$. Then, for a given antenna configuration with an associated $\theta_{12}$ value we sample a value of $\varepsilon$ from $P^\text{R}_{2}(\varepsilon,\theta_{12},L)$ using inversion sampling. For technical reasons, we don't sample an energy value from $0$ to $\infty$ but rather from $\varepsilon_{\rm min}=0.1$ GeV to $\varepsilon_{\rm max}=(150, 300)$ GeV, for ALICE and ATLAS respectively. The former improves the numerical convergence of the $t$-integration while the latter represents the maximum energy that a jet can loose and still contribute to the fiducial cross section. We have checked that the results are not sensitive to order one variations of these cuts.  
\section{Simulation details}\label{app:simulation}
To ensure the reproducibility of our results we provide in this appendix the run cards that we have used to produce the figures in the main text. We only highlight the modified parameters, all default settings are omitted. We used the following versions of the codes: \texttt{PYTHIA8.3}~\cite{Bierlich:2022pfr}, \texttt{MADGRAPH2.9}~\cite{Alwall:2014hca} and \texttt{POWHEG-BOX-V2}~\cite{Alioli:2010xa}. 
\begin{tcolorbox}[colback=violet!5, colframe=violet!80, title=MADGRAPH run\_card.dat, width=\columnwidth]
! Nb. of points per itegration channel \\
  0.015	= req\_acc\_fo ! Required accuracy \\

! Collider type and energy      \\                              
1	= lpp1 ! beam 1 type \\
1	= lpp2 ! beam 2 type \\
2510.0	= ebeam1 ! beam 1 energy in GeV \\
2510.0	= ebeam2 ! beam 2 energy in GeV \\
! PDF choice: this automatically fixes also $\alpha_s$(MZ) and its evol.  \\
lhapdf	= pdlabel ! PDF set \\
13100	= lhaid \\
! Renormalization and factorization scales  \\       
True	= fixed\_ren\_scale ! use fixed ren scale \\
False	= fixed\_fac\_scale ! use fixed fac scale \\
36.0	= mur\_ref\_fixed ! fixed ren reference scale \\
91.118	= muf\_ref\_fixed ! fixed fact reference scale \\
2	= dynamical\_scale\_choice \\

! Scale dependence and PDF uncertainty \\
False	= reweight\_scale ! Reweight to get scale variation \\
False	= reweight\_pdf ! Reweight to get PDF uncertainty \\

! Cuts on the jets (for ALICE) \\    
 -1.0	= jetalgo ! FastJet (1=kT, 0=C/A, -1=anti-kT) \\
 0.4	= jetradius ! The radius for the jet algorithm \\
40.0	= ptj ! Min jet transverse momentum \\
5.0	= etaj ! Max jet abs(pseudo-rap)  
\end{tcolorbox}
\begin{tcolorbox}[colback=gray!10, colframe=black!80, title={General PYTHIA8 settings}, width=\columnwidth]
! Settings used in the main program
\\
Main:numberOfEvents=10000000 !Nb. of events 
\\
\\
! Beam parameter settings.
\\
Beams:idA=2212                   \,\,\,! first proton beam
\\
Beams:idB=2212                   \,\,\,\,! second proton beam
\\
Beams:eCM=5020                   ! CM energy of collision
\\
\\
! Settings for the hard-process generation
\\
HardQCD:all = on                   
\\
!PhaseSpace:pTHatMin=70           \,! for ALICE
\\
PhaseSpace:pTHatMin=120           !  for ATLAS
\\
HadronLevel:all = on
\\
\\
! Shower model
\\
PartonShowers:Model = 1  ! 1:simple shower
\\
\\
! PDF Selection 
\\
 PDF:pSet=LHAPDF6:EPPS16nlo\_CT14nlo\_Pb208 ! For PbPb events
\\
! PDF:pSet = LHAPDF6:CT14nlo ! For pp Events
\end{tcolorbox}
\begin{tcolorbox}[colback=red!5, colframe=red!80, title=POWHEG-BOX LHE generation settings, width=\columnwidth]
numevts 10000000         ! number of events to be generated \\
ih1 1                  ! hadron 1 (1 for protons, -1 for antiprotons) \\
ih2 1                  ! hadron 2 (1 for protons, -1 for antiprotons) \\
ebeam1 2510d0          ! energy of beam 1 \\
ebeam2 2510d0          ! energy of beam 2 \\

! Generation cut: min. kt in underlying Born \\
bornktmin 70d0  ! ALICE    \\
bornktmin 120d0  ! ATLAS    \\

! To be set only if using LHA pdfs \\
! 13100 CT14nlo \\
! 901300 EPPS16nlo\_CT14nlo\_Pb208 \\
 lhans1  901300      ! pdf hadron 1  \\
 lhans2  901300      ! pdf hadron 2  \\
! lhans1  13100      ! pdf hadron 1  \\
! lhans2  13100      ! pdf hadron 2  \\

! Nb. calls for the importance sampling grid \\
ncall1 100000          \\
! No. iterations for grid \\
itmx1 5               \\
! No. calls for the upper bounding \\
ncall2 20000          \\ 
! No. iterations for grid \\ 
itmx2 5            \\

foldcsi 5            ! No. folds on csi integration \\
foldy   5            ! No. folds on  y  integration \\
foldphi 2            ! No. folds on phi integration \\

! No. calls to the upper bounding norms \\
nubound 500000       
\end{tcolorbox}

\begin{tcolorbox}[colback=blue!5, colframe=blue!80, title=Interface between POWHEG-BOX and PYTHIA, width=\columnwidth]
! Choose the shower model to match the POWHEG events to. \\
PartonShowers:model = 1 ! 1:simple shower\\ 

! Input file. \\
Beams:frameType = 4\\
Beams:LHEF = LHEF.lhe\\

! Number of outgoing particles of POWHEG Born level process \\
POWHEG:nFinal = 2 \\

! How vetoing is performed:\\
POWHEG:veto = 1 ! Showers are started at the kinematical limit (pThard $>$ pTemt)\\

! Nb. of verified accepted emissions \\
POWHEG:vetoCount = 100\\

! Selection of pThard: \\
POWHEG:pThard = 2  ! the $p_T$ of all final-state partons is tested against all other incoming and outgoing partons, with the minimal value chosen\\

! Selection of pTemt:\\
POWHEG:pTemt = 0 ! pTemt is the $p_T$ of the emitted parton w.r.t. radiating parton\\

! Selection of emitted parton for FSR\\
POWHEG:emitted = 0 ! Pythia definition of emitted\\

! $p_T$ definitions\\
POWHEG:pTdef = 1 ! POWHEG ISR $p_T$ and FSR $d_{ij}$ definitions\\

! MPI vetoing\\
POWHEG:MPIveto = 0 ! No MPI vetoing is done\\

! QED vetoing\\
POWHEG:QEDveto = 2 ! QED vetoing is done for pTemt $> 0$. If a photon is found with $p_T>$pThard from the Born level process, the event is accepted and no further veto of this event is allowed (for any pTemt).
\end{tcolorbox}

\bibliography{references}

\begin{thebibliography}{90}%
\makeatletter
\providecommand \@ifxundefined [1]{%
 \@ifx{#1\undefined}
}%
\providecommand \@ifnum [1]{%
 \ifnum #1\expandafter \@firstoftwo
 \else \expandafter \@secondoftwo
 \fi
}%
\providecommand \@ifx [1]{%
 \ifx #1\expandafter \@firstoftwo
 \else \expandafter \@secondoftwo
 \fi
}%
\providecommand \natexlab [1]{#1}%
\providecommand \enquote  [1]{``#1''}%
\providecommand \bibnamefont  [1]{#1}%
\providecommand \bibfnamefont [1]{#1}%
\providecommand \citenamefont [1]{#1}%
\providecommand \href@noop [0]{\@secondoftwo}%
\providecommand \href [0]{\begingroup \@sanitize@url \@href}%
\providecommand \@href[1]{\@@startlink{#1}\@@href}%
\providecommand \@@href[1]{\endgroup#1\@@endlink}%
\providecommand \@sanitize@url [0]{\catcode `\\12\catcode `\$12\catcode
  `\&12\catcode `\#12\catcode `\^12\catcode `\_12\catcode `\%12\relax}%
\providecommand \@@startlink[1]{}%
\providecommand \@@endlink[0]{}%
\providecommand \url  [0]{\begingroup\@sanitize@url \@url }%
\providecommand \@url [1]{\endgroup\@href {#1}{\urlprefix }}%
\providecommand \urlprefix  [0]{URL }%
\providecommand \Eprint [0]{\href }%
\providecommand \doibase [0]{http://dx.doi.org/}%
\providecommand \selectlanguage [0]{\@gobble}%
\providecommand \bibinfo  [0]{\@secondoftwo}%
\providecommand \bibfield  [0]{\@secondoftwo}%
\providecommand \translation [1]{[#1]}%
\providecommand \BibitemOpen [0]{}%
\providecommand \bibitemStop [0]{}%
\providecommand \bibitemNoStop [0]{.\EOS\space}%
\providecommand \EOS [0]{\spacefactor3000\relax}%
\providecommand \BibitemShut  [1]{\csname bibitem#1\endcsname}%
\let\auto@bib@innerbib\@empty
\bibitem [{\citenamefont {Busza}\ \emph {et~al.}(2018)\citenamefont {Busza},
  \citenamefont {Rajagopal},\ and\ \citenamefont {van~der
  Schee}}]{Busza:2018rrf}%
  \BibitemOpen
  \bibfield  {author} {\bibinfo {author} {\bibfnamefont {W.}~\bibnamefont
  {Busza}}, \bibinfo {author} {\bibfnamefont {K.}~\bibnamefont {Rajagopal}}, \
  and\ \bibinfo {author} {\bibfnamefont {W.}~\bibnamefont {van~der Schee}},\
  }\href {\doibase 10.1146/annurev-nucl-101917-020852} {\bibfield  {journal}
  {\bibinfo  {journal} {Ann. Rev. Nucl. Part. Sci.}\ }\textbf {\bibinfo
  {volume} {68}},\ \bibinfo {pages} {339} (\bibinfo {year} {2018})},\ \Eprint
  {http://arxiv.org/abs/1802.04801} {arXiv:1802.04801 [hep-ph]} \BibitemShut
  {NoStop}%
\bibitem [{\citenamefont {Wang}\ and\ \citenamefont
  {Wiedemann}(2025)}]{Wang:2025lct}%
  \BibitemOpen
  \bibfield  {author} {\bibinfo {author} {\bibfnamefont {X.-N.}\ \bibnamefont
  {Wang}}\ and\ \bibinfo {author} {\bibfnamefont {U.~A.}\ \bibnamefont
  {Wiedemann}}\ }(\bibinfo {year} {2025})\ \Eprint
  {http://arxiv.org/abs/2508.18794} {arXiv:2508.18794 [hep-ph]} \BibitemShut
  {NoStop}%
\bibitem [{\citenamefont {Abelev}\ \emph {et~al.}(2014)\citenamefont {Abelev}
  \emph {et~al.}}]{ALICE:2013dpt}%
  \BibitemOpen
  \bibfield  {author} {\bibinfo {author} {\bibfnamefont {B.}~\bibnamefont
  {Abelev}} \emph {et~al.} (\bibinfo {collaboration} {ALICE}),\ }\href
  {\doibase 10.1007/JHEP03(2014)013} {\bibfield  {journal} {\bibinfo  {journal}
  {JHEP}\ }\textbf {\bibinfo {volume} {03}},\ \bibinfo {pages} {013} (\bibinfo
  {year} {2014})},\ \Eprint {http://arxiv.org/abs/1311.0633} {arXiv:1311.0633
  [nucl-ex]} \BibitemShut {NoStop}%
\bibitem [{\citenamefont {Aad}\ \emph {et~al.}(2015)\citenamefont {Aad} \emph
  {et~al.}}]{ATLAS:2014ipv}%
  \BibitemOpen
  \bibfield  {author} {\bibinfo {author} {\bibfnamefont {G.}~\bibnamefont
  {Aad}} \emph {et~al.} (\bibinfo {collaboration} {ATLAS}),\ }\href {\doibase
  10.1103/PhysRevLett.114.072302} {\bibfield  {journal} {\bibinfo  {journal}
  {Phys. Rev. Lett.}\ }\textbf {\bibinfo {volume} {114}},\ \bibinfo {pages}
  {072302} (\bibinfo {year} {2015})},\ \Eprint {http://arxiv.org/abs/1411.2357}
  {arXiv:1411.2357 [hep-ex]} \BibitemShut {NoStop}%
\bibitem [{\citenamefont {Khachatryan}\ \emph {et~al.}(2017)\citenamefont
  {Khachatryan} \emph {et~al.}}]{CMS:2016uxf}%
  \BibitemOpen
  \bibfield  {author} {\bibinfo {author} {\bibfnamefont {V.}~\bibnamefont
  {Khachatryan}} \emph {et~al.} (\bibinfo {collaboration} {CMS}),\ }\href
  {\doibase 10.1103/PhysRevC.96.015202} {\bibfield  {journal} {\bibinfo
  {journal} {Phys. Rev. C}\ }\textbf {\bibinfo {volume} {96}},\ \bibinfo
  {pages} {015202} (\bibinfo {year} {2017})},\ \Eprint
  {http://arxiv.org/abs/1609.05383} {arXiv:1609.05383 [nucl-ex]} \BibitemShut
  {NoStop}%
\bibitem [{\citenamefont {Adam}\ \emph {et~al.}(2020)\citenamefont {Adam} \emph
  {et~al.}}]{STAR:2020xiv}%
  \BibitemOpen
  \bibfield  {author} {\bibinfo {author} {\bibfnamefont {J.}~\bibnamefont
  {Adam}} \emph {et~al.} (\bibinfo {collaboration} {STAR}),\ }\href {\doibase
  10.1103/PhysRevC.102.054913} {\bibfield  {journal} {\bibinfo  {journal}
  {Phys. Rev. C}\ }\textbf {\bibinfo {volume} {102}},\ \bibinfo {pages}
  {054913} (\bibinfo {year} {2020})},\ \Eprint
  {http://arxiv.org/abs/2006.00582} {arXiv:2006.00582 [nucl-ex]} \BibitemShut
  {NoStop}%
\bibitem [{\citenamefont {Aad}\ \emph {et~al.}(2010)\citenamefont {Aad} \emph
  {et~al.}}]{ATLAS:2010isq}%
  \BibitemOpen
  \bibfield  {author} {\bibinfo {author} {\bibfnamefont {G.}~\bibnamefont
  {Aad}} \emph {et~al.} (\bibinfo {collaboration} {ATLAS}),\ }\href {\doibase
  10.1103/PhysRevLett.105.252303} {\bibfield  {journal} {\bibinfo  {journal}
  {Phys. Rev. Lett.}\ }\textbf {\bibinfo {volume} {105}},\ \bibinfo {pages}
  {252303} (\bibinfo {year} {2010})},\ \Eprint {http://arxiv.org/abs/1011.6182}
  {arXiv:1011.6182 [hep-ex]} \BibitemShut {NoStop}%
\bibitem [{\citenamefont {Chatrchyan}\ \emph {et~al.}(2011)\citenamefont
  {Chatrchyan} \emph {et~al.}}]{CMS:2011iwn}%
  \BibitemOpen
  \bibfield  {author} {\bibinfo {author} {\bibfnamefont {S.}~\bibnamefont
  {Chatrchyan}} \emph {et~al.} (\bibinfo {collaboration} {CMS}),\ }\href
  {\doibase 10.1103/PhysRevC.84.024906} {\bibfield  {journal} {\bibinfo
  {journal} {Phys. Rev. C}\ }\textbf {\bibinfo {volume} {84}},\ \bibinfo
  {pages} {024906} (\bibinfo {year} {2011})},\ \Eprint
  {http://arxiv.org/abs/1102.1957} {arXiv:1102.1957 [nucl-ex]} \BibitemShut
  {NoStop}%
\bibitem [{\citenamefont {Adamczyk}\ \emph {et~al.}(2017)\citenamefont
  {Adamczyk} \emph {et~al.}}]{STAR:2016dfv}%
  \BibitemOpen
  \bibfield  {author} {\bibinfo {author} {\bibfnamefont {L.}~\bibnamefont
  {Adamczyk}} \emph {et~al.} (\bibinfo {collaboration} {STAR}),\ }\href
  {\doibase 10.1103/PhysRevLett.119.062301} {\bibfield  {journal} {\bibinfo
  {journal} {Phys. Rev. Lett.}\ }\textbf {\bibinfo {volume} {119}},\ \bibinfo
  {pages} {062301} (\bibinfo {year} {2017})},\ \Eprint
  {http://arxiv.org/abs/1609.03878} {arXiv:1609.03878 [nucl-ex]} \BibitemShut
  {NoStop}%
\bibitem [{\citenamefont {Cunqueiro}\ and\ \citenamefont
  {Sickles}(2022)}]{Cunqueiro:2021wls}%
  \BibitemOpen
  \bibfield  {author} {\bibinfo {author} {\bibfnamefont {L.}~\bibnamefont
  {Cunqueiro}}\ and\ \bibinfo {author} {\bibfnamefont {A.~M.}\ \bibnamefont
  {Sickles}},\ }\href {\doibase 10.1016/j.ppnp.2022.103940} {\bibfield
  {journal} {\bibinfo  {journal} {Prog. Part. Nucl. Phys.}\ }\textbf {\bibinfo
  {volume} {124}},\ \bibinfo {pages} {103940} (\bibinfo {year} {2022})},\
  \Eprint {http://arxiv.org/abs/2110.14490} {arXiv:2110.14490 [nucl-ex]}
  \BibitemShut {NoStop}%
\bibitem [{\citenamefont {Apolin{\'a}rio}\ \emph {et~al.}(2022)\citenamefont
  {Apolin{\'a}rio}, \citenamefont {Lee},\ and\ \citenamefont
  {Winn}}]{Apolinario:2022vzg}%
  \BibitemOpen
  \bibfield  {author} {\bibinfo {author} {\bibfnamefont {L.}~\bibnamefont
  {Apolin{\'a}rio}}, \bibinfo {author} {\bibfnamefont {Y.-J.}\ \bibnamefont
  {Lee}}, \ and\ \bibinfo {author} {\bibfnamefont {M.}~\bibnamefont {Winn}},\
  }\href {\doibase 10.1016/j.ppnp.2022.103990} {\bibfield  {journal} {\bibinfo
  {journal} {Prog. Part. Nucl. Phys.}\ }\textbf {\bibinfo {volume} {127}},\
  \bibinfo {pages} {103990} (\bibinfo {year} {2022})},\ \Eprint
  {http://arxiv.org/abs/2203.16352} {arXiv:2203.16352 [hep-ph]} \BibitemShut
  {NoStop}%
\bibitem [{\citenamefont {Apolin{\'a}rio}\ \emph {et~al.}(2024)\citenamefont
  {Apolin{\'a}rio}, \citenamefont {Chien},\ and\ \citenamefont
  {Cunqueiro~Mendez}}]{Apolinario:2024equ}%
  \BibitemOpen
  \bibfield  {author} {\bibinfo {author} {\bibfnamefont {L.}~\bibnamefont
  {Apolin{\'a}rio}}, \bibinfo {author} {\bibfnamefont {Y.-T.}\ \bibnamefont
  {Chien}}, \ and\ \bibinfo {author} {\bibfnamefont {L.}~\bibnamefont
  {Cunqueiro~Mendez}},\ }\href {\doibase 10.1142/9789811294679_0002} {\bibfield
   {journal} {\bibinfo  {journal} {Int. J. Mod. Phys. E}\ }\textbf {\bibinfo
  {volume} {33}},\ \bibinfo {pages} {2430003} (\bibinfo {year}
  {2024})}\BibitemShut {NoStop}%
\bibitem [{\citenamefont {Dasgupta}\ \emph {et~al.}(2013)\citenamefont
  {Dasgupta}, \citenamefont {Fregoso}, \citenamefont {Marzani},\ and\
  \citenamefont {Salam}}]{Dasgupta:2013ihk}%
  \BibitemOpen
  \bibfield  {author} {\bibinfo {author} {\bibfnamefont {M.}~\bibnamefont
  {Dasgupta}}, \bibinfo {author} {\bibfnamefont {A.}~\bibnamefont {Fregoso}},
  \bibinfo {author} {\bibfnamefont {S.}~\bibnamefont {Marzani}}, \ and\
  \bibinfo {author} {\bibfnamefont {G.~P.}\ \bibnamefont {Salam}},\ }\href
  {\doibase 10.1007/JHEP09(2013)029} {\bibfield  {journal} {\bibinfo  {journal}
  {JHEP}\ }\textbf {\bibinfo {volume} {09}},\ \bibinfo {pages} {029} (\bibinfo
  {year} {2013})},\ \Eprint {http://arxiv.org/abs/1307.0007} {arXiv:1307.0007
  [hep-ph]} \BibitemShut {NoStop}%
\bibitem [{\citenamefont {Larkoski}\ \emph {et~al.}(2014)\citenamefont
  {Larkoski}, \citenamefont {Marzani}, \citenamefont {Soyez},\ and\
  \citenamefont {Thaler}}]{Larkoski:2014wba}%
  \BibitemOpen
  \bibfield  {author} {\bibinfo {author} {\bibfnamefont {A.~J.}\ \bibnamefont
  {Larkoski}}, \bibinfo {author} {\bibfnamefont {S.}~\bibnamefont {Marzani}},
  \bibinfo {author} {\bibfnamefont {G.}~\bibnamefont {Soyez}}, \ and\ \bibinfo
  {author} {\bibfnamefont {J.}~\bibnamefont {Thaler}},\ }\href {\doibase
  10.1007/JHEP05(2014)146} {\bibfield  {journal} {\bibinfo  {journal} {JHEP}\
  }\textbf {\bibinfo {volume} {05}},\ \bibinfo {pages} {146} (\bibinfo {year}
  {2014})},\ \Eprint {http://arxiv.org/abs/1402.2657} {arXiv:1402.2657
  [hep-ph]} \BibitemShut {NoStop}%
\bibitem [{\citenamefont {Mehtar-Tani}\ \emph {et~al.}(2020)\citenamefont
  {Mehtar-Tani}, \citenamefont {Soto-Ontoso},\ and\ \citenamefont
  {Tywoniuk}}]{Mehtar-Tani:2019rrk}%
  \BibitemOpen
  \bibfield  {author} {\bibinfo {author} {\bibfnamefont {Y.}~\bibnamefont
  {Mehtar-Tani}}, \bibinfo {author} {\bibfnamefont {A.}~\bibnamefont
  {Soto-Ontoso}}, \ and\ \bibinfo {author} {\bibfnamefont {K.}~\bibnamefont
  {Tywoniuk}},\ }\href {\doibase 10.1103/PhysRevD.101.034004} {\bibfield
  {journal} {\bibinfo  {journal} {Phys. Rev. D}\ }\textbf {\bibinfo {volume}
  {101}},\ \bibinfo {pages} {034004} (\bibinfo {year} {2020})},\ \Eprint
  {http://arxiv.org/abs/1911.00375} {arXiv:1911.00375 [hep-ph]} \BibitemShut
  {NoStop}%
\bibitem [{\citenamefont {Sirunyan}\ \emph {et~al.}(2018)\citenamefont
  {Sirunyan} \emph {et~al.}}]{CMS:2017qlm}%
  \BibitemOpen
  \bibfield  {author} {\bibinfo {author} {\bibfnamefont {A.~M.}\ \bibnamefont
  {Sirunyan}} \emph {et~al.} (\bibinfo {collaboration} {CMS}),\ }\href
  {\doibase 10.1103/PhysRevLett.120.142302} {\bibfield  {journal} {\bibinfo
  {journal} {Phys. Rev. Lett.}\ }\textbf {\bibinfo {volume} {120}},\ \bibinfo
  {pages} {142302} (\bibinfo {year} {2018})},\ \Eprint
  {http://arxiv.org/abs/1708.09429} {arXiv:1708.09429 [nucl-ex]} \BibitemShut
  {NoStop}%
\bibitem [{\citenamefont {Acharya}\ \emph {et~al.}(2020)\citenamefont {Acharya}
  \emph {et~al.}}]{ALICE:2019ykw}%
  \BibitemOpen
  \bibfield  {author} {\bibinfo {author} {\bibfnamefont {S.}~\bibnamefont
  {Acharya}} \emph {et~al.} (\bibinfo {collaboration} {ALICE}),\ }\href
  {\doibase 10.1016/j.physletb.2020.135227} {\bibfield  {journal} {\bibinfo
  {journal} {Phys. Lett. B}\ }\textbf {\bibinfo {volume} {802}},\ \bibinfo
  {pages} {135227} (\bibinfo {year} {2020})},\ \Eprint
  {http://arxiv.org/abs/1905.02512} {arXiv:1905.02512 [nucl-ex]} \BibitemShut
  {NoStop}%
\bibitem [{\citenamefont {Abdallah}\ \emph {et~al.}(2022)\citenamefont
  {Abdallah} \emph {et~al.}}]{STAR:2021kjt}%
  \BibitemOpen
  \bibfield  {author} {\bibinfo {author} {\bibfnamefont {M.~S.}\ \bibnamefont
  {Abdallah}} \emph {et~al.} (\bibinfo {collaboration} {STAR}),\ }\href
  {\doibase 10.1103/PhysRevC.105.044906} {\bibfield  {journal} {\bibinfo
  {journal} {Phys. Rev. C}\ }\textbf {\bibinfo {volume} {105}},\ \bibinfo
  {pages} {044906} (\bibinfo {year} {2022})},\ \Eprint
  {http://arxiv.org/abs/2109.09793} {arXiv:2109.09793 [nucl-ex]} \BibitemShut
  {NoStop}%
\bibitem [{\citenamefont {Acharya}\ \emph {et~al.}(2022)\citenamefont {Acharya}
  \emph {et~al.}}]{ALargeIonColliderExperiment:2021mqf}%
  \BibitemOpen
  \bibfield  {author} {\bibinfo {author} {\bibfnamefont {S.}~\bibnamefont
  {Acharya}} \emph {et~al.} (\bibinfo {collaboration} {A Large Ion Collider
  Experiment, ALICE}),\ }\href {\doibase 10.1103/PhysRevLett.128.102001}
  {\bibfield  {journal} {\bibinfo  {journal} {Phys. Rev. Lett.}\ }\textbf
  {\bibinfo {volume} {128}},\ \bibinfo {pages} {102001} (\bibinfo {year}
  {2022})},\ \Eprint {http://arxiv.org/abs/2107.12984} {arXiv:2107.12984
  [nucl-ex]} \BibitemShut {NoStop}%
\bibitem [{\citenamefont {Acharya}\ \emph
  {et~al.}(2025{\natexlab{a}})\citenamefont {Acharya} \emph
  {et~al.}}]{ALICE:2024fip}%
  \BibitemOpen
  \bibfield  {author} {\bibinfo {author} {\bibfnamefont {S.}~\bibnamefont
  {Acharya}} \emph {et~al.} (\bibinfo {collaboration} {ALICE}),\ }\href
  {\doibase 10.1103/PhysRevLett.135.031901} {\bibfield  {journal} {\bibinfo
  {journal} {Phys. Rev. Lett.}\ }\textbf {\bibinfo {volume} {135}},\ \bibinfo
  {pages} {031901} (\bibinfo {year} {2025}{\natexlab{a}})},\ \Eprint
  {http://arxiv.org/abs/2409.12837} {arXiv:2409.12837 [nucl-ex]} \BibitemShut
  {NoStop}%
\bibitem [{\citenamefont {Acharya}\ \emph
  {et~al.}(2025{\natexlab{b}})\citenamefont {Acharya} \emph
  {et~al.}}]{ALICE:2024jtb}%
  \BibitemOpen
  \bibfield  {author} {\bibinfo {author} {\bibfnamefont {S.}~\bibnamefont
  {Acharya}} \emph {et~al.} (\bibinfo {collaboration} {ALICE}),\ }\href
  {\doibase 10.1016/j.physletb.2025.139409} {\bibfield  {journal} {\bibinfo
  {journal} {Phys. Lett. B}\ }\textbf {\bibinfo {volume} {864}},\ \bibinfo
  {pages} {139409} (\bibinfo {year} {2025}{\natexlab{b}})},\ \Eprint
  {http://arxiv.org/abs/2411.03106} {arXiv:2411.03106 [nucl-ex]} \BibitemShut
  {NoStop}%
\bibitem [{\citenamefont {Hayrapetyan}\ \emph {et~al.}(2025)\citenamefont
  {Hayrapetyan} \emph {et~al.}}]{CMS:2024zjn}%
  \BibitemOpen
  \bibfield  {author} {\bibinfo {author} {\bibfnamefont {A.}~\bibnamefont
  {Hayrapetyan}} \emph {et~al.} (\bibinfo {collaboration} {CMS}),\ }\href
  {\doibase 10.1016/j.physletb.2024.139088} {\bibfield  {journal} {\bibinfo
  {journal} {Phys. Lett. B}\ }\textbf {\bibinfo {volume} {861}},\ \bibinfo
  {pages} {139088} (\bibinfo {year} {2025})},\ \Eprint
  {http://arxiv.org/abs/2405.02737} {arXiv:2405.02737 [nucl-ex]} \BibitemShut
  {NoStop}%
\bibitem [{\citenamefont {Aad}\ \emph {et~al.}(2023{\natexlab{a}})\citenamefont
  {Aad} \emph {et~al.}}]{ATLAS:2022vii}%
  \BibitemOpen
  \bibfield  {author} {\bibinfo {author} {\bibfnamefont {G.}~\bibnamefont
  {Aad}} \emph {et~al.} (\bibinfo {collaboration} {ATLAS}),\ }\href {\doibase
  10.1103/PhysRevC.107.054909} {\bibfield  {journal} {\bibinfo  {journal}
  {Phys. Rev. C}\ }\textbf {\bibinfo {volume} {107}},\ \bibinfo {pages}
  {054909} (\bibinfo {year} {2023}{\natexlab{a}})},\ \Eprint
  {http://arxiv.org/abs/2211.11470} {arXiv:2211.11470 [nucl-ex]} \BibitemShut
  {NoStop}%
\bibitem [{\citenamefont {Aad}\ \emph {et~al.}(2023{\natexlab{b}})\citenamefont
  {Aad} \emph {et~al.}}]{ATLAS:2023hso}%
  \BibitemOpen
  \bibfield  {author} {\bibinfo {author} {\bibfnamefont {G.}~\bibnamefont
  {Aad}} \emph {et~al.} (\bibinfo {collaboration} {ATLAS}),\ }\href {\doibase
  10.1103/PhysRevLett.131.172301} {\bibfield  {journal} {\bibinfo  {journal}
  {Phys. Rev. Lett.}\ }\textbf {\bibinfo {volume} {131}},\ \bibinfo {pages}
  {172301} (\bibinfo {year} {2023}{\natexlab{b}})},\ \Eprint
  {http://arxiv.org/abs/2301.05606} {arXiv:2301.05606 [nucl-ex]} \BibitemShut
  {NoStop}%
\bibitem [{\citenamefont {Aad}\ \emph {et~al.}(2025)\citenamefont {Aad} \emph
  {et~al.}}]{ATLAS:2025lfb}%
  \BibitemOpen
  \bibfield  {author} {\bibinfo {author} {\bibfnamefont {G.}~\bibnamefont
  {Aad}} \emph {et~al.} (\bibinfo {collaboration} {ATLAS}),\ }\href@noop {} {\
  (\bibinfo {year} {2025})},\ \Eprint {http://arxiv.org/abs/2504.04805}
  {arXiv:2504.04805 [nucl-ex]} \BibitemShut {NoStop}%
\bibitem [{\citenamefont {{The CMS Collaboration}}(2025)}]{CMS:2025gdw}%
  \BibitemOpen
  \bibfield  {author} {\bibinfo {author} {\bibnamefont {{The CMS
  Collaboration}}},\ }\href {https://cds.cern.ch/record/2924711} {\  (\bibinfo
  {year} {2025})}\BibitemShut {NoStop}%
\bibitem [{\citenamefont {Cunqueiro}\ \emph {et~al.}(2024)\citenamefont
  {Cunqueiro}, \citenamefont {Pablos}, \citenamefont {Soto-Ontoso},
  \citenamefont {Spousta}, \citenamefont {Takacs},\ and\ \citenamefont
  {Verweij}}]{Cunqueiro:2023vxl}%
  \BibitemOpen
  \bibfield  {author} {\bibinfo {author} {\bibfnamefont {L.}~\bibnamefont
  {Cunqueiro}}, \bibinfo {author} {\bibfnamefont {D.}~\bibnamefont {Pablos}},
  \bibinfo {author} {\bibfnamefont {A.}~\bibnamefont {Soto-Ontoso}}, \bibinfo
  {author} {\bibfnamefont {M.}~\bibnamefont {Spousta}}, \bibinfo {author}
  {\bibfnamefont {A.}~\bibnamefont {Takacs}}, \ and\ \bibinfo {author}
  {\bibfnamefont {M.}~\bibnamefont {Verweij}},\ }\href {\doibase
  10.1103/PhysRevD.110.014015} {\bibfield  {journal} {\bibinfo  {journal}
  {Phys. Rev. D}\ }\textbf {\bibinfo {volume} {110}},\ \bibinfo {pages}
  {014015} (\bibinfo {year} {2024})},\ \Eprint
  {http://arxiv.org/abs/2311.07643} {arXiv:2311.07643 [hep-ph]} \BibitemShut
  {NoStop}%
\bibitem [{\citenamefont {Spousta}\ and\ \citenamefont
  {Cole}(2016)}]{Spousta:2015fca}%
  \BibitemOpen
  \bibfield  {author} {\bibinfo {author} {\bibfnamefont {M.}~\bibnamefont
  {Spousta}}\ and\ \bibinfo {author} {\bibfnamefont {B.}~\bibnamefont {Cole}},\
  }\href {\doibase 10.1140/epjc/s10052-016-3896-0} {\bibfield  {journal}
  {\bibinfo  {journal} {Eur. Phys. J. C}\ }\textbf {\bibinfo {volume} {76}},\
  \bibinfo {pages} {50} (\bibinfo {year} {2016})},\ \Eprint
  {http://arxiv.org/abs/1504.05169} {arXiv:1504.05169 [hep-ph]} \BibitemShut
  {NoStop}%
\bibitem [{\citenamefont {Ringer}\ \emph {et~al.}(2020)\citenamefont {Ringer},
  \citenamefont {Xiao},\ and\ \citenamefont {Yuan}}]{Ringer:2019rfk}%
  \BibitemOpen
  \bibfield  {author} {\bibinfo {author} {\bibfnamefont {F.}~\bibnamefont
  {Ringer}}, \bibinfo {author} {\bibfnamefont {B.-W.}\ \bibnamefont {Xiao}}, \
  and\ \bibinfo {author} {\bibfnamefont {F.}~\bibnamefont {Yuan}},\ }\href
  {\doibase 10.1016/j.physletb.2020.135634} {\bibfield  {journal} {\bibinfo
  {journal} {Phys. Lett. B}\ }\textbf {\bibinfo {volume} {808}},\ \bibinfo
  {pages} {135634} (\bibinfo {year} {2020})},\ \Eprint
  {http://arxiv.org/abs/1907.12541} {arXiv:1907.12541 [hep-ph]} \BibitemShut
  {NoStop}%
\bibitem [{\citenamefont {Chien}\ and\ \citenamefont
  {Vitev}(2017)}]{Chien:2016led}%
  \BibitemOpen
  \bibfield  {author} {\bibinfo {author} {\bibfnamefont {Y.-T.}\ \bibnamefont
  {Chien}}\ and\ \bibinfo {author} {\bibfnamefont {I.}~\bibnamefont {Vitev}},\
  }\href {\doibase 10.1103/PhysRevLett.119.112301} {\bibfield  {journal}
  {\bibinfo  {journal} {Phys. Rev. Lett.}\ }\textbf {\bibinfo {volume} {119}},\
  \bibinfo {pages} {112301} (\bibinfo {year} {2017})},\ \Eprint
  {http://arxiv.org/abs/1608.07283} {arXiv:1608.07283 [hep-ph]} \BibitemShut
  {NoStop}%
\bibitem [{\citenamefont {Wang}\ \emph {et~al.}(2023)\citenamefont {Wang},
  \citenamefont {Kang}, \citenamefont {Zhang}, \citenamefont {Shen},
  \citenamefont {Dai}, \citenamefont {Zhang},\ and\ \citenamefont
  {Wang}}]{Wang:2022yrp}%
  \BibitemOpen
  \bibfield  {author} {\bibinfo {author} {\bibfnamefont {L.}~\bibnamefont
  {Wang}}, \bibinfo {author} {\bibfnamefont {J.-W.}\ \bibnamefont {Kang}},
  \bibinfo {author} {\bibfnamefont {Q.}~\bibnamefont {Zhang}}, \bibinfo
  {author} {\bibfnamefont {S.}~\bibnamefont {Shen}}, \bibinfo {author}
  {\bibfnamefont {W.}~\bibnamefont {Dai}}, \bibinfo {author} {\bibfnamefont
  {B.-W.}\ \bibnamefont {Zhang}}, \ and\ \bibinfo {author} {\bibfnamefont
  {E.}~\bibnamefont {Wang}},\ }\href {\doibase 10.1088/0256-307X/40/3/032101}
  {\bibfield  {journal} {\bibinfo  {journal} {Chin. Phys. Lett.}\ }\textbf
  {\bibinfo {volume} {40}},\ \bibinfo {pages} {032101} (\bibinfo {year}
  {2023})},\ \Eprint {http://arxiv.org/abs/2211.13674} {arXiv:2211.13674
  [nucl-th]} \BibitemShut {NoStop}%
\bibitem [{\citenamefont {Milhano}\ \emph {et~al.}(2018)\citenamefont
  {Milhano}, \citenamefont {Wiedemann},\ and\ \citenamefont
  {Zapp}}]{Milhano:2017nzm}%
  \BibitemOpen
  \bibfield  {author} {\bibinfo {author} {\bibfnamefont {G.}~\bibnamefont
  {Milhano}}, \bibinfo {author} {\bibfnamefont {U.~A.}\ \bibnamefont
  {Wiedemann}}, \ and\ \bibinfo {author} {\bibfnamefont {K.~C.}\ \bibnamefont
  {Zapp}},\ }\href {\doibase 10.1016/j.physletb.2018.01.029} {\bibfield
  {journal} {\bibinfo  {journal} {Phys. Lett. B}\ }\textbf {\bibinfo {volume}
  {779}},\ \bibinfo {pages} {409} (\bibinfo {year} {2018})},\ \Eprint
  {http://arxiv.org/abs/1707.04142} {arXiv:1707.04142 [hep-ph]} \BibitemShut
  {NoStop}%
\bibitem [{\citenamefont {Mehtar-Tani}\ \emph {et~al.}(2011)\citenamefont
  {Mehtar-Tani}, \citenamefont {Salgado},\ and\ \citenamefont
  {Tywoniuk}}]{Mehtar-Tani:2010ebp}%
  \BibitemOpen
  \bibfield  {author} {\bibinfo {author} {\bibfnamefont {Y.}~\bibnamefont
  {Mehtar-Tani}}, \bibinfo {author} {\bibfnamefont {C.~A.}\ \bibnamefont
  {Salgado}}, \ and\ \bibinfo {author} {\bibfnamefont {K.}~\bibnamefont
  {Tywoniuk}},\ }\href {\doibase 10.1103/PhysRevLett.106.122002} {\bibfield
  {journal} {\bibinfo  {journal} {Phys. Rev. Lett.}\ }\textbf {\bibinfo
  {volume} {106}},\ \bibinfo {pages} {122002} (\bibinfo {year} {2011})},\
  \Eprint {http://arxiv.org/abs/1009.2965} {arXiv:1009.2965 [hep-ph]}
  \BibitemShut {NoStop}%
\bibitem [{\citenamefont {Mehtar-Tani}\ \emph {et~al.}(2012)\citenamefont
  {Mehtar-Tani}, \citenamefont {Salgado},\ and\ \citenamefont
  {Tywoniuk}}]{Mehtar-Tani:2011hma}%
  \BibitemOpen
  \bibfield  {author} {\bibinfo {author} {\bibfnamefont {Y.}~\bibnamefont
  {Mehtar-Tani}}, \bibinfo {author} {\bibfnamefont {C.~A.}\ \bibnamefont
  {Salgado}}, \ and\ \bibinfo {author} {\bibfnamefont {K.}~\bibnamefont
  {Tywoniuk}},\ }\href {\doibase 10.1016/j.physletb.2011.12.042} {\bibfield
  {journal} {\bibinfo  {journal} {Phys. Lett. B}\ }\textbf {\bibinfo {volume}
  {707}},\ \bibinfo {pages} {156} (\bibinfo {year} {2012})},\ \Eprint
  {http://arxiv.org/abs/1102.4317} {arXiv:1102.4317 [hep-ph]} \BibitemShut
  {NoStop}%
\bibitem [{\citenamefont {Casalderrey-Solana}\ and\ \citenamefont
  {Iancu}(2011)}]{Casalderrey-Solana:2011ule}%
  \BibitemOpen
  \bibfield  {author} {\bibinfo {author} {\bibfnamefont {J.}~\bibnamefont
  {Casalderrey-Solana}}\ and\ \bibinfo {author} {\bibfnamefont
  {E.}~\bibnamefont {Iancu}},\ }\href {\doibase 10.1007/JHEP08(2011)015}
  {\bibfield  {journal} {\bibinfo  {journal} {JHEP}\ }\textbf {\bibinfo
  {volume} {08}},\ \bibinfo {pages} {015} (\bibinfo {year} {2011})},\ \Eprint
  {http://arxiv.org/abs/1105.1760} {arXiv:1105.1760 [hep-ph]} \BibitemShut
  {NoStop}%
\bibitem [{\citenamefont {Mehtar-Tani}\ and\ \citenamefont
  {Tywoniuk}(2013)}]{Mehtar-Tani:2011vlz}%
  \BibitemOpen
  \bibfield  {author} {\bibinfo {author} {\bibfnamefont {Y.}~\bibnamefont
  {Mehtar-Tani}}\ and\ \bibinfo {author} {\bibfnamefont {K.}~\bibnamefont
  {Tywoniuk}},\ }\href {\doibase 10.1007/JHEP01(2013)031} {\bibfield  {journal}
  {\bibinfo  {journal} {JHEP}\ }\textbf {\bibinfo {volume} {01}},\ \bibinfo
  {pages} {031} (\bibinfo {year} {2013})},\ \Eprint
  {http://arxiv.org/abs/1105.1346} {arXiv:1105.1346 [hep-ph]} \BibitemShut
  {NoStop}%
\bibitem [{\citenamefont {Casalderrey-Solana}\ \emph
  {et~al.}(2013)\citenamefont {Casalderrey-Solana}, \citenamefont
  {Mehtar-Tani}, \citenamefont {Salgado},\ and\ \citenamefont
  {Tywoniuk}}]{Casalderrey-Solana:2012evi}%
  \BibitemOpen
  \bibfield  {author} {\bibinfo {author} {\bibfnamefont {J.}~\bibnamefont
  {Casalderrey-Solana}}, \bibinfo {author} {\bibfnamefont {Y.}~\bibnamefont
  {Mehtar-Tani}}, \bibinfo {author} {\bibfnamefont {C.~A.}\ \bibnamefont
  {Salgado}}, \ and\ \bibinfo {author} {\bibfnamefont {K.}~\bibnamefont
  {Tywoniuk}},\ }\href {\doibase 10.1016/j.physletb.2013.07.046} {\bibfield
  {journal} {\bibinfo  {journal} {Phys. Lett. B}\ }\textbf {\bibinfo {volume}
  {725}},\ \bibinfo {pages} {357} (\bibinfo {year} {2013})},\ \Eprint
  {http://arxiv.org/abs/1210.7765} {arXiv:1210.7765 [hep-ph]} \BibitemShut
  {NoStop}%
\bibitem [{\citenamefont {Apolin\'ario}\ \emph {et~al.}(2015)\citenamefont
  {Apolin\'ario}, \citenamefont {Armesto}, \citenamefont {Milhano},\ and\
  \citenamefont {Salgado}}]{Apolinario:2014csa}%
  \BibitemOpen
  \bibfield  {author} {\bibinfo {author} {\bibfnamefont {L.}~\bibnamefont
  {Apolin\'ario}}, \bibinfo {author} {\bibfnamefont {N.}~\bibnamefont
  {Armesto}}, \bibinfo {author} {\bibfnamefont {J.~G.}\ \bibnamefont
  {Milhano}}, \ and\ \bibinfo {author} {\bibfnamefont {C.~A.}\ \bibnamefont
  {Salgado}},\ }\href {\doibase 10.1007/JHEP02(2015)119} {\bibfield  {journal}
  {\bibinfo  {journal} {JHEP}\ }\textbf {\bibinfo {volume} {02}},\ \bibinfo
  {pages} {119} (\bibinfo {year} {2015})},\ \Eprint
  {http://arxiv.org/abs/1407.0599} {arXiv:1407.0599 [hep-ph]} \BibitemShut
  {NoStop}%
\bibitem [{\citenamefont {Abreu}\ \emph {et~al.}(2025)\citenamefont {Abreu},
  \citenamefont {Mayo~L{\'o}pez}, \citenamefont {Milhano},\ and\ \citenamefont
  {Soto-Ontoso}}]{Abreu:2024wka}%
  \BibitemOpen
  \bibfield  {author} {\bibinfo {author} {\bibfnamefont {S.}~\bibnamefont
  {Abreu}}, \bibinfo {author} {\bibfnamefont {X.}~\bibnamefont
  {Mayo~L{\'o}pez}}, \bibinfo {author} {\bibfnamefont {G.}~\bibnamefont
  {Milhano}}, \ and\ \bibinfo {author} {\bibfnamefont {A.}~\bibnamefont
  {Soto-Ontoso}},\ }\href {\doibase 10.1007/JHEP03(2025)216} {\bibfield
  {journal} {\bibinfo  {journal} {JHEP}\ }\textbf {\bibinfo {volume} {03}},\
  \bibinfo {pages} {216} (\bibinfo {year} {2025})},\ \Eprint
  {http://arxiv.org/abs/2410.24135} {arXiv:2410.24135 [hep-ph]} \BibitemShut
  {NoStop}%
\bibitem [{\citenamefont {Mehtar-Tani}\ and\ \citenamefont
  {Tywoniuk}(2017)}]{Mehtar-Tani:2016aco}%
  \BibitemOpen
  \bibfield  {author} {\bibinfo {author} {\bibfnamefont {Y.}~\bibnamefont
  {Mehtar-Tani}}\ and\ \bibinfo {author} {\bibfnamefont {K.}~\bibnamefont
  {Tywoniuk}},\ }\href {\doibase 10.1007/JHEP04(2017)125} {\bibfield  {journal}
  {\bibinfo  {journal} {JHEP}\ }\textbf {\bibinfo {volume} {04}},\ \bibinfo
  {pages} {125} (\bibinfo {year} {2017})},\ \Eprint
  {http://arxiv.org/abs/1610.08930} {arXiv:1610.08930 [hep-ph]} \BibitemShut
  {NoStop}%
\bibitem [{\citenamefont {Caucal}\ \emph {et~al.}(2022)\citenamefont {Caucal},
  \citenamefont {Soto-Ontoso},\ and\ \citenamefont {Takacs}}]{Caucal:2021cfb}%
  \BibitemOpen
  \bibfield  {author} {\bibinfo {author} {\bibfnamefont {P.}~\bibnamefont
  {Caucal}}, \bibinfo {author} {\bibfnamefont {A.}~\bibnamefont {Soto-Ontoso}},
  \ and\ \bibinfo {author} {\bibfnamefont {A.}~\bibnamefont {Takacs}},\ }\href
  {\doibase 10.1103/PhysRevD.105.114046} {\bibfield  {journal} {\bibinfo
  {journal} {Phys. Rev. D}\ }\textbf {\bibinfo {volume} {105}},\ \bibinfo
  {pages} {114046} (\bibinfo {year} {2022})},\ \Eprint
  {http://arxiv.org/abs/2111.14768} {arXiv:2111.14768 [hep-ph]} \BibitemShut
  {NoStop}%
\bibitem [{\citenamefont {Pablos}\ and\ \citenamefont
  {Soto-Ontoso}(2023)}]{Pablos:2022mrx}%
  \BibitemOpen
  \bibfield  {author} {\bibinfo {author} {\bibfnamefont {D.}~\bibnamefont
  {Pablos}}\ and\ \bibinfo {author} {\bibfnamefont {A.}~\bibnamefont
  {Soto-Ontoso}},\ }\href {\doibase 10.1103/PhysRevD.107.094003} {\bibfield
  {journal} {\bibinfo  {journal} {Phys. Rev. D}\ }\textbf {\bibinfo {volume}
  {107}},\ \bibinfo {pages} {094003} (\bibinfo {year} {2023})},\ \Eprint
  {http://arxiv.org/abs/2210.07901} {arXiv:2210.07901 [hep-ph]} \BibitemShut
  {NoStop}%
\bibitem [{\citenamefont {Caucal}\ \emph {et~al.}(2018)\citenamefont {Caucal},
  \citenamefont {Iancu}, \citenamefont {Mueller},\ and\ \citenamefont
  {Soyez}}]{Caucal:2018dla}%
  \BibitemOpen
  \bibfield  {author} {\bibinfo {author} {\bibfnamefont {P.}~\bibnamefont
  {Caucal}}, \bibinfo {author} {\bibfnamefont {E.}~\bibnamefont {Iancu}},
  \bibinfo {author} {\bibfnamefont {A.~H.}\ \bibnamefont {Mueller}}, \ and\
  \bibinfo {author} {\bibfnamefont {G.}~\bibnamefont {Soyez}},\ }\href
  {\doibase 10.1103/PhysRevLett.120.232001} {\bibfield  {journal} {\bibinfo
  {journal} {Phys. Rev. Lett.}\ }\textbf {\bibinfo {volume} {120}},\ \bibinfo
  {pages} {232001} (\bibinfo {year} {2018})},\ \Eprint
  {http://arxiv.org/abs/1801.09703} {arXiv:1801.09703 [hep-ph]} \BibitemShut
  {NoStop}%
\bibitem [{\citenamefont {Caucal}\ \emph {et~al.}(2019)\citenamefont {Caucal},
  \citenamefont {Iancu},\ and\ \citenamefont {Soyez}}]{Caucal:2019uvr}%
  \BibitemOpen
  \bibfield  {author} {\bibinfo {author} {\bibfnamefont {P.}~\bibnamefont
  {Caucal}}, \bibinfo {author} {\bibfnamefont {E.}~\bibnamefont {Iancu}}, \
  and\ \bibinfo {author} {\bibfnamefont {G.}~\bibnamefont {Soyez}},\ }\href
  {\doibase 10.1007/JHEP10(2019)273} {\bibfield  {journal} {\bibinfo  {journal}
  {JHEP}\ }\textbf {\bibinfo {volume} {10}},\ \bibinfo {pages} {273} (\bibinfo
  {year} {2019})},\ \Eprint {http://arxiv.org/abs/1907.04866} {arXiv:1907.04866
  [hep-ph]} \BibitemShut {NoStop}%
\bibitem [{\citenamefont {Casalderrey-Solana}\ \emph
  {et~al.}(2014)\citenamefont {Casalderrey-Solana}, \citenamefont {Gulhan},
  \citenamefont {Milhano}, \citenamefont {Pablos},\ and\ \citenamefont
  {Rajagopal}}]{Casalderrey-Solana:2014bpa}%
  \BibitemOpen
  \bibfield  {author} {\bibinfo {author} {\bibfnamefont {J.}~\bibnamefont
  {Casalderrey-Solana}}, \bibinfo {author} {\bibfnamefont {D.~C.}\ \bibnamefont
  {Gulhan}}, \bibinfo {author} {\bibfnamefont {J.~G.}\ \bibnamefont {Milhano}},
  \bibinfo {author} {\bibfnamefont {D.}~\bibnamefont {Pablos}}, \ and\ \bibinfo
  {author} {\bibfnamefont {K.}~\bibnamefont {Rajagopal}},\ }\href {\doibase
  10.1007/JHEP09(2015)175} {\bibfield  {journal} {\bibinfo  {journal} {JHEP}\
  }\textbf {\bibinfo {volume} {10}},\ \bibinfo {pages} {019} (\bibinfo {year}
  {2014})},\ \bibinfo {note} {[Erratum: JHEP 09, 175 (2015)]},\ \Eprint
  {http://arxiv.org/abs/1405.3864} {arXiv:1405.3864 [hep-ph]} \BibitemShut
  {NoStop}%
\bibitem [{\citenamefont {Hulcher}\ \emph {et~al.}(2018)\citenamefont
  {Hulcher}, \citenamefont {Pablos},\ and\ \citenamefont
  {Rajagopal}}]{Hulcher:2017cpt}%
  \BibitemOpen
  \bibfield  {author} {\bibinfo {author} {\bibfnamefont {Z.}~\bibnamefont
  {Hulcher}}, \bibinfo {author} {\bibfnamefont {D.}~\bibnamefont {Pablos}}, \
  and\ \bibinfo {author} {\bibfnamefont {K.}~\bibnamefont {Rajagopal}},\ }\href
  {\doibase 10.1007/JHEP03(2018)010} {\bibfield  {journal} {\bibinfo  {journal}
  {JHEP}\ }\textbf {\bibinfo {volume} {03}},\ \bibinfo {pages} {010} (\bibinfo
  {year} {2018})},\ \Eprint {http://arxiv.org/abs/1707.05245} {arXiv:1707.05245
  [hep-ph]} \BibitemShut {NoStop}%
\bibitem [{\citenamefont {Casalderrey-Solana}\ \emph
  {et~al.}(2020)\citenamefont {Casalderrey-Solana}, \citenamefont {Milhano},
  \citenamefont {Pablos},\ and\ \citenamefont
  {Rajagopal}}]{Casalderrey-Solana:2019ubu}%
  \BibitemOpen
  \bibfield  {author} {\bibinfo {author} {\bibfnamefont {J.}~\bibnamefont
  {Casalderrey-Solana}}, \bibinfo {author} {\bibfnamefont {G.}~\bibnamefont
  {Milhano}}, \bibinfo {author} {\bibfnamefont {D.}~\bibnamefont {Pablos}}, \
  and\ \bibinfo {author} {\bibfnamefont {K.}~\bibnamefont {Rajagopal}},\ }\href
  {\doibase 10.1007/JHEP01(2020)044} {\bibfield  {journal} {\bibinfo  {journal}
  {JHEP}\ }\textbf {\bibinfo {volume} {01}},\ \bibinfo {pages} {044} (\bibinfo
  {year} {2020})},\ \Eprint {http://arxiv.org/abs/1907.11248} {arXiv:1907.11248
  [hep-ph]} \BibitemShut {NoStop}%
\bibitem [{\citenamefont {Kudinoor}\ \emph {et~al.}(2025)\citenamefont
  {Kudinoor}, \citenamefont {Pablos},\ and\ \citenamefont
  {Rajagopal}}]{Kudinoor:2025gao}%
  \BibitemOpen
  \bibfield  {author} {\bibinfo {author} {\bibfnamefont {A.~S.}\ \bibnamefont
  {Kudinoor}}, \bibinfo {author} {\bibfnamefont {D.}~\bibnamefont {Pablos}}, \
  and\ \bibinfo {author} {\bibfnamefont {K.}~\bibnamefont {Rajagopal}},\
  }\href@noop {} {\  (\bibinfo {year} {2025})},\ \Eprint
  {http://arxiv.org/abs/2509.08881} {arXiv:2509.08881 [hep-ph]} \BibitemShut
  {NoStop}%
\bibitem [{\citenamefont {Tachibana}\ \emph {et~al.}(2024)\citenamefont
  {Tachibana} \emph {et~al.}}]{JETSCAPE:2023hqn}%
  \BibitemOpen
  \bibfield  {author} {\bibinfo {author} {\bibfnamefont {Y.}~\bibnamefont
  {Tachibana}} \emph {et~al.} (\bibinfo {collaboration} {JETSCAPE}),\ }\href
  {\doibase 10.1103/PhysRevC.110.044907} {\bibfield  {journal} {\bibinfo
  {journal} {Phys. Rev. C}\ }\textbf {\bibinfo {volume} {110}},\ \bibinfo
  {pages} {044907} (\bibinfo {year} {2024})},\ \Eprint
  {http://arxiv.org/abs/2301.02485} {arXiv:2301.02485 [hep-ph]} \BibitemShut
  {NoStop}%
\bibitem [{\citenamefont {Nason}(2004)}]{Nason:2004rx}%
  \BibitemOpen
  \bibfield  {author} {\bibinfo {author} {\bibfnamefont {P.}~\bibnamefont
  {Nason}},\ }\href {\doibase 10.1088/1126-6708/2004/11/040} {\bibfield
  {journal} {\bibinfo  {journal} {JHEP}\ }\textbf {\bibinfo {volume} {11}},\
  \bibinfo {pages} {040} (\bibinfo {year} {2004})},\ \Eprint
  {http://arxiv.org/abs/hep-ph/0409146} {arXiv:hep-ph/0409146} \BibitemShut
  {NoStop}%
\bibitem [{\citenamefont {Alioli}\ \emph {et~al.}(2010)\citenamefont {Alioli},
  \citenamefont {Nason}, \citenamefont {Oleari},\ and\ \citenamefont
  {Re}}]{Alioli:2010xd}%
  \BibitemOpen
  \bibfield  {author} {\bibinfo {author} {\bibfnamefont {S.}~\bibnamefont
  {Alioli}}, \bibinfo {author} {\bibfnamefont {P.}~\bibnamefont {Nason}},
  \bibinfo {author} {\bibfnamefont {C.}~\bibnamefont {Oleari}}, \ and\ \bibinfo
  {author} {\bibfnamefont {E.}~\bibnamefont {Re}},\ }\href {\doibase
  10.1007/JHEP06(2010)043} {\bibfield  {journal} {\bibinfo  {journal} {JHEP}\
  }\textbf {\bibinfo {volume} {06}},\ \bibinfo {pages} {043} (\bibinfo {year}
  {2010})},\ \Eprint {http://arxiv.org/abs/1002.2581} {arXiv:1002.2581
  [hep-ph]} \BibitemShut {NoStop}%
\bibitem [{\citenamefont {Alioli}\ \emph {et~al.}(2011)\citenamefont {Alioli},
  \citenamefont {Hamilton}, \citenamefont {Nason}, \citenamefont {Oleari},\
  and\ \citenamefont {Re}}]{Alioli:2010xa}%
  \BibitemOpen
  \bibfield  {author} {\bibinfo {author} {\bibfnamefont {S.}~\bibnamefont
  {Alioli}}, \bibinfo {author} {\bibfnamefont {K.}~\bibnamefont {Hamilton}},
  \bibinfo {author} {\bibfnamefont {P.}~\bibnamefont {Nason}}, \bibinfo
  {author} {\bibfnamefont {C.}~\bibnamefont {Oleari}}, \ and\ \bibinfo {author}
  {\bibfnamefont {E.}~\bibnamefont {Re}},\ }\href {\doibase
  10.1007/JHEP04(2011)081} {\bibfield  {journal} {\bibinfo  {journal} {JHEP}\
  }\textbf {\bibinfo {volume} {04}},\ \bibinfo {pages} {081} (\bibinfo {year}
  {2011})},\ \Eprint {http://arxiv.org/abs/1012.3380} {arXiv:1012.3380
  [hep-ph]} \BibitemShut {NoStop}%
\bibitem [{\citenamefont {Bierlich}\ \emph {et~al.}(2022)\citenamefont
  {Bierlich} \emph {et~al.}}]{Bierlich:2022pfr}%
  \BibitemOpen
  \bibfield  {author} {\bibinfo {author} {\bibfnamefont {C.}~\bibnamefont
  {Bierlich}} \emph {et~al.},\ }\href {\doibase 10.21468/SciPostPhysCodeb.8}
  {\bibfield  {journal} {\bibinfo  {journal} {SciPost Phys. Codeb.}\ }\textbf
  {\bibinfo {volume} {2022}},\ \bibinfo {pages} {8} (\bibinfo {year} {2022})},\
  \Eprint {http://arxiv.org/abs/2203.11601} {arXiv:2203.11601 [hep-ph]}
  \BibitemShut {NoStop}%
\bibitem [{\citenamefont {Mehtar-Tani}\ and\ \citenamefont
  {Tywoniuk}(2018{\natexlab{a}})}]{Mehtar-Tani:2017ypq}%
  \BibitemOpen
  \bibfield  {author} {\bibinfo {author} {\bibfnamefont {Y.}~\bibnamefont
  {Mehtar-Tani}}\ and\ \bibinfo {author} {\bibfnamefont {K.}~\bibnamefont
  {Tywoniuk}},\ }\href {\doibase 10.1016/j.nuclphysa.2018.09.041} {\bibfield
  {journal} {\bibinfo  {journal} {Nucl. Phys. A}\ }\textbf {\bibinfo {volume}
  {979}},\ \bibinfo {pages} {165} (\bibinfo {year} {2018}{\natexlab{a}})},\
  \Eprint {http://arxiv.org/abs/1706.06047} {arXiv:1706.06047 [hep-ph]}
  \BibitemShut {NoStop}%
\bibitem [{\citenamefont {Mehtar-Tani}\ \emph {et~al.}(2021)\citenamefont
  {Mehtar-Tani}, \citenamefont {Pablos},\ and\ \citenamefont
  {Tywoniuk}}]{Mehtar-Tani:2021fud}%
  \BibitemOpen
  \bibfield  {author} {\bibinfo {author} {\bibfnamefont {Y.}~\bibnamefont
  {Mehtar-Tani}}, \bibinfo {author} {\bibfnamefont {D.}~\bibnamefont {Pablos}},
  \ and\ \bibinfo {author} {\bibfnamefont {K.}~\bibnamefont {Tywoniuk}},\
  }\href {\doibase 10.1103/PhysRevLett.127.252301} {\bibfield  {journal}
  {\bibinfo  {journal} {Phys. Rev. Lett.}\ }\textbf {\bibinfo {volume} {127}},\
  \bibinfo {pages} {252301} (\bibinfo {year} {2021})},\ \Eprint
  {http://arxiv.org/abs/2101.01742} {arXiv:2101.01742 [hep-ph]} \BibitemShut
  {NoStop}%
\bibitem [{\citenamefont {Cacciari}\ \emph {et~al.}(2008)\citenamefont
  {Cacciari}, \citenamefont {Salam},\ and\ \citenamefont
  {Soyez}}]{Cacciari:2008gp}%
  \BibitemOpen
  \bibfield  {author} {\bibinfo {author} {\bibfnamefont {M.}~\bibnamefont
  {Cacciari}}, \bibinfo {author} {\bibfnamefont {G.~P.}\ \bibnamefont {Salam}},
  \ and\ \bibinfo {author} {\bibfnamefont {G.}~\bibnamefont {Soyez}},\ }\href
  {\doibase 10.1088/1126-6708/2008/04/063} {\bibfield  {journal} {\bibinfo
  {journal} {JHEP}\ }\textbf {\bibinfo {volume} {04}},\ \bibinfo {pages} {063}
  (\bibinfo {year} {2008})},\ \Eprint {http://arxiv.org/abs/0802.1189}
  {arXiv:0802.1189 [hep-ph]} \BibitemShut {NoStop}%
\bibitem [{\citenamefont {Cacciari}\ \emph {et~al.}(2012)\citenamefont
  {Cacciari}, \citenamefont {Salam},\ and\ \citenamefont
  {Soyez}}]{Cacciari:2011ma}%
  \BibitemOpen
  \bibfield  {author} {\bibinfo {author} {\bibfnamefont {M.}~\bibnamefont
  {Cacciari}}, \bibinfo {author} {\bibfnamefont {G.~P.}\ \bibnamefont {Salam}},
  \ and\ \bibinfo {author} {\bibfnamefont {G.}~\bibnamefont {Soyez}},\ }\href
  {\doibase 10.1140/epjc/s10052-012-1896-2} {\bibfield  {journal} {\bibinfo
  {journal} {Eur. Phys. J. C}\ }\textbf {\bibinfo {volume} {72}},\ \bibinfo
  {pages} {1896} (\bibinfo {year} {2012})},\ \Eprint
  {http://arxiv.org/abs/1111.6097} {arXiv:1111.6097 [hep-ph]} \BibitemShut
  {NoStop}%
\bibitem [{\citenamefont {Dokshitzer}\ \emph {et~al.}(1997)\citenamefont
  {Dokshitzer}, \citenamefont {Leder}, \citenamefont {Moretti},\ and\
  \citenamefont {Webber}}]{Dokshitzer:1997in}%
  \BibitemOpen
  \bibfield  {author} {\bibinfo {author} {\bibfnamefont {Y.~L.}\ \bibnamefont
  {Dokshitzer}}, \bibinfo {author} {\bibfnamefont {G.~D.}\ \bibnamefont
  {Leder}}, \bibinfo {author} {\bibfnamefont {S.}~\bibnamefont {Moretti}}, \
  and\ \bibinfo {author} {\bibfnamefont {B.~R.}\ \bibnamefont {Webber}},\
  }\href {\doibase 10.1088/1126-6708/1997/08/001} {\bibfield  {journal}
  {\bibinfo  {journal} {JHEP}\ }\textbf {\bibinfo {volume} {08}},\ \bibinfo
  {pages} {001} (\bibinfo {year} {1997})},\ \Eprint
  {http://arxiv.org/abs/hep-ph/9707323} {arXiv:hep-ph/9707323} \BibitemShut
  {NoStop}%
\bibitem [{\citenamefont {Wobisch}\ and\ \citenamefont
  {Wengler}(1998)}]{Wobisch:1998wt}%
  \BibitemOpen
  \bibfield  {author} {\bibinfo {author} {\bibfnamefont {M.}~\bibnamefont
  {Wobisch}}\ and\ \bibinfo {author} {\bibfnamefont {T.}~\bibnamefont
  {Wengler}},\ }in\ \href@noop {} {\emph {\bibinfo {booktitle} {{Workshop on
  Monte Carlo Generators for HERA Physics (Plenary Starting Meeting)}}}}\
  (\bibinfo {year} {1998})\ pp.\ \bibinfo {pages} {270--279},\ \Eprint
  {http://arxiv.org/abs/hep-ph/9907280} {arXiv:hep-ph/9907280} \BibitemShut
  {NoStop}%
\bibitem [{\citenamefont {Andres}\ \emph
  {et~al.}(2025{\natexlab{a}})\citenamefont {Andres}, \citenamefont
  {Apolin{\'a}rio}, \citenamefont {Armesto}, \citenamefont {Cordeiro},
  \citenamefont {Dominguez},\ and\ \citenamefont {Milhano}}]{Andres:2024egc}%
  \BibitemOpen
  \bibfield  {author} {\bibinfo {author} {\bibfnamefont {C.}~\bibnamefont
  {Andres}}, \bibinfo {author} {\bibfnamefont {L.}~\bibnamefont
  {Apolin{\'a}rio}}, \bibinfo {author} {\bibfnamefont {N.}~\bibnamefont
  {Armesto}}, \bibinfo {author} {\bibfnamefont {A.}~\bibnamefont {Cordeiro}},
  \bibinfo {author} {\bibfnamefont {F.}~\bibnamefont {Dominguez}}, \ and\
  \bibinfo {author} {\bibfnamefont {J.~G.}\ \bibnamefont {Milhano}},\ }\href
  {\doibase 10.1007/JHEP08(2025)160} {\bibfield  {journal} {\bibinfo  {journal}
  {JHEP}\ }\textbf {\bibinfo {volume} {08}},\ \bibinfo {pages} {160} (\bibinfo
  {year} {2025}{\natexlab{a}})},\ \Eprint {http://arxiv.org/abs/2409.13536}
  {arXiv:2409.13536 [hep-ph]} \BibitemShut {NoStop}%
\bibitem [{\citenamefont {Alwall}\ \emph {et~al.}(2014)\citenamefont {Alwall},
  \citenamefont {Frederix}, \citenamefont {Frixione}, \citenamefont {Hirschi},
  \citenamefont {Maltoni}, \citenamefont {Mattelaer}, \citenamefont {Shao},
  \citenamefont {Stelzer}, \citenamefont {Torrielli},\ and\ \citenamefont
  {Zaro}}]{Alwall:2014hca}%
  \BibitemOpen
  \bibfield  {author} {\bibinfo {author} {\bibfnamefont {J.}~\bibnamefont
  {Alwall}}, \bibinfo {author} {\bibfnamefont {R.}~\bibnamefont {Frederix}},
  \bibinfo {author} {\bibfnamefont {S.}~\bibnamefont {Frixione}}, \bibinfo
  {author} {\bibfnamefont {V.}~\bibnamefont {Hirschi}}, \bibinfo {author}
  {\bibfnamefont {F.}~\bibnamefont {Maltoni}}, \bibinfo {author} {\bibfnamefont
  {O.}~\bibnamefont {Mattelaer}}, \bibinfo {author} {\bibfnamefont {H.~S.}\
  \bibnamefont {Shao}}, \bibinfo {author} {\bibfnamefont {T.}~\bibnamefont
  {Stelzer}}, \bibinfo {author} {\bibfnamefont {P.}~\bibnamefont {Torrielli}},
  \ and\ \bibinfo {author} {\bibfnamefont {M.}~\bibnamefont {Zaro}},\ }\href
  {\doibase 10.1007/JHEP07(2014)079} {\bibfield  {journal} {\bibinfo  {journal}
  {JHEP}\ }\textbf {\bibinfo {volume} {07}},\ \bibinfo {pages} {079} (\bibinfo
  {year} {2014})},\ \Eprint {http://arxiv.org/abs/1405.0301} {arXiv:1405.0301
  [hep-ph]} \BibitemShut {NoStop}%
\bibitem [{\citenamefont {Dulat}\ \emph {et~al.}(2016)\citenamefont {Dulat},
  \citenamefont {Hou}, \citenamefont {Gao}, \citenamefont {Guzzi},
  \citenamefont {Huston}, \citenamefont {Nadolsky}, \citenamefont {Pumplin},
  \citenamefont {Schmidt}, \citenamefont {Stump},\ and\ \citenamefont
  {Yuan}}]{Dulat:2015mca}%
  \BibitemOpen
  \bibfield  {author} {\bibinfo {author} {\bibfnamefont {S.}~\bibnamefont
  {Dulat}}, \bibinfo {author} {\bibfnamefont {T.-J.}\ \bibnamefont {Hou}},
  \bibinfo {author} {\bibfnamefont {J.}~\bibnamefont {Gao}}, \bibinfo {author}
  {\bibfnamefont {M.}~\bibnamefont {Guzzi}}, \bibinfo {author} {\bibfnamefont
  {J.}~\bibnamefont {Huston}}, \bibinfo {author} {\bibfnamefont
  {P.}~\bibnamefont {Nadolsky}}, \bibinfo {author} {\bibfnamefont
  {J.}~\bibnamefont {Pumplin}}, \bibinfo {author} {\bibfnamefont
  {C.}~\bibnamefont {Schmidt}}, \bibinfo {author} {\bibfnamefont
  {D.}~\bibnamefont {Stump}}, \ and\ \bibinfo {author} {\bibfnamefont {C.~P.}\
  \bibnamefont {Yuan}},\ }\href {\doibase 10.1103/PhysRevD.93.033006}
  {\bibfield  {journal} {\bibinfo  {journal} {Phys. Rev. D}\ }\textbf {\bibinfo
  {volume} {93}},\ \bibinfo {pages} {033006} (\bibinfo {year} {2016})},\
  \Eprint {http://arxiv.org/abs/1506.07443} {arXiv:1506.07443 [hep-ph]}
  \BibitemShut {NoStop}%
\bibitem [{\citenamefont {Mehtar-Tani}\ and\ \citenamefont
  {Tywoniuk}(2018{\natexlab{b}})}]{Mehtar-Tani:2017web}%
  \BibitemOpen
  \bibfield  {author} {\bibinfo {author} {\bibfnamefont {Y.}~\bibnamefont
  {Mehtar-Tani}}\ and\ \bibinfo {author} {\bibfnamefont {K.}~\bibnamefont
  {Tywoniuk}},\ }\href {\doibase 10.1103/PhysRevD.98.051501} {\bibfield
  {journal} {\bibinfo  {journal} {Phys. Rev. D}\ }\textbf {\bibinfo {volume}
  {98}},\ \bibinfo {pages} {051501} (\bibinfo {year} {2018}{\natexlab{b}})},\
  \Eprint {http://arxiv.org/abs/1707.07361} {arXiv:1707.07361 [hep-ph]}
  \BibitemShut {NoStop}%
\bibitem [{\citenamefont {Takacs}\ and\ \citenamefont
  {Tywoniuk}(2021)}]{Takacs:2021bpv}%
  \BibitemOpen
  \bibfield  {author} {\bibinfo {author} {\bibfnamefont {A.}~\bibnamefont
  {Takacs}}\ and\ \bibinfo {author} {\bibfnamefont {K.}~\bibnamefont
  {Tywoniuk}},\ }\href {\doibase 10.1007/JHEP10(2021)038} {\bibfield  {journal}
  {\bibinfo  {journal} {JHEP}\ }\textbf {\bibinfo {volume} {10}},\ \bibinfo
  {pages} {038} (\bibinfo {year} {2021})},\ \Eprint
  {http://arxiv.org/abs/2103.14676} {arXiv:2103.14676 [hep-ph]} \BibitemShut
  {NoStop}%
\bibitem [{\citenamefont {Mehtar-Tani}\ \emph {et~al.}(2024)\citenamefont
  {Mehtar-Tani}, \citenamefont {Pablos},\ and\ \citenamefont
  {Tywoniuk}}]{Mehtar-Tani:2024jtd}%
  \BibitemOpen
  \bibfield  {author} {\bibinfo {author} {\bibfnamefont {Y.}~\bibnamefont
  {Mehtar-Tani}}, \bibinfo {author} {\bibfnamefont {D.}~\bibnamefont {Pablos}},
  \ and\ \bibinfo {author} {\bibfnamefont {K.}~\bibnamefont {Tywoniuk}},\
  }\href {\doibase 10.1103/PhysRevD.110.014009} {\bibfield  {journal} {\bibinfo
   {journal} {Phys. Rev. D}\ }\textbf {\bibinfo {volume} {110}},\ \bibinfo
  {pages} {014009} (\bibinfo {year} {2024})},\ \Eprint
  {http://arxiv.org/abs/2402.07869} {arXiv:2402.07869 [hep-ph]} \BibitemShut
  {NoStop}%
\bibitem [{\citenamefont {Pablos}\ and\ \citenamefont
  {Takacs}(2025)}]{Pablos:2025cli}%
  \BibitemOpen
  \bibfield  {author} {\bibinfo {author} {\bibfnamefont {D.}~\bibnamefont
  {Pablos}}\ and\ \bibinfo {author} {\bibfnamefont {A.}~\bibnamefont
  {Takacs}},\ }\href@noop {} {\  (\bibinfo {year} {2025})},\ \Eprint
  {http://arxiv.org/abs/2509.19430} {arXiv:2509.19430 [hep-ph]} \BibitemShut
  {NoStop}%
\bibitem [{\citenamefont {Acton}\ \emph {et~al.}(1993)\citenamefont {Acton}
  \emph {et~al.}}]{OPAL:1993uun}%
  \BibitemOpen
  \bibfield  {author} {\bibinfo {author} {\bibfnamefont {P.~D.}\ \bibnamefont
  {Acton}} \emph {et~al.} (\bibinfo {collaboration} {OPAL}),\ }\href {\doibase
  10.1007/BF01557696} {\bibfield  {journal} {\bibinfo  {journal} {Z. Phys. C}\
  }\textbf {\bibinfo {volume} {58}},\ \bibinfo {pages} {387} (\bibinfo {year}
  {1993})}\BibitemShut {NoStop}%
\bibitem [{ALI(2024)}]{ALICE-PUBLIC-2024-004}%
  \BibitemOpen
  \href {https://cds.cern.ch/record/2910744} {\  (\bibinfo {year}
  {2024})}\BibitemShut {NoStop}%
\bibitem [{\citenamefont {Eskola}\ \emph {et~al.}(2017)\citenamefont {Eskola},
  \citenamefont {Paakkinen}, \citenamefont {Paukkunen},\ and\ \citenamefont
  {Salgado}}]{Eskola:2016oht}%
  \BibitemOpen
  \bibfield  {author} {\bibinfo {author} {\bibfnamefont {K.~J.}\ \bibnamefont
  {Eskola}}, \bibinfo {author} {\bibfnamefont {P.}~\bibnamefont {Paakkinen}},
  \bibinfo {author} {\bibfnamefont {H.}~\bibnamefont {Paukkunen}}, \ and\
  \bibinfo {author} {\bibfnamefont {C.~A.}\ \bibnamefont {Salgado}},\ }\href
  {\doibase 10.1140/epjc/s10052-017-4725-9} {\bibfield  {journal} {\bibinfo
  {journal} {Eur. Phys. J. C}\ }\textbf {\bibinfo {volume} {77}},\ \bibinfo
  {pages} {163} (\bibinfo {year} {2017})},\ \Eprint
  {http://arxiv.org/abs/1612.05741} {arXiv:1612.05741 [hep-ph]} \BibitemShut
  {NoStop}%
\bibitem [{\citenamefont {Ehlers}\ \emph {et~al.}(2025)\citenamefont {Ehlers}
  \emph {et~al.}}]{JETSCAPE:2024cqe}%
  \BibitemOpen
  \bibfield  {author} {\bibinfo {author} {\bibfnamefont {R.}~\bibnamefont
  {Ehlers}} \emph {et~al.} (\bibinfo {collaboration} {JETSCAPE}),\ }\href
  {\doibase 10.1103/PhysRevC.111.054913} {\bibfield  {journal} {\bibinfo
  {journal} {Phys. Rev. C}\ }\textbf {\bibinfo {volume} {111}},\ \bibinfo
  {pages} {054913} (\bibinfo {year} {2025})},\ \Eprint
  {http://arxiv.org/abs/2408.08247} {arXiv:2408.08247 [hep-ph]} \BibitemShut
  {NoStop}%
\bibitem [{\citenamefont {Acharya}\ \emph {et~al.}(2024)\citenamefont {Acharya}
  \emph {et~al.}}]{ALICE:2023waz}%
  \BibitemOpen
  \bibfield  {author} {\bibinfo {author} {\bibfnamefont {S.}~\bibnamefont
  {Acharya}} \emph {et~al.} (\bibinfo {collaboration} {ALICE}),\ }\href
  {\doibase 10.1016/j.physletb.2023.138412} {\bibfield  {journal} {\bibinfo
  {journal} {Phys. Lett. B}\ }\textbf {\bibinfo {volume} {849}},\ \bibinfo
  {pages} {138412} (\bibinfo {year} {2024})},\ \Eprint
  {http://arxiv.org/abs/2303.00592} {arXiv:2303.00592 [nucl-ex]} \BibitemShut
  {NoStop}%
\bibitem [{\citenamefont {Chekhovsky}\ \emph {et~al.}(2025)\citenamefont
  {Chekhovsky} \emph {et~al.}}]{CMS:2025ydi}%
  \BibitemOpen
  \bibfield  {author} {\bibinfo {author} {\bibfnamefont {V.}~\bibnamefont
  {Chekhovsky}} \emph {et~al.} (\bibinfo {collaboration} {CMS}),\ }\href
  {\doibase 10.1016/j.physletb.2025.139556} {\bibfield  {journal} {\bibinfo
  {journal} {Phys. Lett. B}\ }\textbf {\bibinfo {volume} {866}},\ \bibinfo
  {pages} {139556} (\bibinfo {year} {2025})},\ \Eprint
  {http://arxiv.org/abs/2503.19993} {arXiv:2503.19993 [nucl-ex]} \BibitemShut
  {NoStop}%
\bibitem [{\citenamefont {Andres}\ \emph {et~al.}(2023)\citenamefont {Andres},
  \citenamefont {Dominguez}, \citenamefont {Kunnawalkam~Elayavalli},
  \citenamefont {Holguin}, \citenamefont {Marquet},\ and\ \citenamefont
  {Moult}}]{Andres:2022ovj}%
  \BibitemOpen
  \bibfield  {author} {\bibinfo {author} {\bibfnamefont {C.}~\bibnamefont
  {Andres}}, \bibinfo {author} {\bibfnamefont {F.}~\bibnamefont {Dominguez}},
  \bibinfo {author} {\bibfnamefont {R.}~\bibnamefont {Kunnawalkam~Elayavalli}},
  \bibinfo {author} {\bibfnamefont {J.}~\bibnamefont {Holguin}}, \bibinfo
  {author} {\bibfnamefont {C.}~\bibnamefont {Marquet}}, \ and\ \bibinfo
  {author} {\bibfnamefont {I.}~\bibnamefont {Moult}},\ }\href {\doibase
  10.1103/PhysRevLett.130.262301} {\bibfield  {journal} {\bibinfo  {journal}
  {Phys. Rev. Lett.}\ }\textbf {\bibinfo {volume} {130}},\ \bibinfo {pages}
  {262301} (\bibinfo {year} {2023})},\ \Eprint
  {http://arxiv.org/abs/2209.11236} {arXiv:2209.11236 [hep-ph]} \BibitemShut
  {NoStop}%
\bibitem [{\citenamefont {Barata}\ \emph
  {et~al.}(2024{\natexlab{a}})\citenamefont {Barata}, \citenamefont {Caucal},
  \citenamefont {Soto-Ontoso},\ and\ \citenamefont {Szafron}}]{Barata:2023bhh}%
  \BibitemOpen
  \bibfield  {author} {\bibinfo {author} {\bibfnamefont {J.}~\bibnamefont
  {Barata}}, \bibinfo {author} {\bibfnamefont {P.}~\bibnamefont {Caucal}},
  \bibinfo {author} {\bibfnamefont {A.}~\bibnamefont {Soto-Ontoso}}, \ and\
  \bibinfo {author} {\bibfnamefont {R.}~\bibnamefont {Szafron}},\ }\href
  {\doibase 10.1007/JHEP11(2024)060} {\bibfield  {journal} {\bibinfo  {journal}
  {JHEP}\ }\textbf {\bibinfo {volume} {11}},\ \bibinfo {pages} {060} (\bibinfo
  {year} {2024}{\natexlab{a}})},\ \Eprint {http://arxiv.org/abs/2312.12527}
  {arXiv:2312.12527 [hep-ph]} \BibitemShut {NoStop}%
\bibitem [{\citenamefont {Barata}\ \emph
  {et~al.}(2024{\natexlab{b}})\citenamefont {Barata}, \citenamefont {Milhano},\
  and\ \citenamefont {Sadofyev}}]{Barata:2023zqg}%
  \BibitemOpen
  \bibfield  {author} {\bibinfo {author} {\bibfnamefont {J.}~\bibnamefont
  {Barata}}, \bibinfo {author} {\bibfnamefont {J.~G.}\ \bibnamefont {Milhano}},
  \ and\ \bibinfo {author} {\bibfnamefont {A.~V.}\ \bibnamefont {Sadofyev}},\
  }\href {\doibase 10.1140/epjc/s10052-024-12514-1} {\bibfield  {journal}
  {\bibinfo  {journal} {Eur. Phys. J. C}\ }\textbf {\bibinfo {volume} {84}},\
  \bibinfo {pages} {174} (\bibinfo {year} {2024}{\natexlab{b}})},\ \Eprint
  {http://arxiv.org/abs/2308.01294} {arXiv:2308.01294 [hep-ph]} \BibitemShut
  {NoStop}%
\bibitem [{\citenamefont {Yang}\ \emph {et~al.}(2024)\citenamefont {Yang},
  \citenamefont {He}, \citenamefont {Moult},\ and\ \citenamefont
  {Wang}}]{Yang:2023dwc}%
  \BibitemOpen
  \bibfield  {author} {\bibinfo {author} {\bibfnamefont {Z.}~\bibnamefont
  {Yang}}, \bibinfo {author} {\bibfnamefont {Y.}~\bibnamefont {He}}, \bibinfo
  {author} {\bibfnamefont {I.}~\bibnamefont {Moult}}, \ and\ \bibinfo {author}
  {\bibfnamefont {X.-N.}\ \bibnamefont {Wang}},\ }\href {\doibase
  10.1103/PhysRevLett.132.011901} {\bibfield  {journal} {\bibinfo  {journal}
  {Phys. Rev. Lett.}\ }\textbf {\bibinfo {volume} {132}},\ \bibinfo {pages}
  {011901} (\bibinfo {year} {2024})},\ \Eprint
  {http://arxiv.org/abs/2310.01500} {arXiv:2310.01500 [hep-ph]} \BibitemShut
  {NoStop}%
\bibitem [{\citenamefont {Fu}\ \emph {et~al.}(2025)\citenamefont {Fu},
  \citenamefont {M{\"u}ller},\ and\ \citenamefont {Sirimanna}}]{Fu:2024pic}%
  \BibitemOpen
  \bibfield  {author} {\bibinfo {author} {\bibfnamefont {Y.}~\bibnamefont
  {Fu}}, \bibinfo {author} {\bibfnamefont {B.}~\bibnamefont {M{\"u}ller}}, \
  and\ \bibinfo {author} {\bibfnamefont {C.}~\bibnamefont {Sirimanna}},\ }\href
  {\doibase 10.1103/qxdv-cmrg} {\bibfield  {journal} {\bibinfo  {journal}
  {Phys. Rev. Lett.}\ }\textbf {\bibinfo {volume} {135}},\ \bibinfo {pages}
  {112302} (\bibinfo {year} {2025})},\ \Eprint
  {http://arxiv.org/abs/2411.04866} {arXiv:2411.04866 [nucl-th]} \BibitemShut
  {NoStop}%
\bibitem [{\citenamefont {Xing}\ \emph {et~al.}(2025)\citenamefont {Xing},
  \citenamefont {Cao}, \citenamefont {Qin},\ and\ \citenamefont
  {Wang}}]{Xing:2024yrb}%
  \BibitemOpen
  \bibfield  {author} {\bibinfo {author} {\bibfnamefont {W.-J.}\ \bibnamefont
  {Xing}}, \bibinfo {author} {\bibfnamefont {S.}~\bibnamefont {Cao}}, \bibinfo
  {author} {\bibfnamefont {G.-Y.}\ \bibnamefont {Qin}}, \ and\ \bibinfo
  {author} {\bibfnamefont {X.-N.}\ \bibnamefont {Wang}},\ }\href {\doibase
  10.1103/PhysRevLett.134.052301} {\bibfield  {journal} {\bibinfo  {journal}
  {Phys. Rev. Lett.}\ }\textbf {\bibinfo {volume} {134}},\ \bibinfo {pages}
  {052301} (\bibinfo {year} {2025})},\ \Eprint
  {http://arxiv.org/abs/2409.12843} {arXiv:2409.12843 [hep-ph]} \BibitemShut
  {NoStop}%
\bibitem [{\citenamefont {Singh}\ and\ \citenamefont
  {Vaidya}(2025)}]{Singh:2024vwb}%
  \BibitemOpen
  \bibfield  {author} {\bibinfo {author} {\bibfnamefont {B.}~\bibnamefont
  {Singh}}\ and\ \bibinfo {author} {\bibfnamefont {V.}~\bibnamefont {Vaidya}},\
  }\href {\doibase 10.1007/JHEP06(2025)071} {\bibfield  {journal} {\bibinfo
  {journal} {JHEP}\ }\textbf {\bibinfo {volume} {06}},\ \bibinfo {pages} {071}
  (\bibinfo {year} {2025})},\ \Eprint {http://arxiv.org/abs/2408.02753}
  {arXiv:2408.02753 [hep-ph]} \BibitemShut {NoStop}%
\bibitem [{\citenamefont {Bossi}\ \emph {et~al.}(2024)\citenamefont {Bossi},
  \citenamefont {Kudinoor}, \citenamefont {Moult}, \citenamefont {Pablos},
  \citenamefont {Rai},\ and\ \citenamefont {Rajagopal}}]{Bossi:2024qho}%
  \BibitemOpen
  \bibfield  {author} {\bibinfo {author} {\bibfnamefont {H.}~\bibnamefont
  {Bossi}}, \bibinfo {author} {\bibfnamefont {A.~S.}\ \bibnamefont {Kudinoor}},
  \bibinfo {author} {\bibfnamefont {I.}~\bibnamefont {Moult}}, \bibinfo
  {author} {\bibfnamefont {D.}~\bibnamefont {Pablos}}, \bibinfo {author}
  {\bibfnamefont {A.}~\bibnamefont {Rai}}, \ and\ \bibinfo {author}
  {\bibfnamefont {K.}~\bibnamefont {Rajagopal}},\ }\href {\doibase
  10.1007/JHEP12(2024)073} {\bibfield  {journal} {\bibinfo  {journal} {JHEP}\
  }\textbf {\bibinfo {volume} {12}},\ \bibinfo {pages} {073} (\bibinfo {year}
  {2024})},\ \Eprint {http://arxiv.org/abs/2407.13818} {arXiv:2407.13818
  [hep-ph]} \BibitemShut {NoStop}%
\bibitem [{\citenamefont {Barata}\ \emph
  {et~al.}(2025{\natexlab{a}})\citenamefont {Barata}, \citenamefont {Kuzmin},
  \citenamefont {Milhano},\ and\ \citenamefont {Sadofyev}}]{Barata:2024ukm}%
  \BibitemOpen
  \bibfield  {author} {\bibinfo {author} {\bibfnamefont {J.}~\bibnamefont
  {Barata}}, \bibinfo {author} {\bibfnamefont {M.~V.}\ \bibnamefont {Kuzmin}},
  \bibinfo {author} {\bibfnamefont {J.~G.}\ \bibnamefont {Milhano}}, \ and\
  \bibinfo {author} {\bibfnamefont {A.~V.}\ \bibnamefont {Sadofyev}},\ }\href
  {\doibase 10.1103/q8pr-djlw} {\bibfield  {journal} {\bibinfo  {journal}
  {Phys. Rev. D}\ }\textbf {\bibinfo {volume} {112}},\ \bibinfo {pages}
  {016005} (\bibinfo {year} {2025}{\natexlab{a}})},\ \Eprint
  {http://arxiv.org/abs/2412.03616} {arXiv:2412.03616 [hep-ph]} \BibitemShut
  {NoStop}%
\bibitem [{\citenamefont {Barata}\ \emph
  {et~al.}(2025{\natexlab{b}})\citenamefont {Barata}, \citenamefont {Brewer},
  \citenamefont {Lee},\ and\ \citenamefont {Silva}}]{Barata:2025uxp}%
  \BibitemOpen
  \bibfield  {author} {\bibinfo {author} {\bibfnamefont {J.}~\bibnamefont
  {Barata}}, \bibinfo {author} {\bibfnamefont {J.}~\bibnamefont {Brewer}},
  \bibinfo {author} {\bibfnamefont {K.}~\bibnamefont {Lee}}, \ and\ \bibinfo
  {author} {\bibfnamefont {J.~M.}\ \bibnamefont {Silva}},\ }\href@noop {} {\
  (\bibinfo {year} {2025}{\natexlab{b}})},\ \Eprint
  {http://arxiv.org/abs/2508.19404} {arXiv:2508.19404 [hep-ph]} \BibitemShut
  {NoStop}%
\bibitem [{\citenamefont {Andres}\ \emph
  {et~al.}(2025{\natexlab{b}})\citenamefont {Andres}, \citenamefont {Holguin},
  \citenamefont {Kimelman}, \citenamefont {Kunnawalkam~Elayavalli},
  \citenamefont {Viinikainen},\ and\ \citenamefont {Yang}}]{Andres:2025yls}%
  \BibitemOpen
  \bibfield  {author} {\bibinfo {author} {\bibfnamefont {C.}~\bibnamefont
  {Andres}}, \bibinfo {author} {\bibfnamefont {J.}~\bibnamefont {Holguin}},
  \bibinfo {author} {\bibfnamefont {B.}~\bibnamefont {Kimelman}}, \bibinfo
  {author} {\bibfnamefont {R.}~\bibnamefont {Kunnawalkam~Elayavalli}}, \bibinfo
  {author} {\bibfnamefont {J.}~\bibnamefont {Viinikainen}}, \ and\ \bibinfo
  {author} {\bibfnamefont {Z.}~\bibnamefont {Yang}},\ }\href@noop {} {\
  (\bibinfo {year} {2025}{\natexlab{b}})},\ \Eprint
  {http://arxiv.org/abs/2512.10026} {arXiv:2512.10026 [hep-ph]} \BibitemShut
  {NoStop}%
\bibitem [{\citenamefont {Apolin{\'a}rio}\ \emph
  {et~al.}(2025{\natexlab{a}})\citenamefont {Apolin{\'a}rio}, \citenamefont
  {Kunnawalkam~Elayavalli}, \citenamefont {Madureira}, \citenamefont {Sheng},
  \citenamefont {Wang},\ and\ \citenamefont {Yang}}]{Apolinario:2025vtx}%
  \BibitemOpen
  \bibfield  {author} {\bibinfo {author} {\bibfnamefont {L.}~\bibnamefont
  {Apolin{\'a}rio}}, \bibinfo {author} {\bibfnamefont {R.}~\bibnamefont
  {Kunnawalkam~Elayavalli}}, \bibinfo {author} {\bibfnamefont {N.~O.}\
  \bibnamefont {Madureira}}, \bibinfo {author} {\bibfnamefont {J.-X.}\
  \bibnamefont {Sheng}}, \bibinfo {author} {\bibfnamefont {X.-N.}\ \bibnamefont
  {Wang}}, \ and\ \bibinfo {author} {\bibfnamefont {Z.}~\bibnamefont {Yang}},\
  }\href {\doibase 10.1103/4sfq-315y} {\bibfield  {journal} {\bibinfo
  {journal} {Phys. Rev. D}\ }\textbf {\bibinfo {volume} {112}},\ \bibinfo
  {pages} {054018} (\bibinfo {year} {2025}{\natexlab{a}})},\ \Eprint
  {http://arxiv.org/abs/2502.11406} {arXiv:2502.11406 [hep-ph]} \BibitemShut
  {NoStop}%
\bibitem [{\citenamefont {Caletti}\ \emph {et~al.}(2022)\citenamefont
  {Caletti}, \citenamefont {Larkoski}, \citenamefont {Marzani},\ and\
  \citenamefont {Reichelt}}]{Caletti:2022hnc}%
  \BibitemOpen
  \bibfield  {author} {\bibinfo {author} {\bibfnamefont {S.}~\bibnamefont
  {Caletti}}, \bibinfo {author} {\bibfnamefont {A.~J.}\ \bibnamefont
  {Larkoski}}, \bibinfo {author} {\bibfnamefont {S.}~\bibnamefont {Marzani}}, \
  and\ \bibinfo {author} {\bibfnamefont {D.}~\bibnamefont {Reichelt}},\ }\href
  {\doibase 10.1140/epjc/s10052-022-10568-7} {\bibfield  {journal} {\bibinfo
  {journal} {Eur. Phys. J. C}\ }\textbf {\bibinfo {volume} {82}},\ \bibinfo
  {pages} {632} (\bibinfo {year} {2022})},\ \Eprint
  {http://arxiv.org/abs/2205.01109} {arXiv:2205.01109 [hep-ph]} \BibitemShut
  {NoStop}%
\bibitem [{\citenamefont {Czakon}\ \emph {et~al.}(2023)\citenamefont {Czakon},
  \citenamefont {Mitov},\ and\ \citenamefont {Poncelet}}]{Czakon:2022wam}%
  \BibitemOpen
  \bibfield  {author} {\bibinfo {author} {\bibfnamefont {M.}~\bibnamefont
  {Czakon}}, \bibinfo {author} {\bibfnamefont {A.}~\bibnamefont {Mitov}}, \
  and\ \bibinfo {author} {\bibfnamefont {R.}~\bibnamefont {Poncelet}},\ }\href
  {\doibase 10.1007/JHEP04(2023)138} {\bibfield  {journal} {\bibinfo  {journal}
  {JHEP}\ }\textbf {\bibinfo {volume} {04}},\ \bibinfo {pages} {138} (\bibinfo
  {year} {2023})},\ \Eprint {http://arxiv.org/abs/2205.11879} {arXiv:2205.11879
  [hep-ph]} \BibitemShut {NoStop}%
\bibitem [{\citenamefont {Gauld}\ \emph {et~al.}(2023)\citenamefont {Gauld},
  \citenamefont {Huss},\ and\ \citenamefont {Stagnitto}}]{Gauld:2022lem}%
  \BibitemOpen
  \bibfield  {author} {\bibinfo {author} {\bibfnamefont {R.}~\bibnamefont
  {Gauld}}, \bibinfo {author} {\bibfnamefont {A.}~\bibnamefont {Huss}}, \ and\
  \bibinfo {author} {\bibfnamefont {G.}~\bibnamefont {Stagnitto}},\ }\href
  {\doibase 10.1103/PhysRevLett.130.161901} {\bibfield  {journal} {\bibinfo
  {journal} {Phys. Rev. Lett.}\ }\textbf {\bibinfo {volume} {130}},\ \bibinfo
  {pages} {161901} (\bibinfo {year} {2023})},\ \bibinfo {note} {[Erratum:
  Phys.Rev.Lett. 132, 159901 (2024)]},\ \Eprint
  {http://arxiv.org/abs/2208.11138} {arXiv:2208.11138 [hep-ph]} \BibitemShut
  {NoStop}%
\bibitem [{\citenamefont {Caola}\ \emph {et~al.}(2023)\citenamefont {Caola},
  \citenamefont {Grabarczyk}, \citenamefont {Hutt}, \citenamefont {Salam},
  \citenamefont {Scyboz},\ and\ \citenamefont {Thaler}}]{Caola:2023wpj}%
  \BibitemOpen
  \bibfield  {author} {\bibinfo {author} {\bibfnamefont {F.}~\bibnamefont
  {Caola}}, \bibinfo {author} {\bibfnamefont {R.}~\bibnamefont {Grabarczyk}},
  \bibinfo {author} {\bibfnamefont {M.~L.}\ \bibnamefont {Hutt}}, \bibinfo
  {author} {\bibfnamefont {G.~P.}\ \bibnamefont {Salam}}, \bibinfo {author}
  {\bibfnamefont {L.}~\bibnamefont {Scyboz}}, \ and\ \bibinfo {author}
  {\bibfnamefont {J.}~\bibnamefont {Thaler}},\ }\href {\doibase
  10.1103/PhysRevD.108.094010} {\bibfield  {journal} {\bibinfo  {journal}
  {Phys. Rev. D}\ }\textbf {\bibinfo {volume} {108}},\ \bibinfo {pages}
  {094010} (\bibinfo {year} {2023})},\ \Eprint
  {http://arxiv.org/abs/2306.07314} {arXiv:2306.07314 [hep-ph]} \BibitemShut
  {NoStop}%
\bibitem [{\citenamefont {Behring}\ \emph {et~al.}(2025)\citenamefont {Behring}
  \emph {et~al.}}]{Behring:2025ilo}%
  \BibitemOpen
  \bibfield  {author} {\bibinfo {author} {\bibfnamefont {A.}~\bibnamefont
  {Behring}} \emph {et~al.},\ }\href {\doibase 10.1007/JHEP09(2025)149}
  {\bibfield  {journal} {\bibinfo  {journal} {JHEP}\ }\textbf {\bibinfo
  {volume} {09}},\ \bibinfo {pages} {149} (\bibinfo {year} {2025})},\ \Eprint
  {http://arxiv.org/abs/2506.13449} {arXiv:2506.13449 [hep-ph]} \BibitemShut
  {NoStop}%
\bibitem [{\citenamefont {Apolin{\'a}rio}\ \emph
  {et~al.}(2025{\natexlab{b}})\citenamefont {Apolin{\'a}rio}, \citenamefont
  {Roux},\ and\ \citenamefont {Zapp}}]{Apolinario:2025fwd}%
  \BibitemOpen
  \bibfield  {author} {\bibinfo {author} {\bibfnamefont {L.}~\bibnamefont
  {Apolin{\'a}rio}}, \bibinfo {author} {\bibfnamefont {C.~L.}\ \bibnamefont
  {Roux}}, \ and\ \bibinfo {author} {\bibfnamefont {K.}~\bibnamefont {Zapp}},\
  }\href@noop {} {\  (\bibinfo {year} {2025}{\natexlab{b}})},\ \Eprint
  {http://arxiv.org/abs/2510.11914} {arXiv:2510.11914 [hep-ph]} \BibitemShut
  {NoStop}%
\end{thebibliography}%
\end{document}